\documentclass{aa}
\usepackage[varg]{txfonts}
\usepackage{graphicx}
\usepackage{natbib}
\usepackage{color}

\newcommand{\Msun}{$\rm M_{\odot}$}
\usepackage[dvipsnames]{xcolor}
\usepackage{subfig}
\usepackage{multirow}
\usepackage{hyperref}

\hypersetup{colorlinks=true,
    linkcolor=blue,
    filecolor=blue,   urlcolor=blue,
    citecolor=blue
    }

\begin{document}

\title{Timing the Milky Way bar formation and the accompanying radial migration episode}
\titlerunning{The bar resonances and metallicity structure of the disc}

\author{Misha Haywood\inst{1}
\and Sergey Khoperskov\inst{2}
\and Valeria Cerqui\inst{1}
\and Paola Di Matteo\inst{1}
\and David Katz\inst{1}
\and Owain Snaith\inst{3}
}

\offprints{M. Haywood, \email{misha.haywood@obspm.fr} et al.}

\institute{GEPI, Observatoire de Paris, CNRS, Universit\'e
  Paris Diderot, 5 place Jules Janssen, 92190 Meudon, France
\and 
Leibniz Institut f\"ur Astrophysik Potsdam (AIP), An der Sternwarte 16, D-14482, Potsdam, Germany
\and
University of Exeter, School of Physics and Astronomy, Stocker Road, Exeter, EX4 4QL, UK
}

\date{Received / Accepted}

\abstract{We derived the metallicity profile of the Milky Way low-$\alpha$ disc population from 2 to 20 kpc from the Galactic centre in 1 Gyr age bins using the astroNN catalogue, and we show that it is highly structured, with a plateau between 4 and 7 kpc and a break at 10-12 kpc. We argue that these features result from the two main bar resonances, the corotation and the outer Lindblad resonance (OLR), respectively. 
We show that the break in the metallicity profile is most visible in stars having 7-8 Gyr, reaching an amplitude of about 0.4 dex, and it is the signpost of the position of the bar OLR. The bar formation was accompanied by an episode of radial migration triggered by it slowing down and it is responsible for spreading old metal-rich stars up to the OLR. The data show that the slowdown of the bar ended 6-7 Gyr ago. Based on numerical simulations that reproduce well the characteristic break observed in the metallicity profile, we argue that this implies that the bar formed in our Galaxy 8-10 Gyr ago. Analysis of the metallicity distribution as a function of radius shows no evidence of significant systematic outward radial migration after this first episode. We argue that the variation of the metallicity dispersion as a function of the guiding radius is dominated by the migration triggered by the bar, but also
that the libration of orbits around the bar resonances induces a mixing that may have a significant impact on the observed metallicity dispersion.
In contrast, the absence of a break in the metallicity profile of populations younger than $\sim$6 Gyr and the flattening of the gradient at younger ages is interpreted as evidence that the strength of the bar has decreased, loosening its barrier effect and allowing the gas and metals on both sides of the OLR to mix, erasing the break. Beyond the OLR, stars younger than 7 Gyr show very small metallicity dispersion, suggesting that no or limited mixing induced by the spiral arms has occurred in the outer disc.
}

\keywords{stars: abundances -- stars: kinematics and dynamics -- Galaxy:
solar neighborhood -- Galaxy: disk -- Galaxy: evolution}

\maketitle

\nolinenumbers

\section{Introduction}\label{sec:intro}

The presence of a bar at the centre of the Milky Way, which had been suspected since the early 1960s \citep{1964IAUS...20..195D,1967IAUS...31..239K,1968Obs....88..254C,1978AJ.....83.1163D}, was finally confirmed by several studies about 30 years later 
\citep{1991Natur.353..140N,1991ApJ...379..631B,1992ApJ...384...81W}. 
Since its discovery, an extensive literature has been dedicated to the study of its shape and orientation \citep[e.g.][]{1999MNRAS.304..512E, 2002MNRAS.330..591B,2012A&A...538A.106R,2010ApJ...721L..28N,2012ApJ...756...22N,2013MNRAS.435.1874W,2015MNRAS.450.4050W,2015MNRAS.448..713P,2017MNRAS.465.1621P,2017MNRAS.471.3988C}, its links to stellar populations \citep{2013MNRAS.430..836N,2013MNRAS.432.2092N,2014A&A...567A.122D,2015A&A...577A...1D,2016PASA...33...27D,2017A&A...606A..47F,2018A&A...616A.180F}, age \citep[e.g.][]{2013A&A...559A..98V,2013A&A...549A.147B,2014ApJ...787L..19N,2016A&A...593A..82H,2017A&A...605A..89B}, its effect on the local kinematics \citep[e.g.][]{1998A&A...335L..61R,2000AJ....119..800D,2001A&A...373..511F,2003AJ....125..785Q,2007A&A...467..145C,2009MNRAS.396L..56M,2009ApJ...700L..78A,2010ApJ...717..617B,2015MNRAS.451..705M,2015MNRAS.452..747M,2016MNRAS.457.1062M,2018MNRAS.477.3945H,2022A&A...663A..38K,2022MNRAS.514..460A}, and on the possible migration of stars in the disc \citep[e.g.][]{2010ApJ...722..112M,2015A&A...578A..58H,2015MNRAS.446..823M,2020A&A...638A.144K,2022ApJ...936L...7C}. However, comparatively few papers have focussed on its large-scale effects other than kinematically. It is not well known, for example, how the bar affects the general structure of the metallicity distribution of stars, or the distribution of the gas or young stellar populations throughout the disc. 
A reason for that has been the difficulty to locate these resonances, due in particular to the uncertainties of the bar rotation pattern speed, with a debate oscillating between studies finding a short, rapidly rotating bar \citep[e.g.][]{2000AJ....119..800D,2014A&A...563A..60A,
2019MNRAS.488.3324F} or, rather, a long, slowly rotating bar \citep[e.g.][]{2013MNRAS.435.1874W,2017MNRAS.465.1621P}.
The most recent measurements, benefiting from the now extensive data available -- \textit{Gaia}, APOGEE, and LAMOST -- seem to converge towards a longer bar and a slower pattern speed \citep[e.g.][]{2018MNRAS.477.3945H,2019MNRAS.488.4552S,2019A&A...626A..41M,2020A&A...634L...8K,2021MNRAS.508..728K,2021MNRAS.500.4710C,2021MNRAS.505.2412C,2022MNRAS.509..844T,2022MNRAS.512.2171C}; however, for more details, readers can refer to \citet{2020MNRAS.497..933H}, \cite{2023MNRAS.520.4779L}, and \cite{2024MNRAS.528.3576V}.

The resonances are the manifestation of the angular momentum redistribution of both stars and gas in a galaxy, and contribute altogether to erase the first-order signatures of the stellar disc formation \citep{2002ARA&A..40..487F}. At the same time, they create new conditions for the next generations of stars by redistributing gas, thus influencing the star formation activity and hence the course of chemical evolution. 
An illustration of this is the formation of outer rings \citep{1996FCPh...17...95B} which are thought to correspond to outer Lindblad resonances (hereafter OLRs) in galaxies, where gas accumulates, due to the torques exerted by the bar,  transferring the angular momentum to the gas and driving it from corotation towards the OLR. The gas outside the OLR also tends to accumulate \citep{1981ApJ...247...77S}. 
As a result of the increased gas density, stars form, creating stellar rings.

The aim of this paper is to search for the possible observational signatures left by the formation and evolution of the bar in the large-scale stellar abundance distribution, and to understand how these signatures can be interpreted using numerical simulations. In the following section, we describe the data used in our study; in Section \ref{sec:metallicity_profile}, we explain how we derived the radial metallicity profiles of the disc as a function of age. In Section \ref{sec:simulations}, we compare the observed distributions with numerical simulations. In Section \ref{sec:discussion} we discuss the implications of our results for the present structure of the disc and its history and in particular the role of the bar. We conclude in the last section.

\section{Data}\label{sec:data}

%%%%%%%%%%%%%%%%%%%%%%%%%%%%%%%%%%%

\begin{figure}
\includegraphics[width=9.5cm]{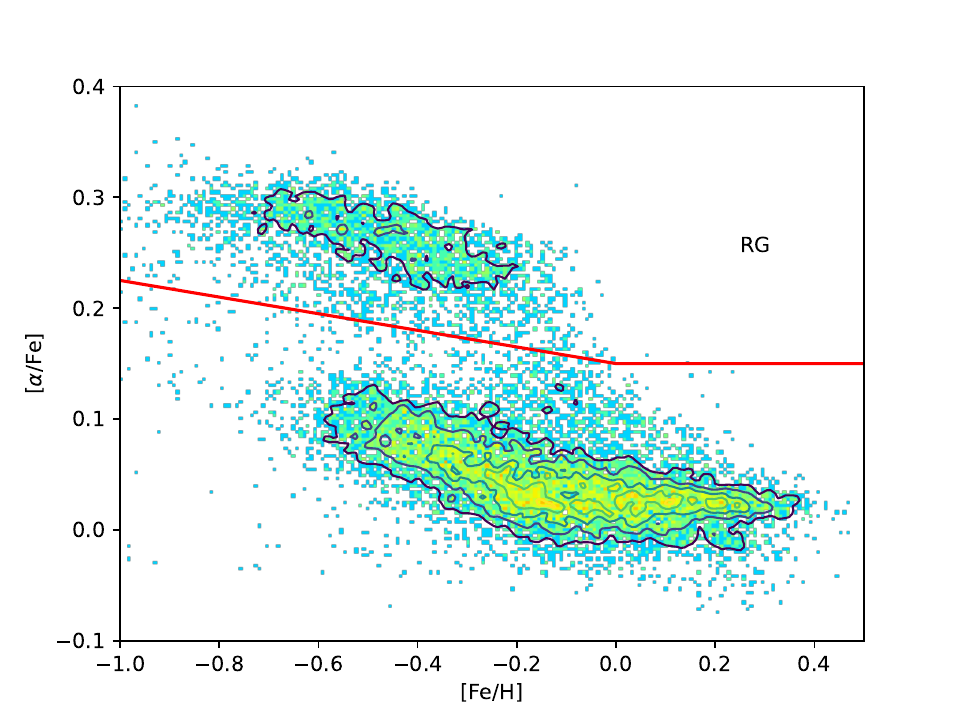}
\caption{[Fe/H]-[$\alpha$/Fe] distributions for red giants within 3~kpc from the Sun. The red lines define our low and high-$\alpha$ samples.}
\label{fig:alphafeh}{}
\end{figure}
%%%%%%%%%%%%%%%%%%%%%%%%%%%%%%%%%%%

\begin{figure*}[ht]
\includegraphics[width=9.5cm]{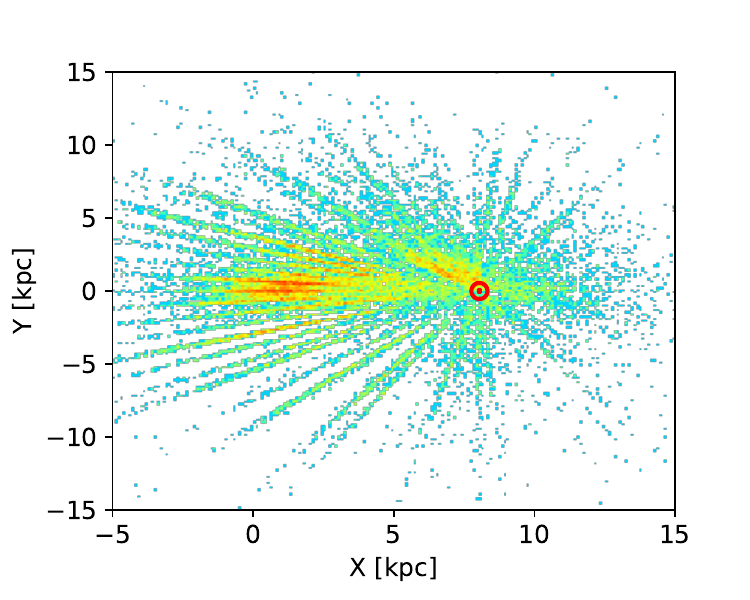}
\includegraphics[width=9.5cm]{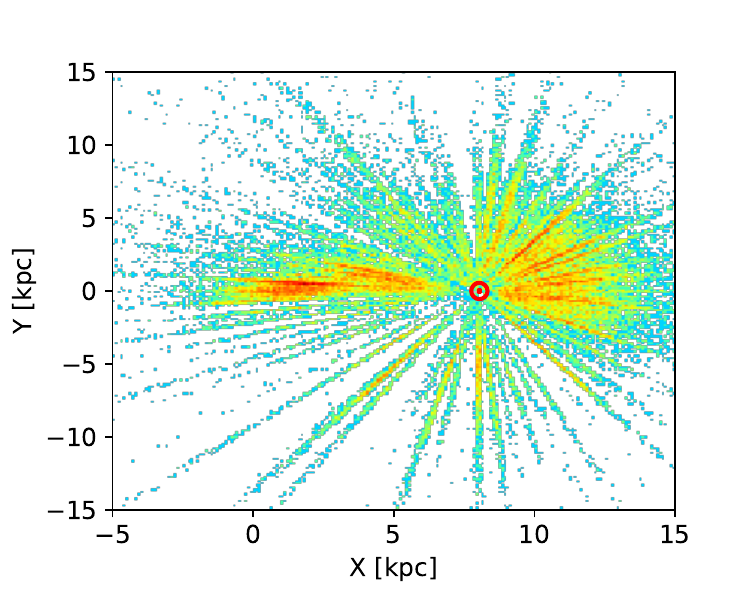}
\includegraphics[width=9.5cm]{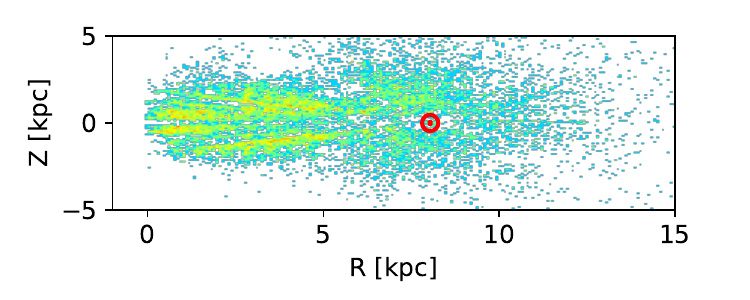}
\includegraphics[width=9.5cm]{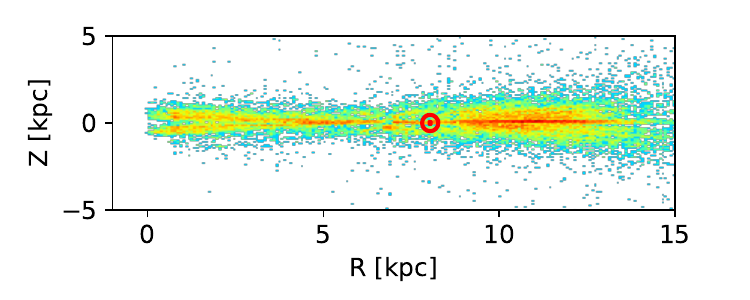}
\caption{XY and RZ stellar density distributions of the red giants in our sample, divided in high $\alpha$ (left) and low $\alpha$ (right).}
\label{fig:xy_rz_distributions}{}
\end{figure*}

We use the APOGEE atmospheric parameters and stellar abundances from the Data Release 17  \citep{2022ApJS..259...35A}. We select 332483 stars flagged as ASPCAPFLAG=0, EXTRATARG=0 and having signal-to-noise ratio greater than 50. 
The resulting selection contains stars with widely different atmospheric parameters (dwarfs, subgiants, red clump, giants).
For reasons detailed in \cite{2023A&A...676A.108C} - and in particular the wide Galactic radius range probed by these stars - we select giants that have log g $<$ 2.2, avoiding red clumps stars, and height above the plane -5$<$z$<$5~kpc, which leaves us with 77248 objects.
The ([Fe/H], [$\alpha$/Fe]) distribution of these stars is shown in Fig. \ref{fig:alphafeh}, the red line shows the adopted separation between high and low-$\alpha$ stars and used throughout the study. [$\alpha$/Fe] abundance ratio is the one provided in the DR17 catalogue.
The X-Y and R-Z plane distributions of these stars, separated in low and high-$\alpha$ populations are shown in Fig.~\ref{fig:xy_rz_distributions}.
 We use a right-handed Cartesian Galactocentric coordinate system (X , Y , Z ), with X pointing towards the Galactic centre, Y in the direction of Galactic rotation and Z towards the north Galactic pole. The density distributions of high- and low-$\alpha$ populations exhibit a notable contrast in their spatial arrangement, as already known: high-alpha stars are predominantly clustered near the Galactic centre, forming a thicker component, whereas low-alpha stars are dispersed more widely throughout the Milky Way (MW) disc. However, they tend to be concentrated very closely to the disc midplane, with some flaring occurring at distances greater than 10-12 kpc.

\begin{figure}
\includegraphics[width=10cm]{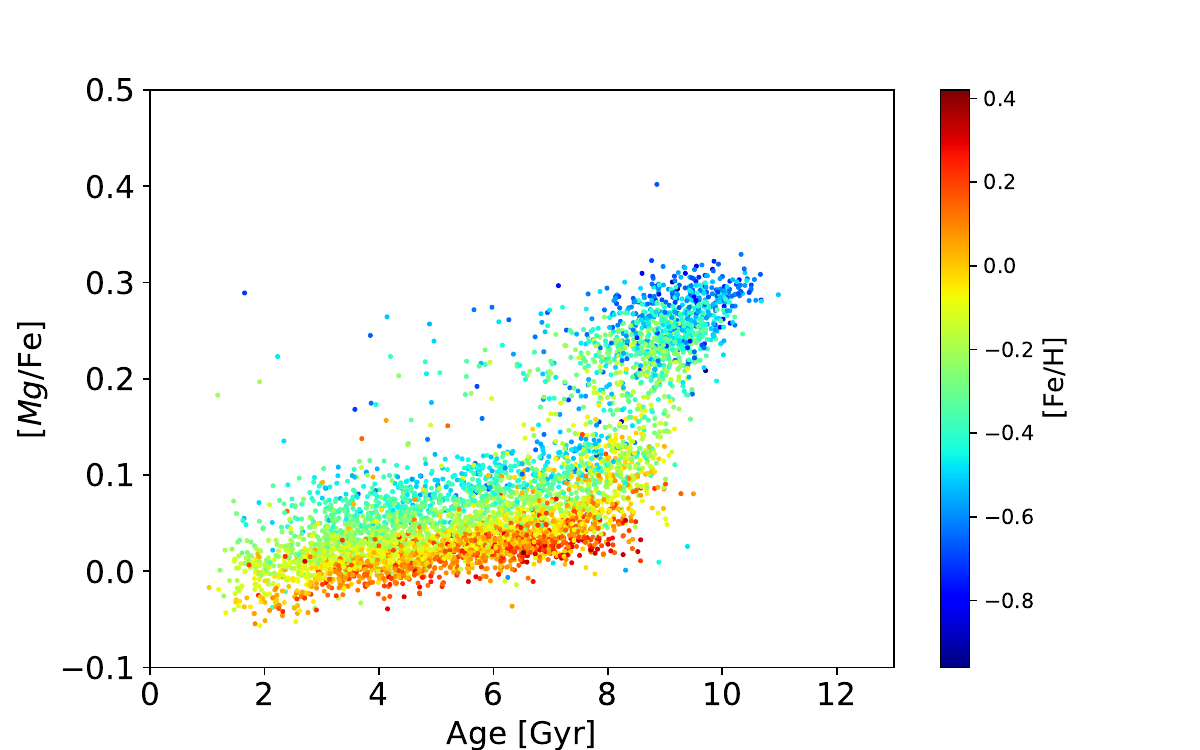}
\includegraphics[width=10cm]{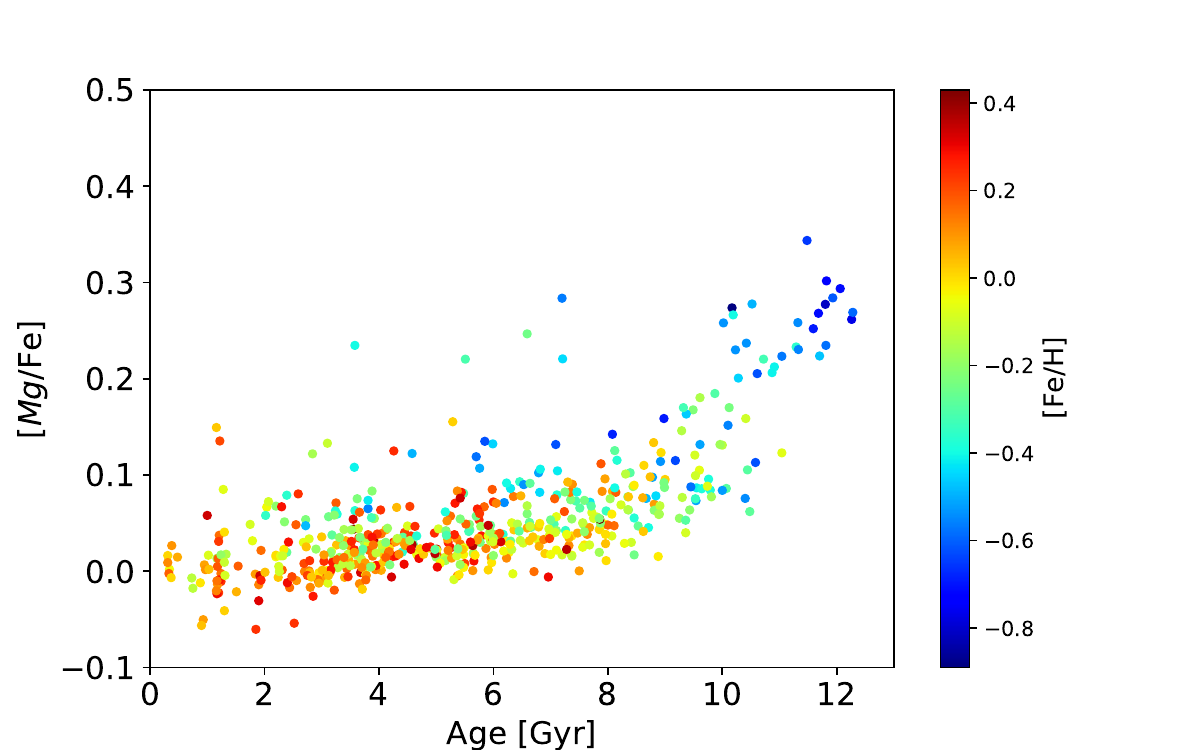}
\caption{Comparison of Age-[Mg/Fe] distributions. Top: Age-[Mg/Fe] distribution for red giants from AstroNN within 3 kpc from the Sun. Bottom: Age-[Mg/Fe] distribution for dwarfs from the Adibekyan sample.}
\label{fig:agealpha}
\end{figure}

We use the age estimates and orbital parameters (including guiding radii) provided by the astroNN value added catalogue \citep{2019MNRAS.483.3255L}\footnote{https://github.com/henrysky/astroNN}. Galactocentric coordinates are calculated assuming the Sun is at 8.18~kpc from the Galactic centre \cite{2019A&A...625L..10G}.
Stellar ages in astroNN are estimated with the procedure explained by \cite{2019MNRAS.489..176M}, using Bayesian neural network trained on asteroseismic ages. 
The median uncertainty reported in \cite{2019MNRAS.489..176M} from applying their procedure to the DR14 is 30\%, while these authors also point out that ages above 10 Gyr are likely to be underestimated, by as much as 3.5 Gyr. 
We check how the age scale provided by astroNN compares with scales based on solar vicinity dwarfs. Figure \ref{fig:agealpha} compares the age-[Mg/Fe] distributions of giants in APOGEE within 3~kpc from the Sun with a sample of dwarfs from \cite{2012A&A...545A..32A}, for which we determined ages using \textit{Gaia} DR3 parallaxes, following the same procedure as in \cite{2013A&A...560A.109H}.
The two plots show similar distributions, with two segments, the flattest and youngest (ages $<8$~Gyr) one made of stars of the thin disc, the steepest and oldest (ages $> 8$~Gyr) one made of stars of the thick disc \citep{2013A&A...560A.109H}. The age scale is different, the extension of the thick disc being limited to 10 Gyr on the astroNN scale, while it goes to 12 Gyr on the dwarf scale. However, the break occurs at a similar epoch ($\sim$ 8 Gyr) being possibly slightly older for the dwarfs.
More importantly, the metallicity structure of the two distributions (metal-poor thin disc objects dominating the upper envelope of the distribution at ages$<$ 8 Gyr), is well visible, indicating that the relative age scale is precise enough to preserve this structure. 
We are therefore confident that the ages provided in astroNN are good enough to study metallicity gradients as a function of age in the thin disc. 

Following other studies \citep[e.g.][]{2023ApJ...954..124I}, we do not correct the metallicity radial profiles for the effects of the selection function. The main reason for this is that we use only local estimates of the metallicity and alpha abundance, and our study is not based on density estimates (which depend on the selection function). 

%%%%%%%%%%%%%%%%%%%%%%%%%%%%%%%%%%%%%%%%
%%%%%%%%%%%%%%%%%%%%%%%%%%%%%%%%%%%%%%%%

%%%%%%%%%%%%%%%%%%%%%%%%%%%%%%%%%%%%%%%%

\section{The metallicity radial profile of the Milky Way disc from APOGEE}\label{sec:metallicity_profile}

Resonances are expected to shape the density, kinematic and chemical structures observed in the Galactic disc \citep[see][]{2022A&A...663A..38K}, and in particular its metallicity profile. For example \cite{2006MNRAS.370.1046V} has shown that at corotation of spiral patterns metallicity gradients are expected to flatten. \cite{2015A&A...578A..58H} has shown that the OLR is able to generate a boundary that limits the outward or inward migration of stars and gas. If this effect can be maintained for a long enough time, it is expected that separate evolution on either side of the OLR could induce a metallicity difference, creating a break in the metallicity profile. 
Using the selection of stars made in the previous section, we now characterize the chemical profile of the Milky Way disc, in search of such signatures. 
All metallicity profiles are obtained using the mode of the metallicity distribution in each distance interval containing at least 50 stars. The mode is obtained from smoothing the metallicity distribution using a kernel density estimation method based on Gaussian functions. The advantage of using the mode of the metallicity distribution is that the metallicity profile is more sensitive to the variation of the main population than using the mean or the median. We refer the reader to the very detailed study by \cite{2023ApJ...954..124I} of the metallicity profile of the disc using APOGEE DR17.

\subsection{General distributions}

%%%%%%%%%%%%%%%%%%%%%%%%%%%%%%%%%%%%%%%%
\begin{figure*}   % alpha_gradient_LH.py  feh_gradient_LH.py
\includegraphics[width=9.5cm]{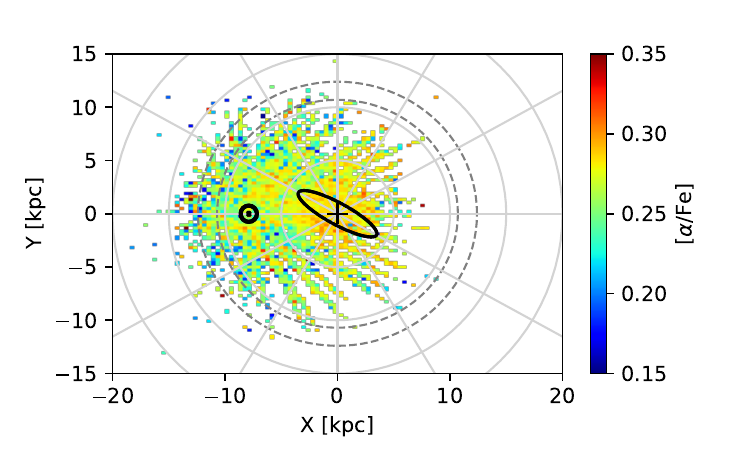}
\includegraphics[width=9.5cm]{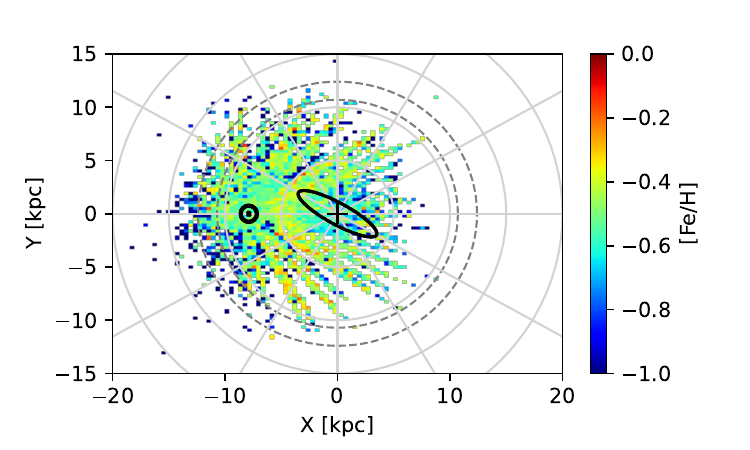}
\includegraphics[width=9.5cm]{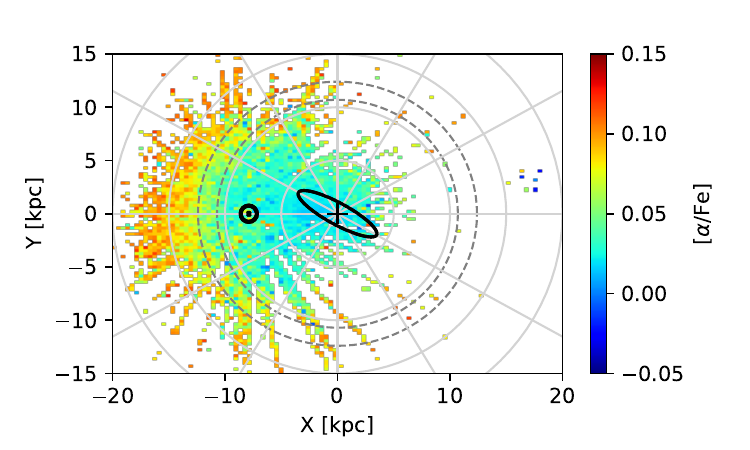}
\includegraphics[width=9.5cm]{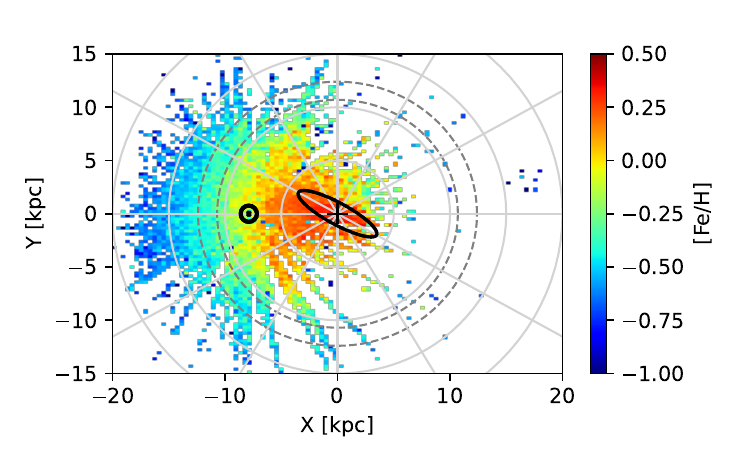}
\caption{[$\alpha$/Fe] and [Fe/H] XY distributions for high-$\alpha$ (top) and low-$\alpha$ (bottom) stars in our sample. 
The two dashed circles bracket the OLR radius according to \cite{2022MNRAS.512.2171C}.}
\label{fig:general_maps}
\end{figure*}

Figure~\ref{fig:general_maps} shows the [$\alpha$/Fe] and [Fe/H] maps, projected onto the X-Y plane, of the high and low-$\alpha$ populations. The two dashed circles show the limits of the interval in which \cite{2022MNRAS.512.2171C} have located the OLR. The high-$\alpha$ population shows essentially featureless distributions. 
The low-$\alpha$ population is more complex, and shows a systematically increasing $\alpha$ abundance ratio and systematic decreasing metallicity as a function of radius. The $\alpha$ abundance ratio is rather uniform within the limit of the OLR, then increases. The decrease in metallicity is more monotonic, but we subsequently see that mixing stars of all ages hides an abrupt change of metallicity at the OLR in the oldest low-$\alpha$ disc populations.
These trends do not seem to depend on the Galactic azimuth covered by the data, which is roughly 90$\degr$. We now quantify the $\alpha$ abundance ratio and metallicity profiles as a function of the guiding radius.

\begin{figure}   % alpha_gradient_LH.py  feh_gradient_LH.py
\includegraphics[width=9.5cm]{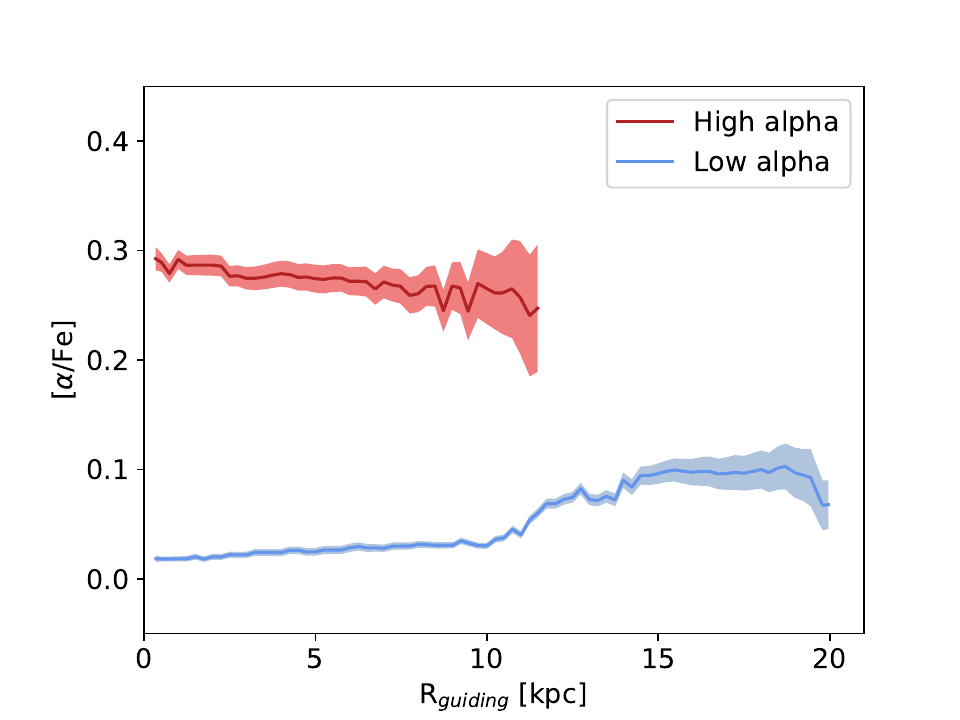}
\includegraphics[width=9.5cm]{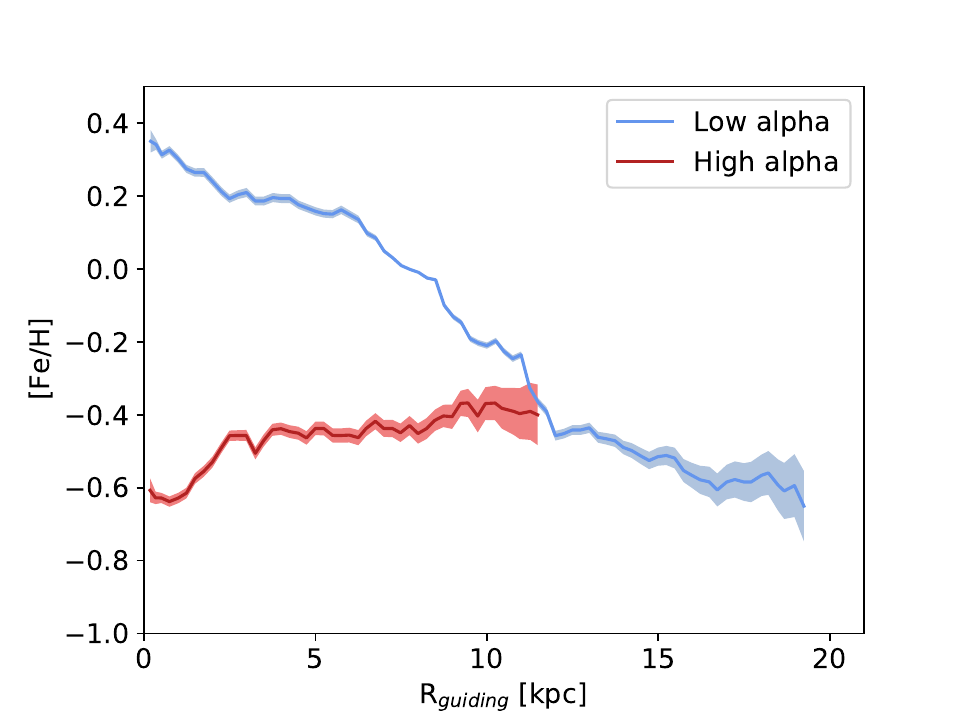}
\caption{[$\alpha$/Fe] and [Fe/H] profiles as a function of R$_{\rm guiding}$ for low (blue curves) and high $\alpha$ (red curves) stars. The shaded area delineates the uncertainty in the profiles derived as described in the text. 
}
\label{fig:feh_alpha_gradient}
\end{figure}

\paragraph{High-$\alpha$ population} We first examine the gradient of the $\alpha$-rich population by selecting stars above the red lines in Fig.~\ref{fig:alphafeh}.  
Fig.~\ref{fig:feh_alpha_gradient} shows the mode of the MDF and $\alpha$DF of these stars  as a function of the distance from the Galactic centre. 
The uncertainty in the metallicity profile is estimated by resampling metallicities and distances 100 times, assuming the errors in these parameters are Gaussians. 
The figure shows an almost flat gradient compatible with no variation in both metallicity and $\alpha$-abundance, see also \cite{2023ApJ...954..124I} for a similar result (their Figure 13). 
These results are also consistent with gradients measured in discs at redshifts compatible with the epoch of the thick disc formation (z$>$1.5), see \cite{2022MNRAS.511.1667T}, and which show that the gradients are essentially flat at these epochs.
A natural explanation for these flat gradients is that of a homogeneous chemical evolution with age in the thick disc. There is indeed a growing evidence of tight age-chemistry relations in this population \citep[e.g.][]{2013A&A...560A.109H,2015A&A...579A...5H,2022Natur.603..599X} and the present sample makes no exception. For example, stars selected to have  0.2$<$[$\alpha$/Fe]$<$0.25 have an age dispersion of only 1.0~Gyr. Removing stars with age$<$~7Gyr, which are not objects of uncertain age determination, but most probably stars that have been rejuvenated by mass transfer \citep[see][]{2023A&A...676A.108C}, the dispersion in age decreases to 0.63~Gyr. Thus APOGEE also suggests that samples covering the entire radial range of the thick disc, have homogeneous chemical properties at any given time.
Samples defined at a given radius, such as the one explored in \cite{2013A&A...560A.109H}, are made of thick disc stars born at various Galactocentric radii, thus tight age-chemistry relations also imply synchronized chemical evolution and good mixing of chemical species. It is thus natural that the radial gradient of the thick disc is flat. 
We note that, on the contrary, \cite{2023MNRAS.525.2208R} find that the metallicity radial gradient of this population must have been steep. This is an effect of the limited range of stellar ages of the thick disc stars in their observed sample: if thick disc stars are restricted to a narrow age range, the only way to explain a large metallicity spread is to assume a steep metallicity gradient of the ISM from which these stars were born.

\paragraph{Low-$\alpha$ population} Fig.~\ref{fig:feh_alpha_gradient} also shows the mode of the MDF and $\alpha$DF of the low-$\alpha$ stars as a function of the distance from the Galactic centre, illustrating a structured metallicity radial profile with several changes of slopes \citep[see also][]{2021A&A...655A.111K}. Two main breaks are visible at R$_{\rm guiding}$ $\sim$ 6~kpc and 12~kpc, delineating three main intervals. 
The change in slope at R$_{guiding}>$11-12~kpc follows a steep drop in metallicity of almost 0.2 dex that occurs within 1~kpc, between 11 and 12~kpc. We  show in the next section that this break is in fact partially smoothed out by the mix of young and old stars in the sample, the break being mostly present in stars older than about 5 Gyr.

\begin{figure}   % alpha_gradient_LH.py  feh_gradient_LH.py
\includegraphics[width=9.5cm]{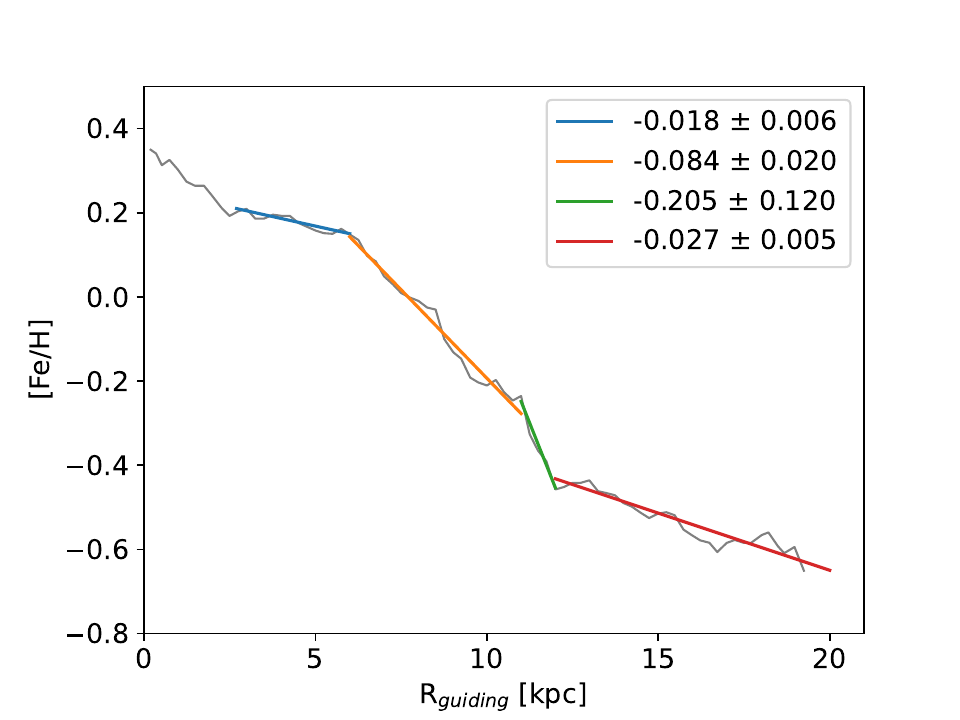}
\includegraphics[width=9.5cm]{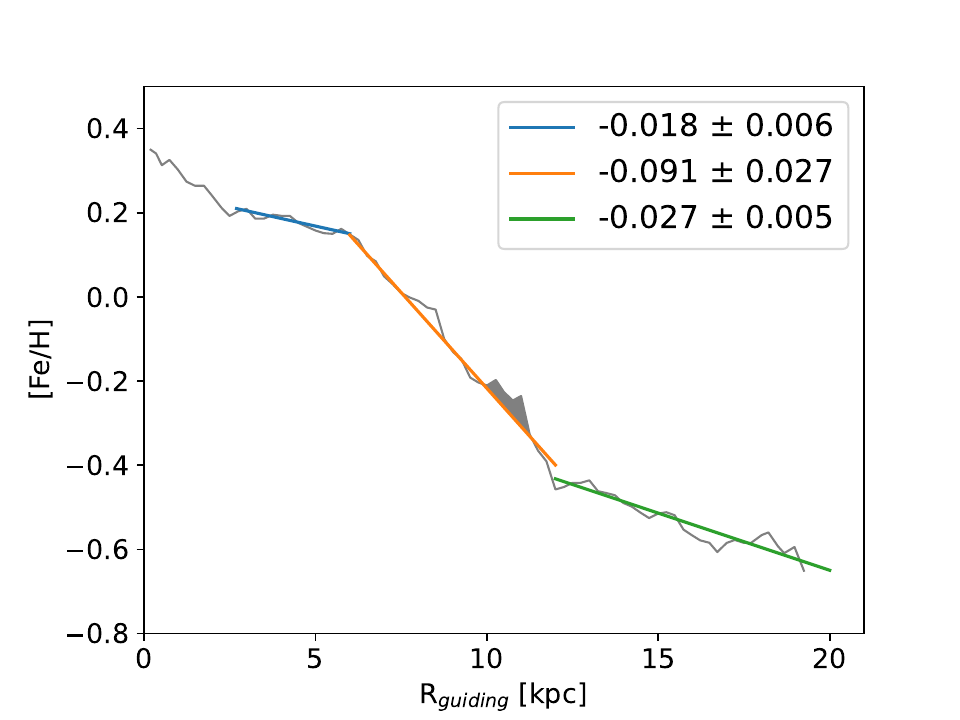}
\caption{Fits to radial metallicity profiles of the low-$\alpha$ population in different Galactocentric intervals. Measured gradients are given on each plot. On the bottom plot, the fit between 6 and 11~kpc is prolonged to 12~kpc. The grey area emphasizes the bump in the metallicity profile due to the accumulation of metal-rich stars up to the OLR. 
}
\label{fig:feh_alpha_gradient_fit}
\end{figure}

A piecewise linear fit to the metallicity gradient (Fig. \ref{fig:feh_alpha_gradient_fit}) gives a gradient of -0.024$\pm$0.008 between 2.8 and 6~kpc and -0.086$\pm$0.02 dex.kpc$^{-1}$ between 6 and 11~kpc, then after a sharp break at 11-12kpc, the gradient flattens out again at -0.024~$\pm$0.0045~dex.kpc$^{-1}$ at larger distances. The value obtained between 6 and 11 kpc is steeper than most values found in the literature. For example \cite{2014A&A...566A..37G} found -0.060$\pm$0.002 $^{-1}$, \cite{2022A&A...659A.167R} found -0.052$\pm$0.002 dex.kpc$^{-1}$ with Classical Cepheids, \cite{2022Univ....8...87S} found -0.064$\pm$0.007 $^{-1}$ on open clusters, \cite{2023A&A...674A..38G} found -0.056$\pm$0.007 $^{-1}$ on field stars. 
These variations are easily explained by the radial extent and age of the tracers used. For example, if we fit our sample over the radial range 7 to 20 kpc, we find -0.053$\pm$0.007 dex.kpc$^{-1}$, very close to the same value found by \cite{2023A&A...674A..38G}. However, Fig. 13 in that study shows a change in the slope of the metallicity profile at R$\sim$ 11~kpc, and a fit restricted to the distance range [6,11] kpc would yield a higher slope value, closer to the one derived here. Also, younger (age$<$ 4 Gyr) tracers give shallower gradients, as we show in the next subsection, explaining the difference between the value obtained with our sample (which contains stars of all ages) and those obtained with Cepheids or open clusters.

The breaks at R$_{guiding}\sim$6 and 11 kpc correspond to the location of the corotation and OLR of the bar mentioned in the introduction. We comment on this in the next section. 
The break at $\sim$11~kpc in the metallicity gradient of the disc has been known since at least \cite{1997AJ....114.2556T} (see also 
\cite{2004AJ....128.1676C,2007A&A...476..217C}, \cite{2005AJ....130..597Y} and \cite{2008A&A...488..943S}) using open clusters. More recently \cite{2020AJ....159..199D} find a slope change at about 14 kpc, while \cite{2022Univ....8...87S} advocate for a change at 12~kpc, while \cite{2022A&A...663A..38K} find a break at 9-10~kpc. We discuss these apparent discrepancies in the next subsection. 
Fig. \ref{fig:feh_alpha_gradient_fit} shows the metallicity profile with a fit made between 6 to 9.7 kpc, but extended to 12~kpc, to emphasize that the metallicity profile in this region can be modelled by a linear fit plus a bump generated by a surplus of metal-rich stars at a radius where the mean metallicity is much lower, as will be shown in the next subsection.

%%%%%%%%%%%%%%%%%%%%%%%%%%%%%%%%%%%%%%%%

\begin{figure} % feh_gradient_age.py 
\includegraphics[width=9.5cm]{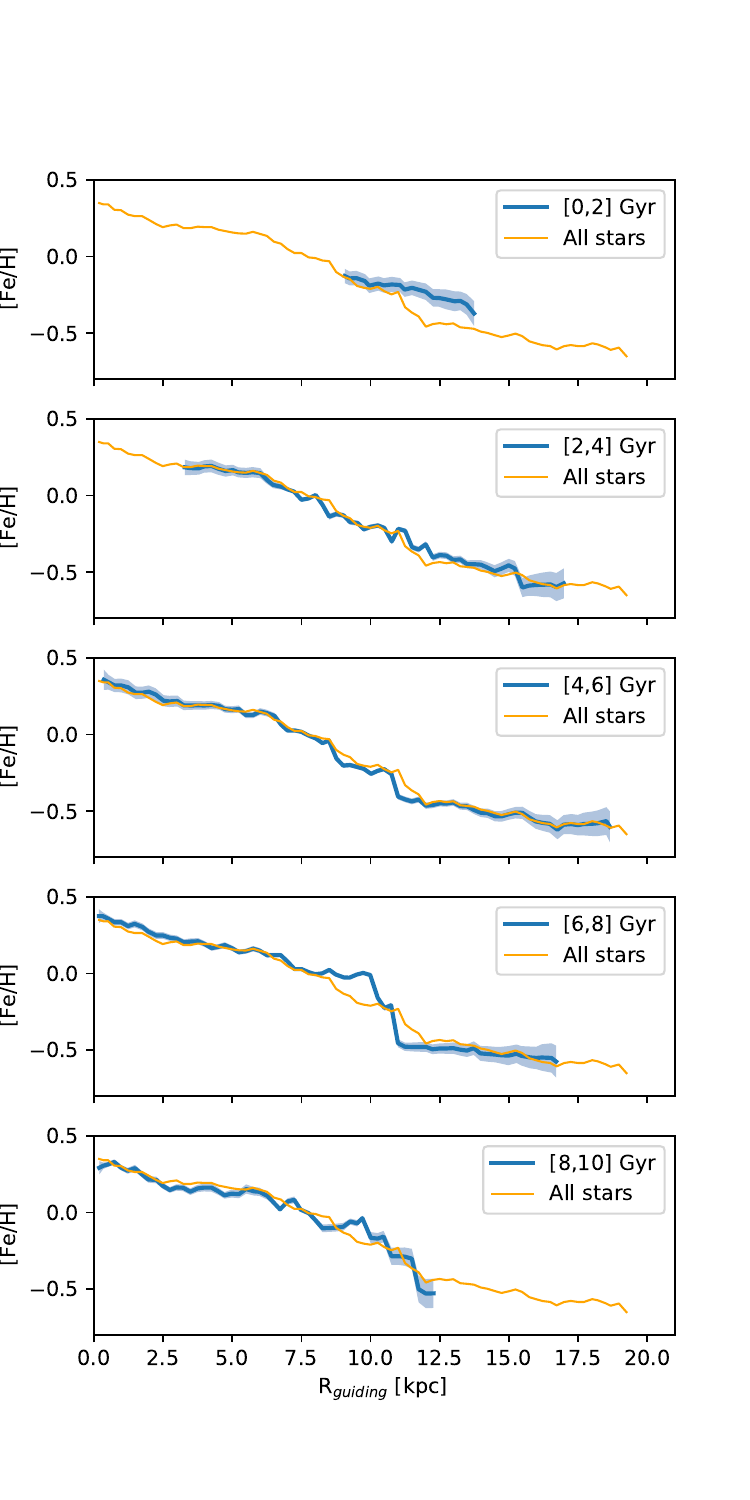}
\caption{Metallicity gradients (in blue) in different age intervals indicated in gigayears in each plot, superimposed on the general gradient (all ages) for the low-$\alpha$ population.}
\label{fig:FeH_gradient_age}
\end{figure}

%%%%%%%%%%%%%%%%%%%%%%%%%%%%%%%%%%%%%%%%
\subsection{Radial profiles as a function of age}

Figure \ref{fig:FeH_gradient_age} shows the radial metallicity profile for stars in 2~Gyr age intervals from 0 to 10 Gyr.
Most of the stars in the age interval from 0 to 2 Gyr are between 9 and 14~kpc. Within this range, a linear fit to the metallicity profile gives a gradient of -0.042 $\pm$0.010 dex.kpc$^{-1}$, which is compatible with the values mentioned previously for gradients measured on classical Cepheids. 

In the age interval from 2 to 4 Gyr, the radial range is much larger, from about 3 to 17 kpc. Excluding the flattened part at R$<$ 5kpc, a linear fit to the observed profile yields a gradient of -0.066 $\pm$ 0.01 dex.kpc$^{-1}$.
Restricting the distance range to 6-12 kpc yields even steeper gradients -0.079$\pm$ 0.016 dex.kpc$^{-1}$, and -0.087$\pm$ 0.023 dex.kpc$^{-1}$ between 6 and 10 kpc. 
The increase of the gradient in the age intervals 0-2 to 2-4 Gyr is compatible with the results obtained on open clusters. For example, \cite{2022MNRAS.509..421N} shows an increase in the slope of the gradient from -0.054 at ages$<$0.4 Gyr to -0.062 dex.kpc$^{-1}$ between 1.9 and 4 Gyr, while \cite{2020AJ....159..199D} obtained -0.048 dex.kpc$^{-1}$ at ages$<$0.4 Gyr and -0.066 between 0.8 and 2 Gyr, and \cite{2023A&A...669A.119M} found an increase from -0.038 between 0.1 and 1 Gyr, to -0.063 between 1 and 3 Gyr and -0.084 dex.kpc$^{-1}$ above 3 Gyr.
We note that although the flattening at R$_{\rm guiding}$ $<$ 7~kpc seen in the general distribution is visible between 2 and 4 Gyr, the break observed at 10-12 kpc in Fig. \ref{fig:feh_alpha_gradient} is essentially absent. 
This is also the case for the youngest (age$<$2 Gyr) open clusters. For example, the step is not visible in \cite{2020AJ....159..199D} (their Fig. 12) and \cite{2022MNRAS.509..421N} (their Fig. 7), where the metallicity decreases monotonically up to 14 kpc, as is the case for the Cepheids \citep{2022A&A...659A.167R,2022MNRAS.510.1894K}.\\

%\subsubsection{Ages $>$ 4 Gyr}
In the 4 to 6 Gyr age range, the break becomes visible at 11~kpc on field giants. Open clusters also show a break, but only at age $>$~2Gyr. For example, on Fig. 7 from \cite{2022MNRAS.509..421N}, a step of about -0.3 dex is clearly visible between 9 and 10kpc. \cite{2020AJ....159..199D} and \cite{2022AJ....164...85M} also show a similar step at ages $>$2 Gyr (see also \cite{2023A&A...669A.119M}, their Fig. 14).
Given that the astroNN ages of young stars may be overestimated according to \cite{2019MNRAS.489..176M}, we may be seeing the same feature in both clusters and field stars. Looking at the plots of the age intervals 6-8 and 8-10 Gyr in Fig. \ref{fig:FeH_gradient_age} we see that this break, which is barely visible in open cluster data (also due to lower statistics), becomes increasingly important in field stars older than 4 Gyr.

The amplitude of the step is the largest at 6-8 Gyr, with a jump in metallicity of almost 0.45 dex between 10 and 11.3 kpc. This is also visible in the age range between 8 and 10 Gyr. 
The break is therefore a feature that characterizes the distribution of old stars, and is dominant at ages $>$ 6 Gyr.
In the 6-8 age range, gradients become flatter on either side of the break, at -0.04 dex.kpc$^{-1}$ at R$<$~10 kpc, and -0.013 dex.kpc$^{-1}$  beyond 11.3~kpc and up to 18 kpc. We subsequently see that this trend is due to the radial migration of metal-rich stars at this epoch which increases the mean metallicity just before the OLR.
Although the break is already visible in the 4-6 Gyr age range, it is possibly enhanced by a contamination from older objects due to uncertainties in the age determination. 
Above 8 Gyr, the break is still visible, but for older, low-alpha stars, the statistics become sparser at radii greater than about 11-12kpc. 

\begin{figure*}
\includegraphics[width=9.25cm]{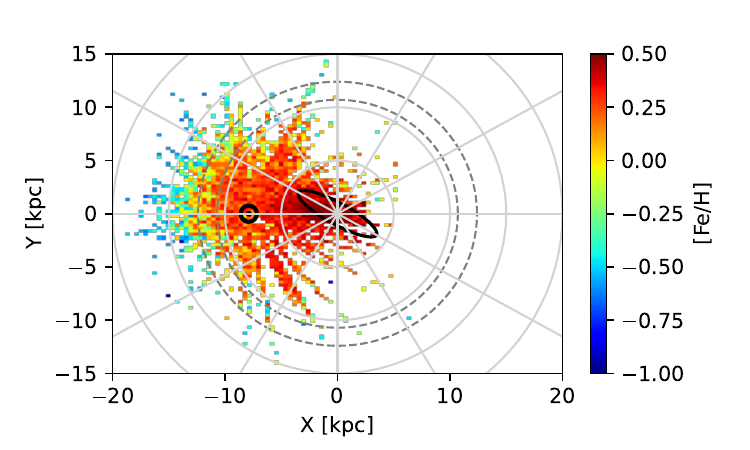}
\includegraphics[width=9.25cm]{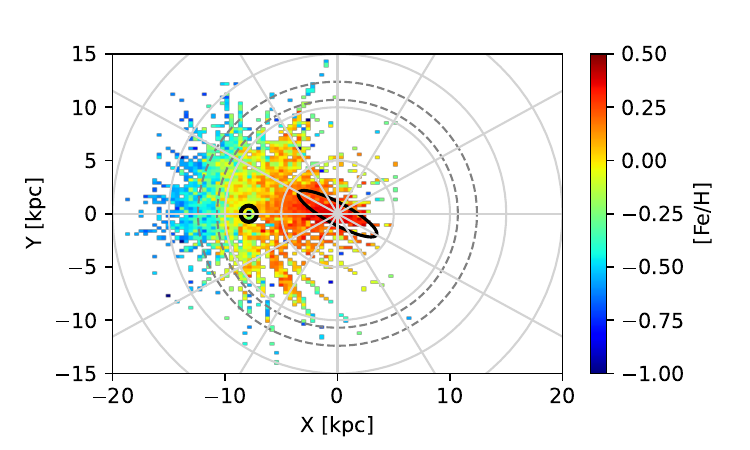}
\caption{Metallicity maps.
Left: XY distributions of the maximum metallicity reached in each pixel by the MDF of stars in the age range 6-8 Gyr.
Right: XY distribution of the mode of the metallicity distribution for the same stars.
}
\label{fig:low_alpha_xy_fehMax}
\end{figure*}

Figure~\ref{fig:low_alpha_xy_fehMax} shows that the break is azimuthally uniform on the angle covered by the APOGEE survey.
The plot on the left shows the maximum metallicity measured on the MDF in each pixel of the XY map for stars in the age range 6-8 Gyr. It illustrates that metal-rich stars have populated the disc uniformly up to the radius of the OLR, with a sharp transition to lower metallicities in the interval where the OLR is located according to \cite{2022MNRAS.512.2171C}. The plot on the right shows the mode of the MDF for the same sample. It illustrates that the break in metallicity is also independent of the azimuthal angle covered by the data (almost 50$^\circ$). The break is clearly visible by the change of colour from yellowish (around solar metallicity) pixels within the inner radius of the OLR ($\sim$10.7~kpc) to cyan in between the two OLR radii ([Fe/H]$\sim$-0.3) to blue ([Fe/H]$\sim$-0.5) outside.

%%%%%%%%%%%%%%%%%%%%%%%%%%%%%%%%%%%%%%%%
%%%%%%%%%%%%%%%%%%%%%%%%%%%%%%%%%%%%%%%%

\begin{figure}
\includegraphics[width=9.5cm]{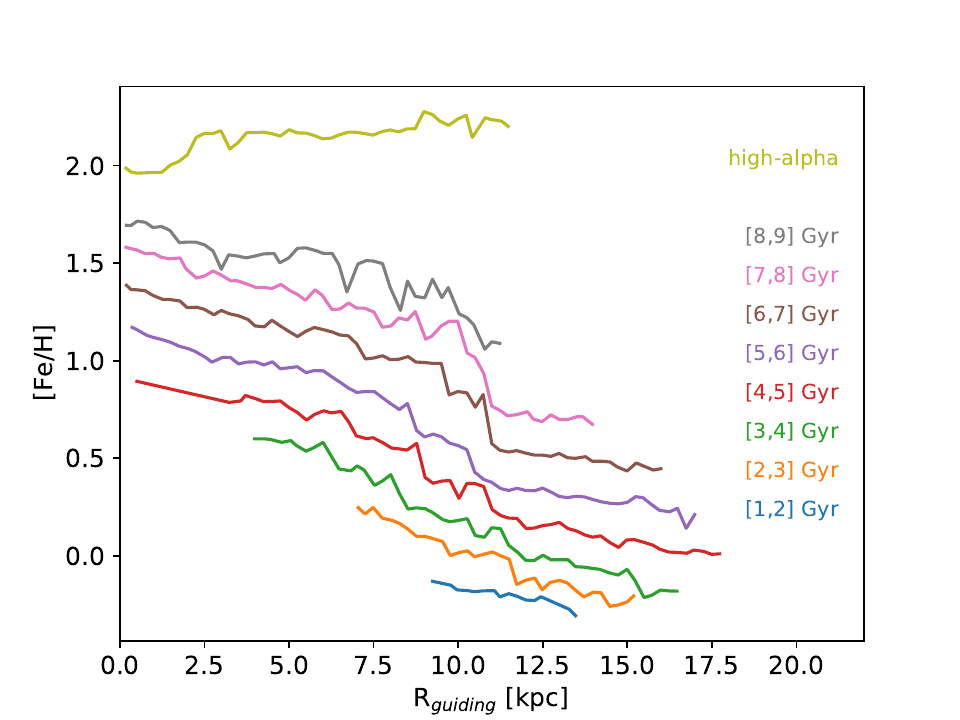}
\includegraphics[width=9.5cm]{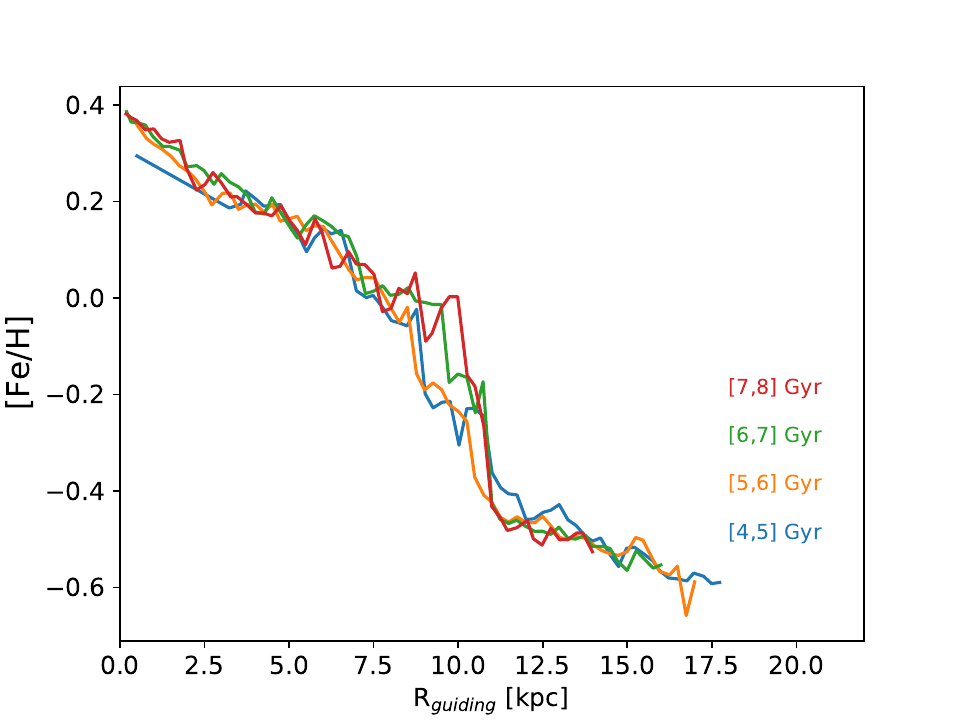}
\caption{Metallicity profiles versus age. Top: Peak metallicity as a function of the guiding radius for low-$\alpha$ stars in different age intervals and high-$\alpha$ stars (upper curve).
Curves have been shifted by +0.2 dex starting after the youngest bin. Bottom: Metallicity profiles for stars from 4 to 8 Gyr in the step of 1 Gyr.}
\label{fig:FeH_gradient_with_age}
\end{figure}

%%%%%%%%%%%%%%%%%%%%%%%%%%%%%%%%%%%%%%%%

Figure \ref{fig:FeH_gradient_with_age} (top) shows a more detailed view of the metallicity profile in 1 Gyr age intervals from 1 to 9 Gyr, complemented by the profile for high-$\alpha$ stars. The vertical metallicity scale is shifted by 0.2 dex to separate the different curves. It can be seen how the break evolves, reaching a maximum amplitude of $\sim$0.4 dex between 6 and 8 Gyr. 
Between 8 and 9 Gyr, the break becomes shallower, but also more radially confined. In the high-$\alpha$ subsample, where the age distribution is dominated by stars between 8 and 10 Gyr, the break is no longer visible. 
Figure \ref{fig:FeH_gradient_with_age} (bottom) shows the strong similarity of the profiles among the oldest populations (from 4 to 8 Gyr). This is particularly true at R$_{guiding}<$ 7-8 and $>$ 11~kpc. 
In between these values, the metallicity break is steeper in the oldest bins (6-8~Gyr) than in the youngest (4-6~Gyr) bins, in all cases, the drop in metallicity is surprisingly large at 0.4-0.5 dex between 8-9 and 11-12~kpc. The step is slightly less steep in the 4-6 Gyr age bins, the metallicity starting to fall at $\sim$8.6~kpc, than for stars in the 6-8 Gyr age range, where the metallicity falls between 9.5 and 10~kpc, corresponding to gradients of -0.2 and -0.3 dex.kpc$^{-1}$. 

The break found at 10-12~kpc is absent in previous studies using APOGEE data. This is because the metallicity indicator usually adopted to perform this kind of analysis is the mean or the median. We show in Figs~\ref{fig:feh_gradient_age_mode} and \ref{fig:median_or_mode} that the mean and the median have a tendency to smooth out structures in the metallicity profiles, the reason for which we adopted the mode. Apart from this aspect, the metallicity profile evolution obtained here is in agreement with other studies. For example, \cite{2017A&A...600A..70A,2023A&A...678A.158A,2023MNRAS.526.2141W} find a gradient of the order of about -0.06 to -0.07 dex.kpc$^{-1}$ for stars younger than 4 Gyr, that flattens to about -0.04 dex.kpc$^{-1}$ for stars older than 6 Gyr. This is what we find for stars within 10~kpc. This is also consistent with evolution of the metallicity profile shown in \cite{2023ApJ...954..124I} of roughly an increase by a factor of 2 of the metallicity gradient between 6 and 10~kpc between stars older than 8~Gyr and those younger than 4~Gyr.

We provide an interpretation of the time variation of these profiles in Section \ref{sec:discussion}.

%%%%%%%%%%%%%%%%%%%%%%%%%%%%%%%%%%%%%%%%
\subsection{Metallicity dispersion and skewness}\label{sec:dispersion_and_skewness}

%%%%%%%%%%%%%%%%%%%%%%%%%%%%%%%%%%%%%%%%
\begin{figure}
\includegraphics[width=9.cm]{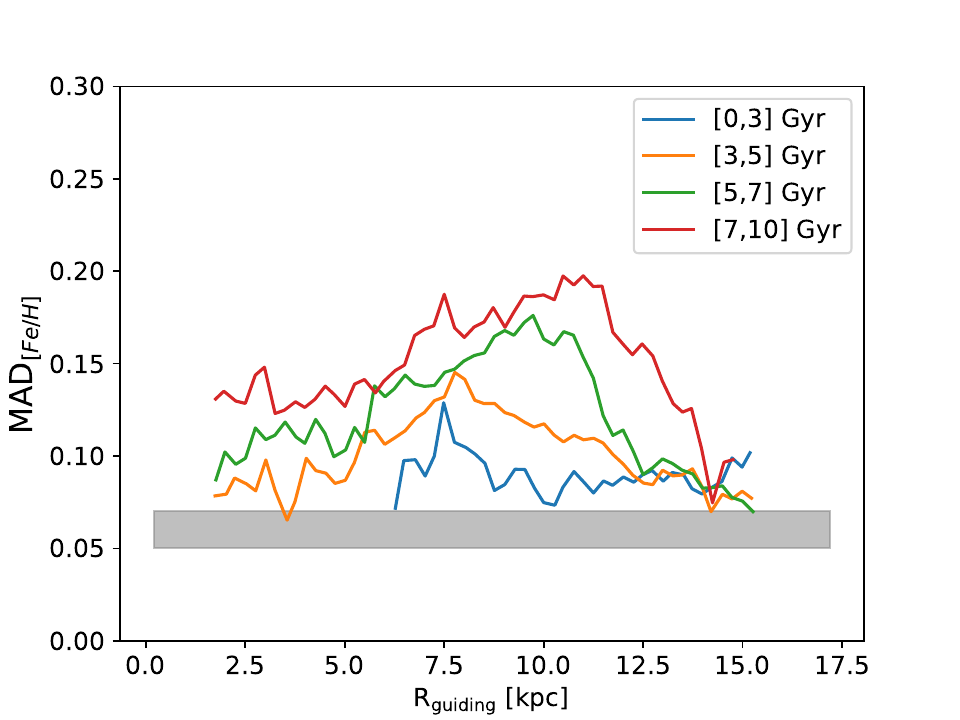}
\includegraphics[width=9.cm]{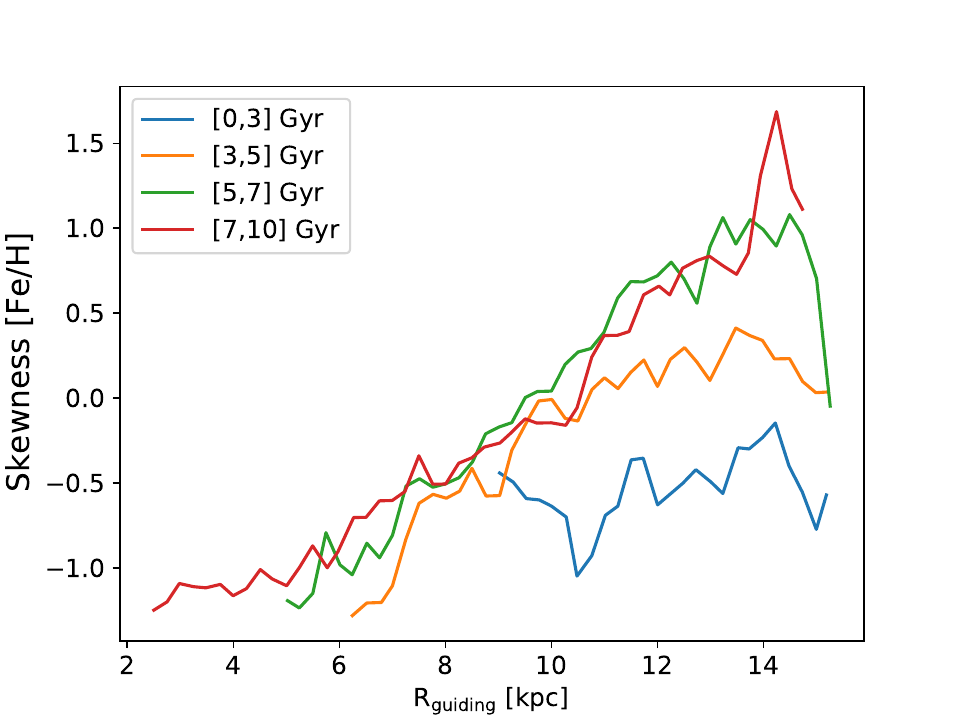}
\caption{Median absolute deviation (top) and skewness (bottom) of the metallicity distributions as a function of the guiding radius. The area in grey on the top plot shows the estimated MAD for young objects -- see Appendix~\ref{B}.}
\label{fig:dispersion_skewness}
\end{figure}
%%%%%%%%%%%%%%%%%%%%%%%%%%%%%%%%%%%%%%%%

Figure~\ref{fig:dispersion_skewness} shows the median absolute deviation (MAD) and skewness of the metallicity distributions, in bins of 0-3, 3-5, 5-7, and 7-10 Gyr, grouping stars in larger age intervals to get better estimates of these quantities. The MAD is used to limit the effect of outliers in the measurement of the dispersion.

The curve for stars younger than 3 Gyr is limited to about 6~kpc, due to the lack of young stars within the bar corotation.
Within this limit (R$<$6~kpc), the MAD for the age bins 3-5 and 5-7 Gyr increases slightly with age.
Beyond corotation (R$>$6~kpc), the four curves clearly separate in increasing order of MAD with age.
The lowest curve (0-3~Gyr) varies between 0.08 and 0.09 dex for most of the radius range, except for a peak near 7.5~kpc. In Appendix~\ref{B}, we summarize estimates of the MAD measured on Cepheids, young giants, open clusters, and neutral gas from samples from the literature.
These estimates give MAD values of $\sim$0.05-0.07~dex for tracers at distances less than 2~kpc.  
The dispersions measured on stars in the age range 0-3~Gyr are therefore a bit higher than these values. 
Finally, it is interesting to observe that the MAD for the age intervals 3-5 and 5-7~Gyr converge, beyond R$\rm _g$$\sim$~12~kpc, to the dispersion observed on stars in the age range 0-3~Gyr. 
A detailed interpretation of this plot is given in Section \ref{sec:migration}.

As observed on the bottom plot of Fig.~\ref{fig:dispersion_skewness}, the skewness behaviour found in \cite{2015ApJ...808..132H} is confirmed, with similar variation with radius. The interesting addition is its variation with time, with the youngest stars (0-3~Gyr) having a noisy (due to the statistics) but more or less constant negative skewness. While the skewness remains more or less similar for stars older than 3 Gyr within the OLR ($<$ 10~kpc), it then separates with increasing (positive) skewness. The two oldest age bins (5-7, 7-10~Gyr) have the same skewness variation with radius.
The fact that the skewness is negative for the 0-3 Gyr stars, which may have been the least affected by migration, suggests that the negative skewness may not be an effect of migration. On the contrary, the positive skewness may be an effect of migration induced by the bar, see Section~\ref{sec:simulation_results}.

\subsection{Metallicity-guiding radius distributions}

The mode (or the median, as is often used for gradients) contains only part of the information and the full distribution of metallicity as a function of the guiding radius can also be useful in understanding which stars are responsible for the different features, and in particular for the break at $\sim$ 11 kpc.

%%%%%%%%%%%%%%%%%%%%%%%%%%%%%%%%%%%%%%%%
\begin{figure*}
\includegraphics[width=4.4cm]{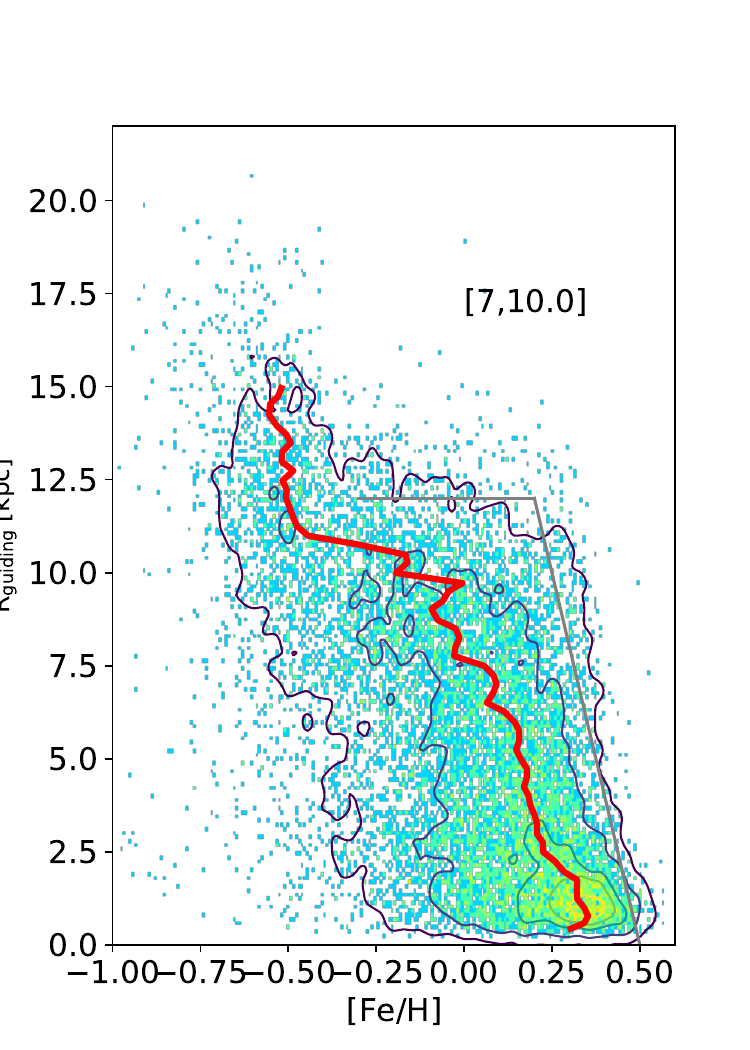}
\includegraphics[width=4.4cm]{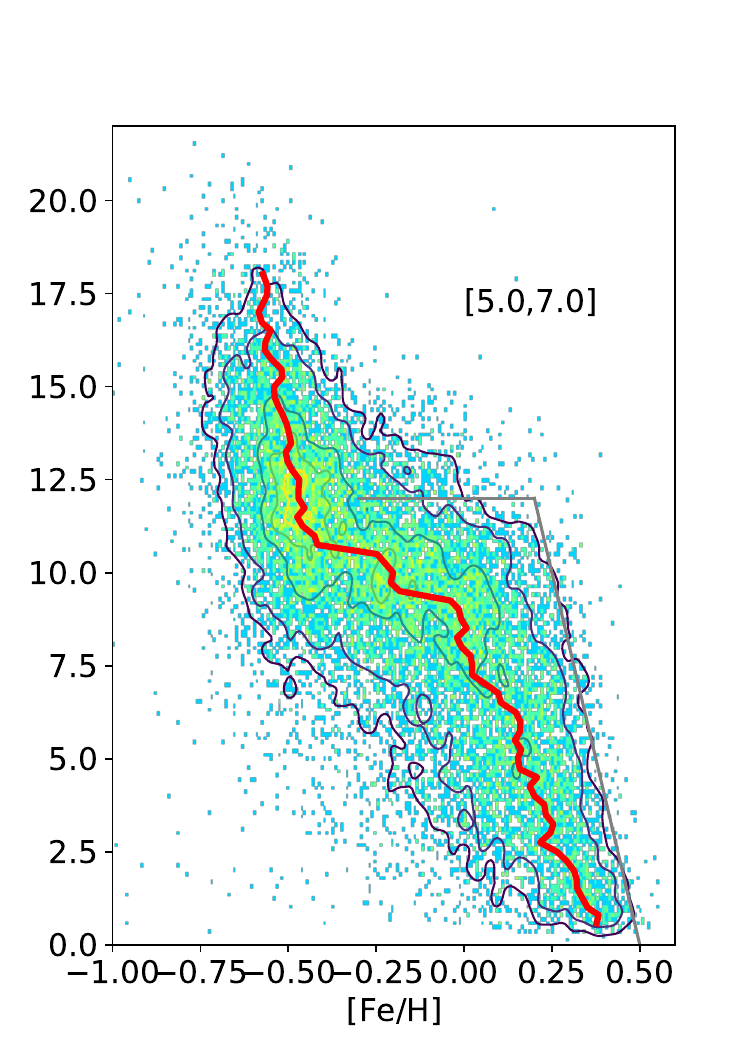}
\includegraphics[width=4.4cm]{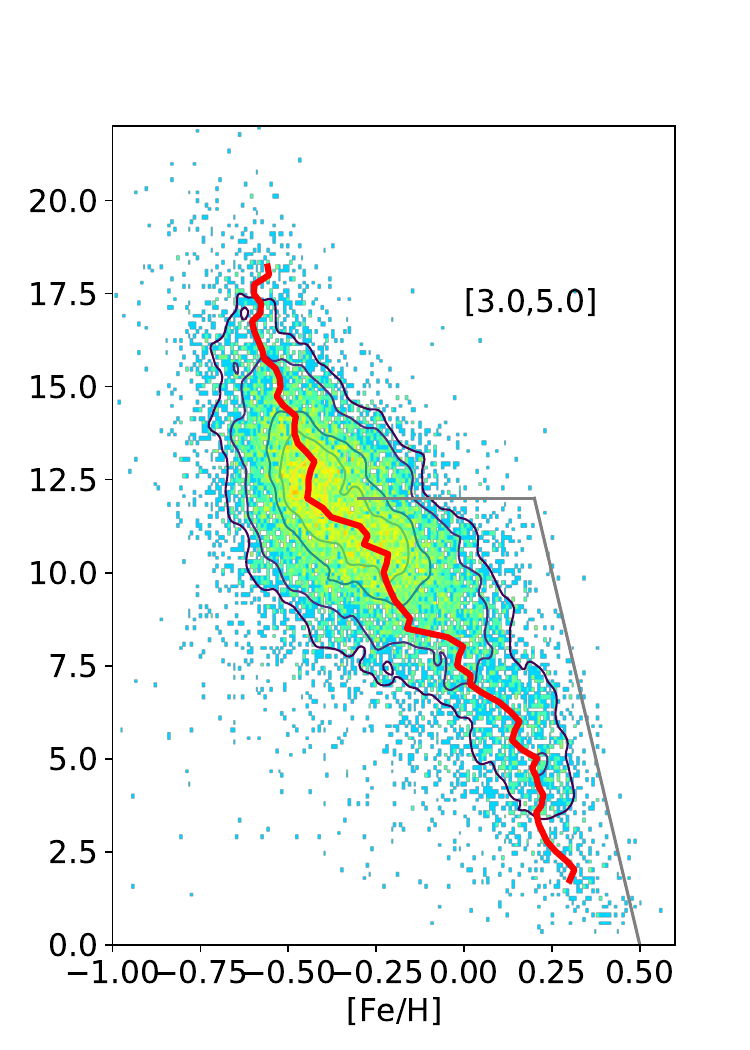}
\includegraphics[width=4.4cm]{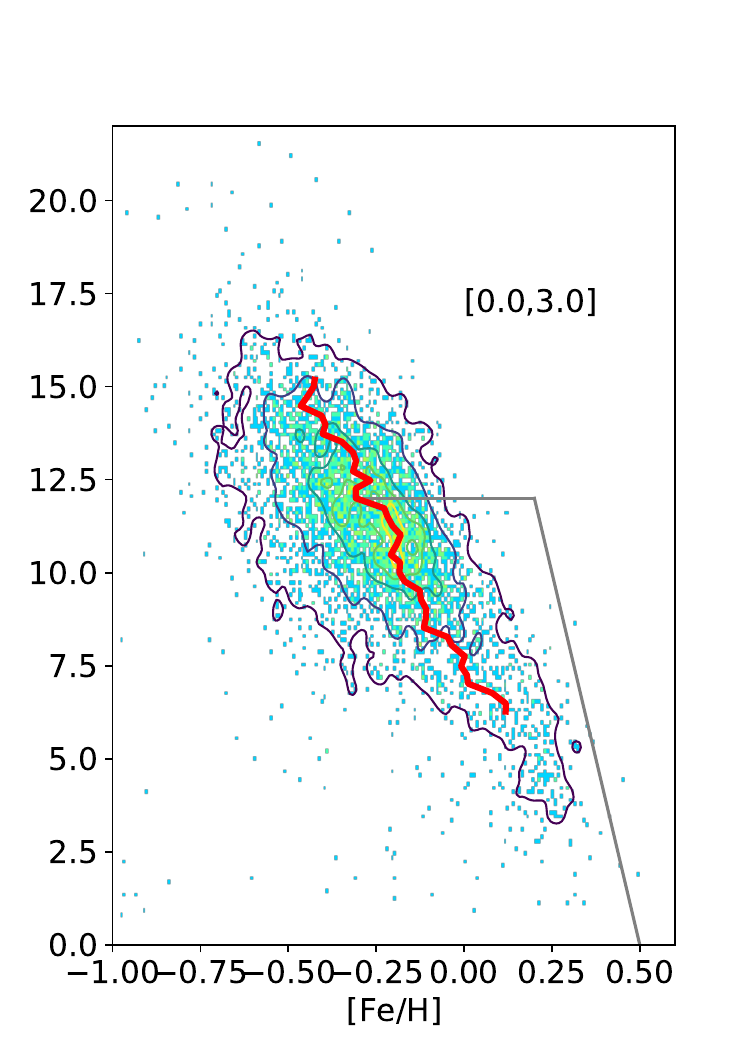}
\caption{R$_{\rm guiding}$ versus metallicity in different age intervals indicated in gigayears in each plot. The grey lines are the same in all plots and serve as a guide to the eye. The red lines are the metallicity gradients measured (using the mode) as in previous figures but measured on stars in the time intervals written on each plot.
}
\label{fig:rmean_FeH_age}
\end{figure*}

Figure \ref{fig:rmean_FeH_age} shows the distribution of guiding radii as a function of metallicity for four different time intervals, for all stars in the sample. 
The following features are visible:\\

-- In the youngest age intervals (0-3 and 3-5 Gyr), stars are concentrated around the Galactic radius interval 10-13~kpc, suggesting that most of the star formation in the last 5 Gyr has occurred in this region. Correlatively, the absence of stars younger than this limit within corotation ($<$6 kpc) is apparent, thus most stars younger than 5 Gyr are observed outside corotation. We note that although we infer these trends from distributions uncorrected for the survey selection function, these results are consistent with \cite{2017MNRAS.471.3057M}, where the density profiles have been corrected to take into account the selection function.
These two plots also show that stars younger than 3 Gyr (but older than about 1~Gyr, since there are only very few stars younger than this limit among giants in astroNN) were probably born around the OLR radius (at $\sim$ 11 kpc).

-- Comparison between the two plots between 7 and 10 Gyr shows that during this time interval, star formation in the disc had already moved from within 5~kpc at age greater than 7 Gyr to 11~kpc for younger stars. 
Stars that are beyond 9-10~kpc cannot have migrated from the inner disc, because the inner disc was already forming stars at a much higher metallicity at this epoch. Hence, together with the fact that the stars which have suffered negligible radial migration (because they are young) are found around the OLR, it suggests that most of the stars formed in the Milky Way in the last 7 Gyr were born at this same radius. 

-- The mean metallicity (red curves) of stars between 8 and 11~kpc is lower in the age interval 3-5 Gyr compared to 5-7~Gyr. This effect may have two origins. The first is the absence of metal-rich stars younger than 5 Gyr, pulling the mean metallicity to lower values. The second origin may come from an episode of dilution occurring at age$<$5 Gyr, where the ISM in this radius interval may have been diluted by more metal-poor gas coming from larger radii.

-- In the oldest age intervals ($>$5~Gyr), the distribution extends towards the Galactic centre. 
It is observed that while the metallicity is roughly correlated with the guiding radii at ages $<$7~Gyr and [Fe/H]$>$-0.3, this is no longer the case at ages greater than 7~Gyr, where we observe a large dispersion in metallicity for a given guiding radius. 

-- In the 7-10 Gyr age interval, stars with metallicity between -0.3 and 0.4 all reach the same outward extension to 12~kpc. Stars found beyond this limit have metallicities below -0.3.

\begin{figure*}{}
\includegraphics[width=4.4cm]{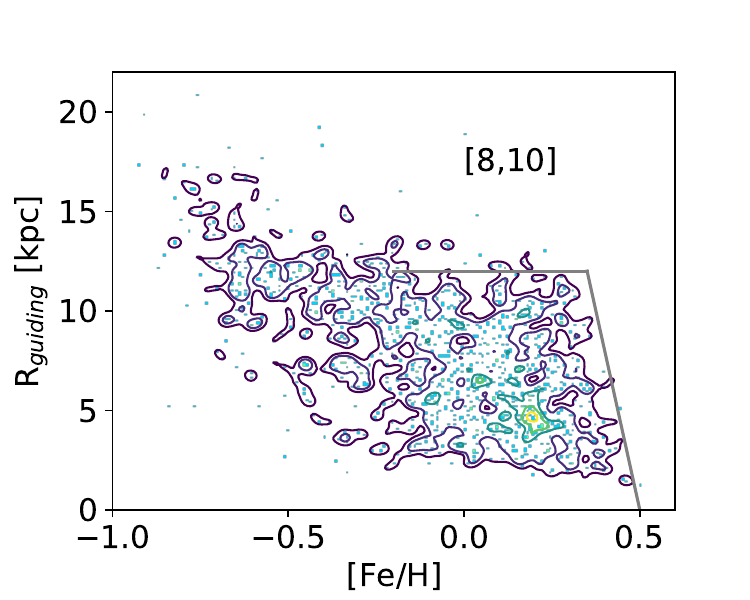}
\includegraphics[width=4.4cm]{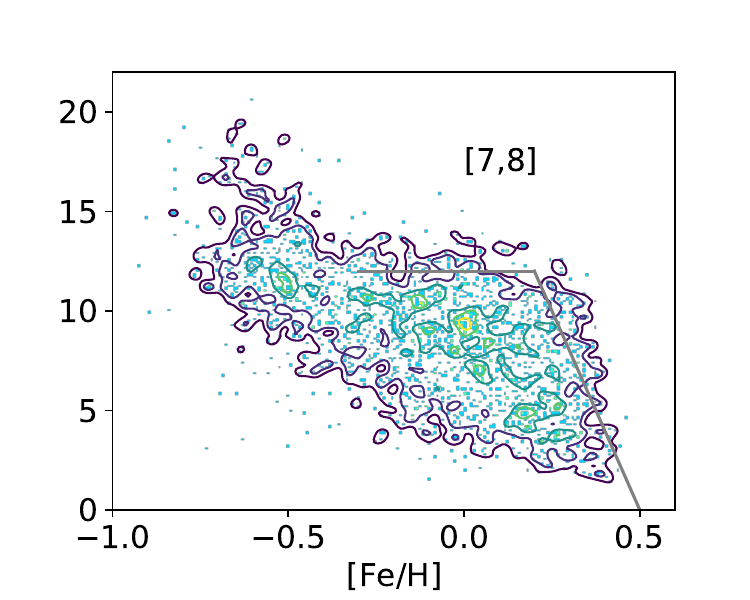}
\includegraphics[width=4.4cm]{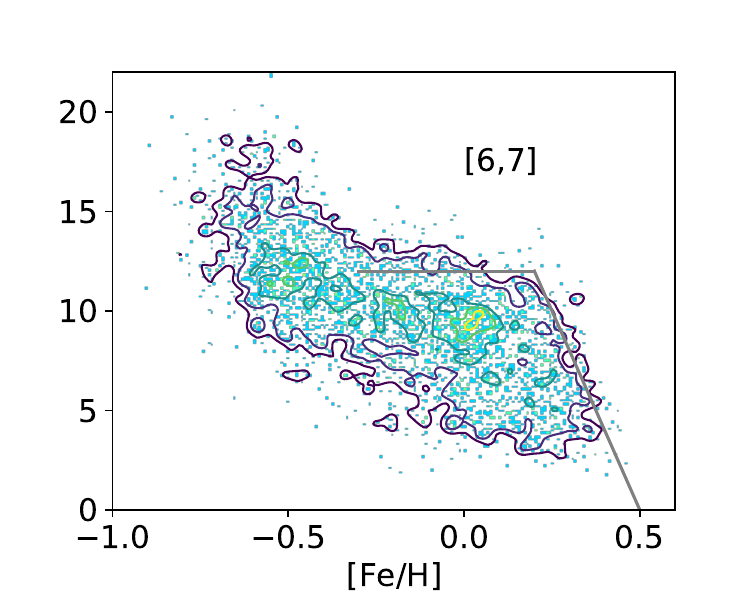}
\includegraphics[width=4.4cm]{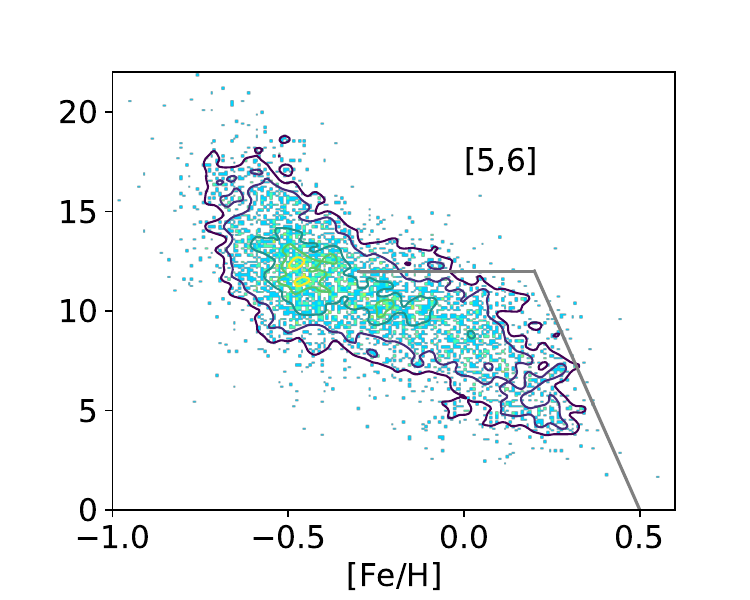}
\caption{R$_{\rm guiding}$ versus metallicity from 5 to 10 Gyr and with an eccentricity lower than 0.15. The grey lines are the same in all plots and serve as guide to the eye. The presence of metal-rich stars up to the OLR limit becomes conspicuous at ages~$>$ 6 Gyr. See text for details. The age intervals are indicated on each plot.
}
\label{fig:rmean_FeH_age5-10}
\end{figure*}

-- The knee/break starts to appear in the age interval 5-7 Gyr, and is clearest at 7-10 Gyr.
This is particularly evident in Fig. \ref{fig:rmean_FeH_age5-10}, which shows a more detailed view of the guiding radius - metallicity plots for stars with eccentricity restricted to $<$0.15, by 1~Gyr intervals, to illustrate the age at which the most metal-rich stars spread to larger guiding radii. It shows that the extension of high metal-rich stars to 12~kpc is already visible in the 6-7 Gyr age range, covering uniformly the guiding radius range from 2.5 to 12.5~kpc between metallicities -0.2 and +0.3. 
However, this feature is dominant for older ages. Due to age uncertainty, it is possible that the presence of metal-rich stars at 12kpc in the 6-7~Gyr age range is partly due to contamination by older stars. 

-- The knee/break, illustrated by the steep gradient measured between 8 and 12 kpc (-0.3dex.kpc$^{-1}$), has been maintained since 7-10~Gyr. This suggests that no significant migration due to the bar over this radial range has taken place during this period. On the contrary, the old age of the break and the fact that stars separated by only 2-3 kpc have such a different metallicity suggest that they were able to evolve separately, or equivalently that gas was not able to mix over this radial range for several gigayears.

%%%%%%%%%%%%%%%%%%%%%%%%%%%%%%%%%%%%%%%%

%%%%%%%%%%%%%%%%%%%%%%%%%%%%%%%%%%%%%%%%
\subsection{Different chemical evolutionary pathways on either side of the OLR}

\begin{figure}
\includegraphics[width=4.25cm]{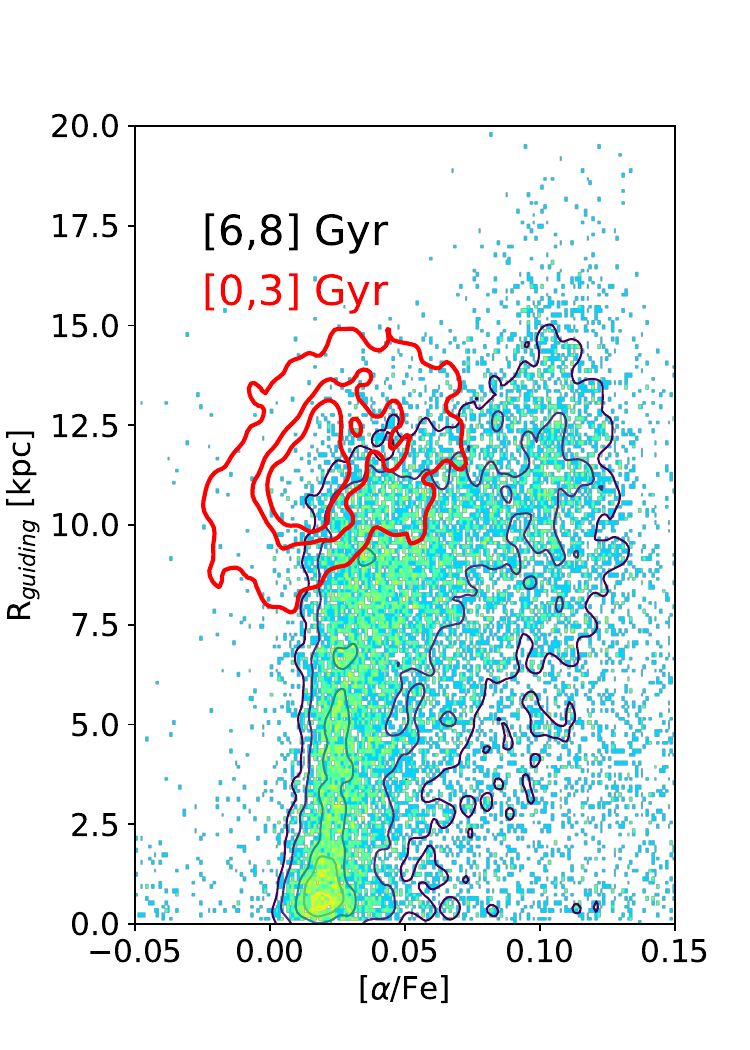}
\includegraphics[width=4.25cm]{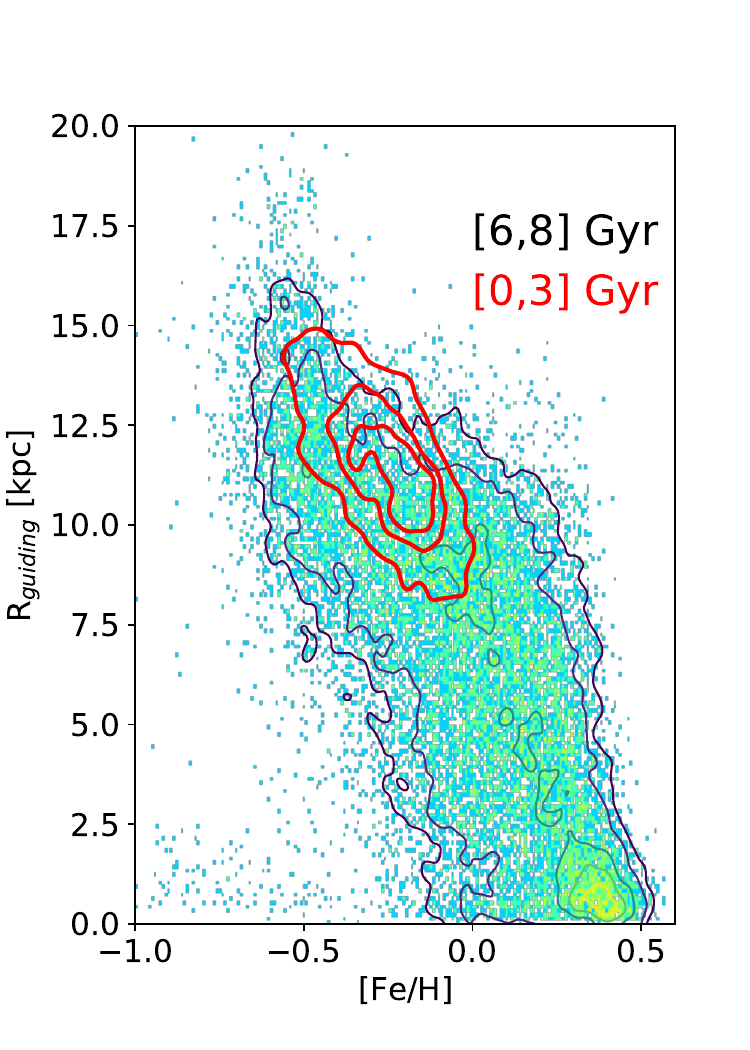}
\caption{
Density contour distribution of the guiding radius as a function of the alpha element abundance (left) and metallicity (right) of stars in the age range 5 to 7 Gyr, with contours for stars with ages below 3 Gyr in red. 
}
\label{fig:rmean_alpha_and_FeH_age3Gyr}
\end{figure}

The change in metallicity (of about 0.4 dex) that occurs within 1~kpc is evidence that, at the epoch where this difference is observed (7-10~Gyr), the two parts of the disc on either side of the OLR must have evolved independently.
Fig. 10 of \cite{2021A&A...655A.111K} shows age-[$\alpha$/Fe] relations sampled in different Galactocentric intervals from 2 to 16~kpc. It illustrates that while the age-$\alpha$ relations vary smoothly with radius  between 2$<$R$<$10~kpc on the one side and R$>$ 10~kpc on the other side, there is a clear gap between these two groups, when passing from stars that are within 10kpc to stars that are beyond 10~kpc. This is due to the even more rapid change of mean $\alpha$ abundance with radius. 
This is illustrated in Fig. \ref{fig:rmean_alpha_and_FeH_age3Gyr}, which shows the distributions of guiding radius as a function of [$\alpha$/Fe] and [Fe/H] in the age interval between 6 and 8 Gyr. 
The left plot shows the sudden change in [$\alpha$/Fe] abundance ratio that occurs between 11 and 12~kpc. Within such a short distance, the peak metallicity decreases by a factor of 0.5 dex, from solar to -0.5, while the [$\alpha$/Fe] abundance ratio increases from 0.03 to 0.11 dex.

The fact that stars with such different chemical properties coexisted for such a long time without mixing means that the regions inside and outside the OLR followed separate evolutionary paths, confirming that the barrier effect \citep{2015A&A...578A..58H} existed immediately after the end of the thick disc phase in our Galaxy. 
The red contours show the position in these two plots of stars having less than 3 Gyr, the metallicities now bridging the two sides of the OLR. This is indicating that the conditions that separated the two parts of the disc until 6-7~Gyr are no longer present, and that the disc regions inside and outside the OLR, that were on different chemical trajectories, merged smoothly less than 2~Gyr ago.

%%%%%%%%%%%%%%%%%%%%%%%%%%%%%%%%%%%%%%%%
%%%%%%%%%%%%%%%%%%%%%%%%%%%%%%%%%%%%%%%%

%%%%%%%%%%%%%%%%%%%%%%%%%%%%%%%%%%%%%%%%%%%%%%%%%%%%%%%%%%%%%%%%%%
%%%%%%%%%%%%%%%%%%%%%%%%%%%      SIMULATIONS 
%%%%%%%%%%%%%%%%%%%%%%%%%%%%%%%%%%%%%%%%%%%%%%%%%%%%%%%%%%%%%%%%%%
\section{N-body simulations}\label{sec:simulations}

 In this section, motivated by the observed effects of the bar resonances in test-particle and N-body simulations \citep{1981ApJ...247...77S,2015A&A...578A..58H,2018A&A...616A..86H,2020A&A...638A.144K}, and the observational results obtained in the previous section, 
we assess whether the structures observed in the metallicity profile could correspond to the expected signatures left by the bar formation and evolution as described in these studies. 

\subsection{Simulation characteristics}

We explore an $N$-body/hydrodynamical simulation of a disc galaxy with a total stellar mass and a rotation curve compatible with those of the Milky Way. Our model starts from a pre-existing axisymmetric stellar disc with gas and where star formation is coupled with chemical evolution. In this model, we focus on the evolutionary phases where a significant fraction of the stellar mass is already in place which, in the context of the MW evolution, would correspond to the epoch soon after the thick disc formation.

Initially, stellar particles are redistributed following a Miyamoto--Nagai density profile~\citep{1975PASJ...27..533M} that has a characteristic scale length of $4$~kpc, vertical thicknesses of $0.2$~kpc and mass of $4.5 \times 10^{10}$~\Msun. Our simulation also includes a live dark matter halo~($5\times 10^6$ particles) whose density distribution follows a Plummer sphere~\citep{1911MNRAS..71..460P}, with a total mass of $6.2\times 10^{11}$~\Msun and a radius of $21$~kpc. The choice of parameters leads to a galaxy mass model with a circular velocity of $\approx 220$~km/s. The gas component is represented by an exponential disc with a scale length of $5$~kpc and a total mass of $1.5 \times 10^{10}$~\Msun. The initial equilibrium state has been generated using the iterative method from AGAMA software~\citep{2019MNRAS.482.1525V}.

In our simulations, a gaseous cell undergoes star formation if: i) the gas mass is $> 2\times10^5$~\Msun, (ii) the temperature $T$ is lower than $100$~K and (iii) is the cell is part of a converging flow. The efficiency of star formation is set to $0.05$, that is, $5\%$ of the gas eligible to form a new star particle per dynamical time. We consider the ISM as a mixture of several species (H, He, Si, Mg, O, Fe, and other metals) which is sufficient for modelling the galactic chemical evolution~\citep[see][]{2021MNRAS.501.5176K} and the newborn stellar particles inherit both kinematics and elemental abundances of their parent gas cells.  

Since our model does not aim to reproduce the earlier phases of the galaxy formation initially we assume a radial metallicity of gas of $\rm [Fe/H]=-0.2$~dex.kpc$^{-1}$ and $\rm [\alpha/Fe]=0.0$. These values are somewhat arbitrary and represent exactly neither thick disc formation nor early thin disc evolution of the MW. However, such simplified initial conditions allow us to better trace the chemical trends caused by the dynamical processes and specific regimes of star formation in a different galactic environment without going deeply into the details of the early phases of galaxy formation, which is of primary interest for cosmological simulations~\citep[see, e.g.][]{2004ApJ...612..894B,2005ApJ...630..298B,2018MNRAS.477.5072M,2019MNRAS.490.4786G,2020MNRAS.491.5435B}.

Following the chemical evolution models by~\cite{2015A&A...578A..87S}, at each time step, for newly formed stars we calculate the amount of gas returned, the mass of the various species of metals, the number of SNII or SNIa for a given initial mass and metallicity, the cumulative yield of various chemical elements, the total metallicity, and the total gas released. Feedback associated with the evolution of massive stars is implemented as an injection of thermal energy in a nearby gas cell proportional to the number of SNII, SNI and AGB stars. The hydrodynamical part also includes gas-metallicity dependent radiative cooling~\citep[see details in][]{2021MNRAS.501.5176K}. 

The simulations were evolved with the $N$-body+Total Variation Diminishing hydrodynamical code~\citep{2014JPhCS.510a2011K}. For the $N$-body system integration and gas self-gravity, we used our parallel version of the TREE-GRAPE code~\citep[][]{2005PASJ...57.1009F} with multithread usage under the SSE and AVX instructions. In recent years we already used and extensively tested our hardware-accelerator-based gravity calculation routine in several galaxy dynamics studies where we obtained accurate results with a good performance~\citep{2018A&A...611L...2K,2018A&A...620A.154K,2018MNRAS.481.3534S,2019A&A...622L...6K}. For the time integration, we used a leapfrog integrator with a fixed step size of $0.1$~Myr. In the simulation, we adopted the standard opening angle $\theta = 0.7$. The dynamics of the ISM is simulated on a Cartesian grid with static mesh refinement and a minimum cell size of $\approx 10$~pc in the galactic plane.

\begin{figure*}
\includegraphics[width=18cm]{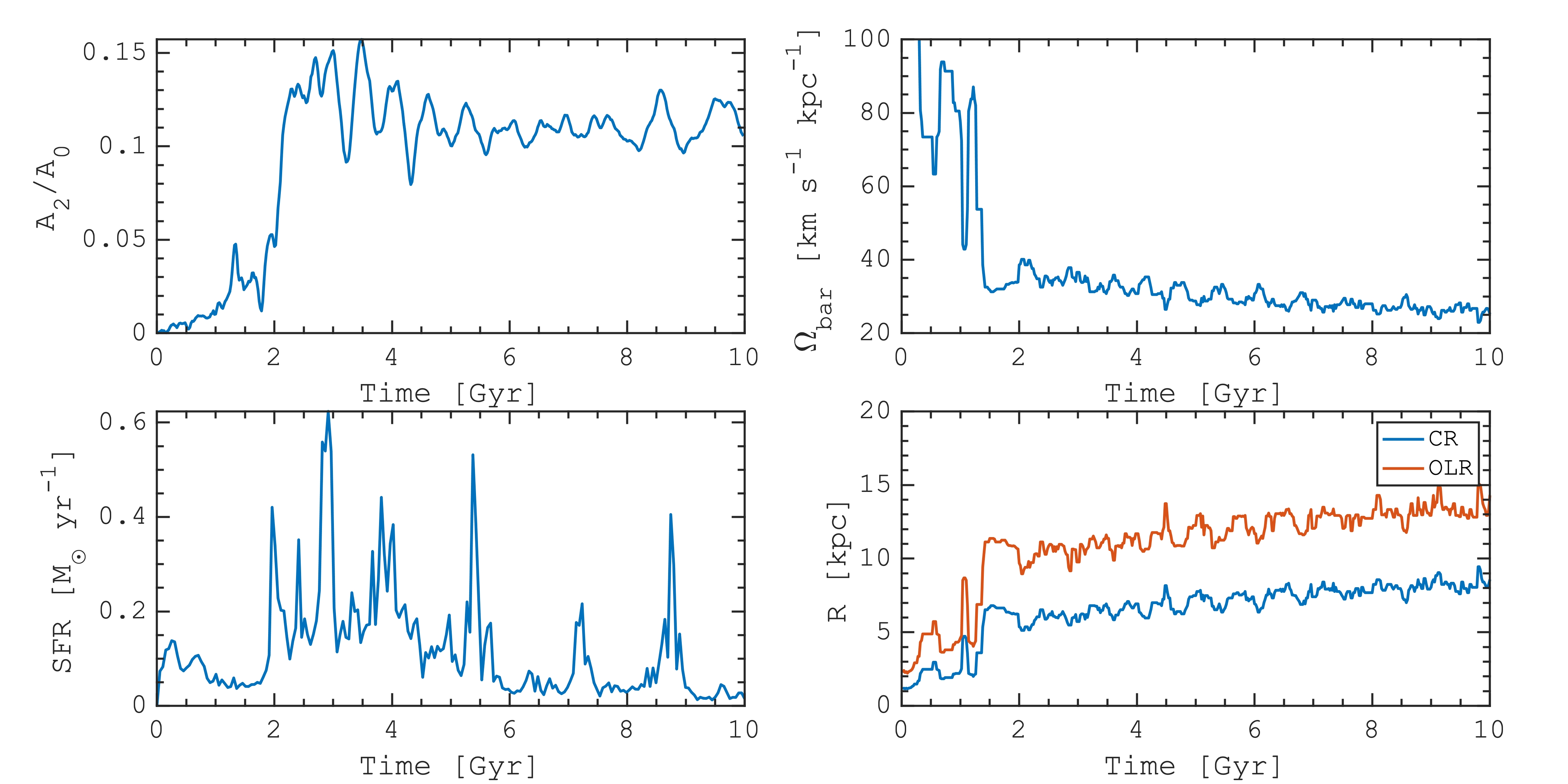}
\caption{Clockwise from top left, various quantities as a function of time: the strength of the bar as measured by the A2 asymmetry, the pattern speed of the bar, the corotation and OLR radii, and the star formation history.
}
\label{fig:simulation_characteristics}
\end{figure*}

Figure \ref{fig:simulation_characteristics} illustrates the time evolution of various characteristics of the bar in the simulation: bar strength as measured by the asymmetry coefficient A2,  the pattern speed, the radii of the corotation and OLR resonance. The corotation radius at the end of the simulation is between 8 and 9kpc, while the OLR radius is between 13 and 14~kpc. The star formation history is also shown. More extensive face-on maps of the simulated galaxy at different epochs are also given in Fig. \ref{fig:simulation_main}.

%%%%%%%%%%%%%%%%%%%%%%%%%%%%%%%%%%%%%%%%

\begin{figure*}
\includegraphics[width=18cm]{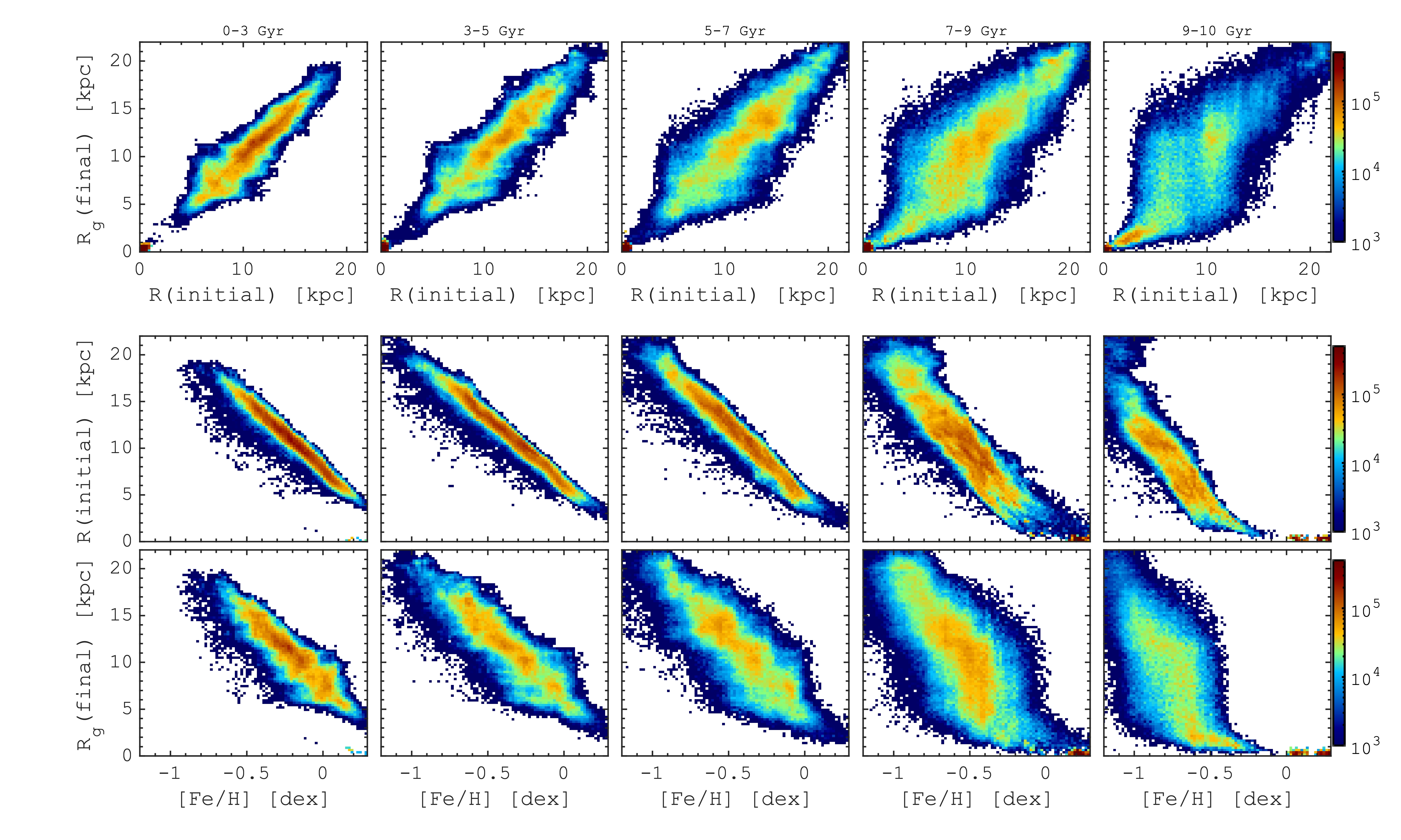}
\caption{Simulation results. Top row: Final guiding radius as a function of the initial radius of the stars, in different age intervals. Middle: Initial guiding radii as a function of metallicity. Bottom: Final guiding radius as a function of the metallicity of the stars.
}
\label{fig:guiding_radii}
\end{figure*}

\begin{figure*}
\includegraphics[width=17cm]{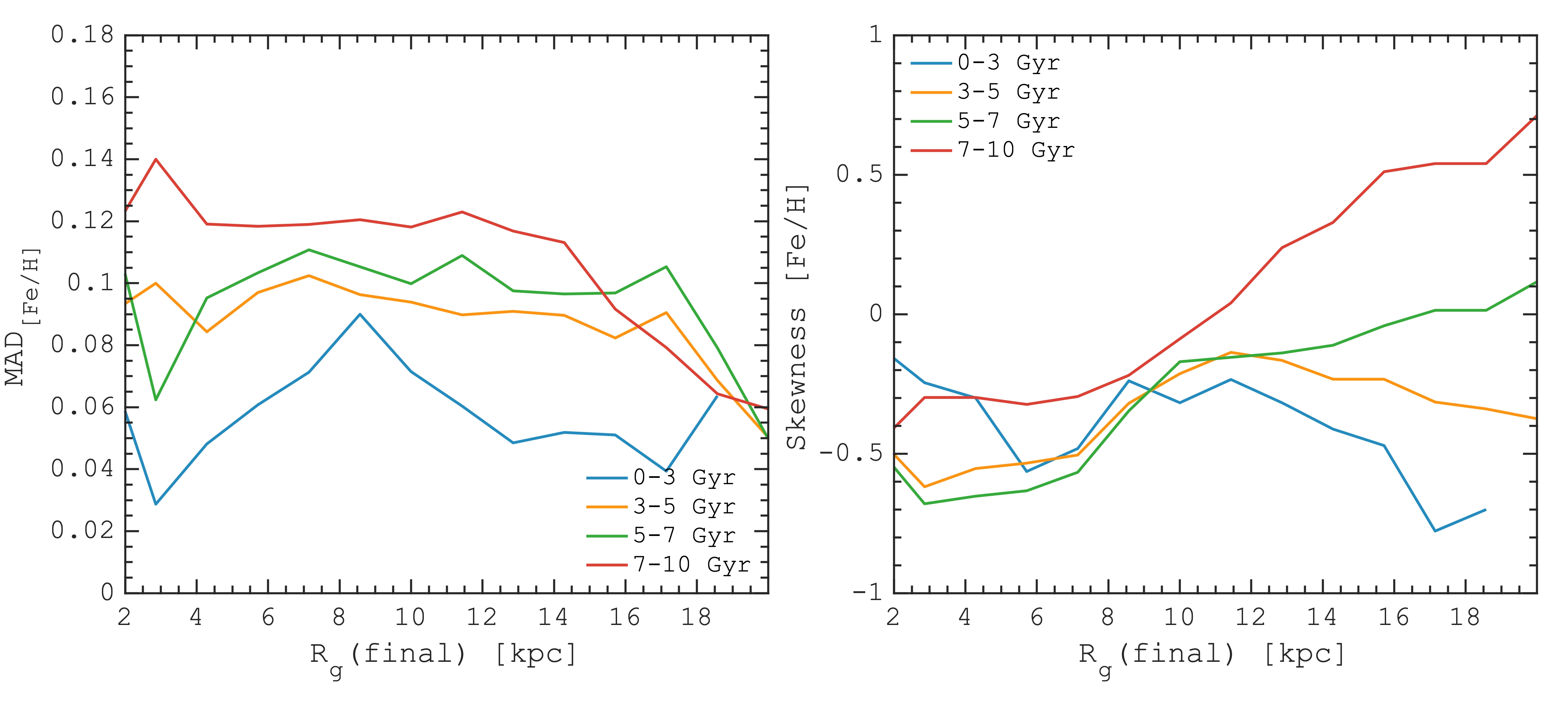}\caption{MAD (left panel) and skewness (right panel) of the metallicity distributions as a function of the guiding radii in the simulation.
}
\label{fig:simulation_dispersion}
\end{figure*}
%%%%%%%%%%%%%%%%%%%%%%%%%%%%%%%%%%%%%%%%

%%%%%%%%%%%%%%%%%%%%%%%%%%%%%%%%%%%%%%%%
\subsection{Results from N-body simulation}\label{sec:simulation_results}

Fig.~\ref{fig:guiding_radii} shows the results of the simulation for different age intervals, the top row showing the final as a function of the initial guiding radii, the second row the initial guiding radii as a function of metallicity, and finally, the last row shows the final guiding radii as a function of metallicity.

 We focus first on the two rightmost plots in the first row. These two plots are similar to the distribution shown in \cite{2020A&A...638A.144K}, and show that a fraction of the stars initially in the inner disc at R$<$7~kpc have migrated to about 13~kpc, which is the radius of the OLR at the end of the simulation.
 Some stars are caught in the bar and some others migrate inwardly from more than 10kpc to R$<$5~kpc. Most of the migration occurs outwards, the two plots showing, as in \cite{2020A&A...638A.144K}, that the stars move at most to the OLR radius.

Similarly to \cite{2020A&A...638A.144K}, Fig.~\ref{fig:guiding_radii} shows that most migration occurs within 2 Gyr after the beginning of the formation of the bar (two rightmost plots). This corresponds to the early slowdown phase of the bar, when the bar stabilizes and the OLR reaches 10~kpc within less than 2 Gyr (see Fig.\ref{fig:simulation_characteristics}). The slowdown continues during the secular phase (after the first 2~Gyr), but at a much slower pace. The term slowdown or early slowdown used in the rest of this article refers to the first 2 Gyr in the simulation.
Similar time scales are observed in \cite{2020A&A...638A.144K} (their Fig. 3). The strength of the bar remains at a maximum for 2.5 Gyr after the migration has ended. 
 
When seen as a function of metallicity (third row), the distribution of final guiding radii resembles strikingly the observed distributions (the overall metal content in the simulation is shifted to lower metallicities because of the simulation lacking of a thick disc phase). The ([Fe/H], $R_{final}$) distribution in the age bin 9-10 Gyr shows the same characteristics as the observations: a rather flat metallicity gradient up to the OLR, then a drop in metallicity, the distribution forming a characteristic knee also found in the data. 
The knee feature disappears in plots showing particles younger than 7 Gyr, illustrating the fact that the systematic migration observed in intervals 7-9 and 9-10 Gyr is terminated. Thus, the knee observed in the rightmost plots is the signpost for the migration episode that occurs during the early slowdown of the bar, and is therefore an indication of the epoch of formation of the bar. 
It is worth noting that the region of flat gradient in the 8-9 Gyr interval (where the metallicity stays constant at about [Fe/H]$\sim$-0.6) is also the region spanned by the corotation (between about 5 and 10~kpc).
Figure 12 of \cite{2015A&A...578A..58H} \citep[see also][]{2012A&A...548A.126M} showed that the outward and inward flux of migrating stars through churning is maximal in the region of bar corotation. \cite{2006MNRAS.370.1046V} and \cite{2022ApJ...935...28W} also showed that spiral and bar corotation flatten metallicity gradients. This also corresponds to the regions of minimal gradient in the data (at R$<$7~kpc).

Figure \ref{fig:simulation_dispersion} quantifies the dispersion and skewness of the simulated metallicity distributions as a function of age and the guiding radius. 
The simulation offers several similarities with the observations:
(1) The dispersion increases with age, meaning that some residual mixing occurs through the variation of angular momentum of stars. The decrease with radius is less pronounced than in the observations, but nonetheless present, in particular at age$<$5 Gyr.
(2) The dispersion of the oldest component (7-10 Gyr) drops after the bar OLR, while in the other age range the dispersion drops only beyond 17-18 kpc. This limit corresponds roughly to the distance at which spiral arms are observed in the simulation. It suggests that the dispersion in the oldest age bin is dominated by the bar-driven migrated stars. 

The different dispersion curves within corotation are less noisy than in the observations. The overall level of dispersion is less important in the simulation than in the observation, which may have different explanations: the intrinsic metallicity dispersion of the ISM is probably higher in the observations ($\sim$0.1 dex compared to 0.06-0.08 dex), the chemical evolution may cover a more limited range of metallicities, radial mixing may be less important, et cetera.

The skewness also shows several interesting patterns in common with the observations. 
(1) The skewness increases at large radii with the age of the stars. The oldest interval shows the highest positive skewness. 

(2) The particles/stars which are the least likely affected by migration show negative skewness. Figure \ref{fig:simulation_mdf} shows that the youngest age bin (1 Gyr), for which migration may be negligible, is negatively skewed. 
(3) For the oldest age bin, the skewness is positive for stars beyond corotation (8-9 kpc), and continues to increase up to the OLR (14-15 kpc).

These results suggest that in the simulation, the evolution of the skewness from negative to positive is mainly driven by the stars that have migrated as a result of the formation of the bar.

\subsection{Quantifying radial migration in simulations}

\begin{figure}
\includegraphics[width=9.5cm]{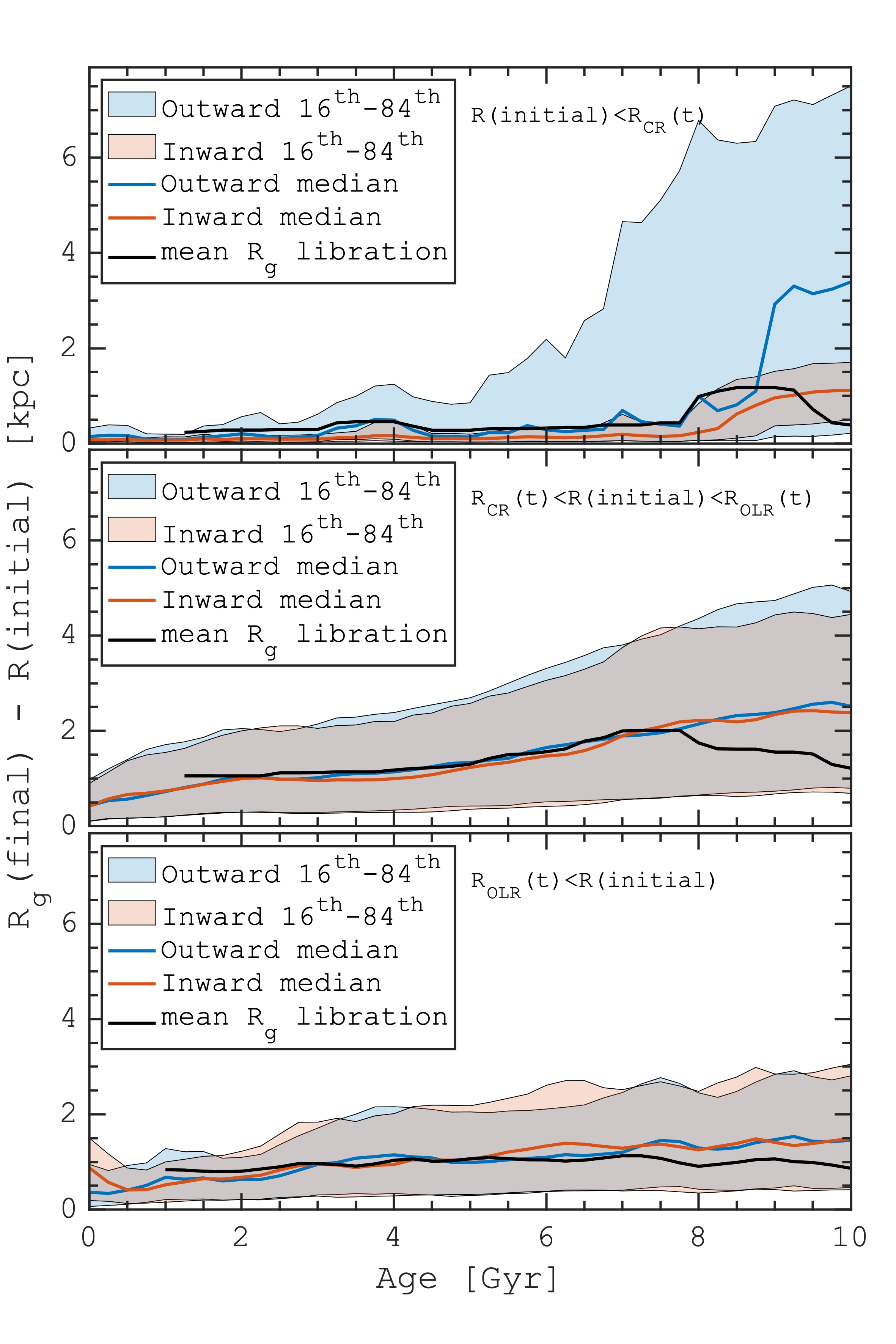}
\caption{Migration distance, as given by the difference between the final and initial guiding radius for stars with their initial guiding radius inside corotation (top plot), between corotation and OLR (middle), and outside the OLR (bottom plot), divided in an inward (red) and outward motion (blue) (median and 16$^{\rm th}$ and 84$^{\rm th}$ percentile.)
The black curves are the mean change in the guiding radius induced by the libration of the orbits.}
\label{fig:migrations}
\end{figure}

Different types of mixing or migration are at play in the simulation and explain the features described in Fig.~\ref{fig:guiding_radii} and \ref{fig:simulation_dispersion}.
Their effects are quantified in Fig. \ref{fig:migrations} where solid lines correspond to the median difference between final and initial guiding radii, and the filled areas show the 16th-84th quantiles of the same quantity. The red and blue colours correspond to the outward and inward migration. Different panels depict the migration rate of stars formed in three radial regions: inside corotation~(top), between corotation and the OLR~(middle) and outside the OLR~(bottom), where the time-dependence of the resonance location is taken into account~(see Fig.~\ref{fig:simulation_characteristics}).

\paragraph{The migration due to the bar} The first type of migration is illustrated on the top panel and shows the systematic, essentially outward~(blue colour), shift of stars that are initially inside the corotation due to the rapid growth and slowdown of the bar in the first 2 Gyr (ages 8 to 10 Gyr). These stars migrate to the largest distances, and since they are the most metal-rich for a given age range, they give rise to the knee in the metallicity-guiding radius distributions~(see bottom panels of  Fig.~\ref{fig:guiding_radii}). It is important to note that there are no spirals inside the bar radius; thus, the only mechanism responsible for the migration of stars formed in this region is the evolution of the bar pattern speed~\citep{2015A&A...578A..58H, 2020A&A...638A.144K}.

The middle plot shows the migration of the stars that originate from the region between the initial corotation and OLR. In this case, the amplitude of radial migration is more limited but, more importantly, symmetrical (the inward and outward migrations extend to similar distances). For stars above 7 Gyr, the median radial migration distance is only about 2~kpc, while below 6 Gyr, it is about 1~kpc. The symmetric migration results in a substantial radial mixing of stars manifested in a broadening of the metallicity-guiding radius distributions~(see bottom panels of Fig.~\ref{fig:guiding_radii}). The bottom panel shows the migration of stars initially beyond the OLR where, as in the previous panel, the migration is symmetrical and limited to about 1~kpc for most stars, decreasing to about 0.5~kpc below 2~Gyr.

\paragraph{Scattering by spiral arms} The most prominent broadening of the metallicity-guiding radius distribution is seen around the resonances~(see Fig.~\ref{fig:guiding_radii}), in agreement with previous studies~\citep{2010ApJ...722..112M,2011A&A...527A.147M,2015A&A...578A..58H}. This intermediate galactic region is where the spiral arms can be found~(see Appendix~\ref{C}), which might suggest that the spiral arms are the source of radial migration in the middle plot of Fig.~\ref{fig:migrations}. Theory, however, suggests that stars migrate if they interact with transient spirals~\citep{2002MNRAS.336..785S,2008ApJ...684L..79R,2012MNRAS.426.2089R}; while, in the case of a barred galaxy, the appearance of spiral density waves is regulated by the bar which makes their evolution more regular~\citep{2020MNRAS.497..933H,2024MNRAS.528.3576V}. This, nevertheless, does not fully exclude the migration of stars formed outside the bar further out due to scattering by spirals - by a process that remains to be determined. It is noteworthy that this process fails to elucidate the observed behaviour (the knee) of the metallicity-guiding radius~(see Fig. \ref{fig:rmean_FeH_age}) -- the distance of migration is much too short to explain this feature: the most metal-rich stars are predominantly formed in the inner galaxy, and are unlikely to be impacted by this migration.

\paragraph{Libration of orbits} Nevertheless, there is a certain academic interest in understanding stellar radial migration outside the bar region, which is poorly investigated in the literature, traditionally targeting mostly spiral-arms-induced scattering. The parameter usually used to quantify the migration rate is the difference between the initial~(birth or at a given time) and the final or instantaneous galactocentric position of stars. This measurement, however, is affected by the epicyclic motion of stars around their genuine orbits~(blurring), which can be dumped using the angular momentum or guiding radii. In such a case, the change of the guiding radius is considered a manifestation of radial migration or churning. However, in a non-axisymmetric barred potential, as in the MW or in our simulation, the angular momentum of individual stars is not conserved~(see \cite{2008gady.book.....B}, Chapter 3.3.2 and also \cite{2007MNRAS.379.1155C}) but rather oscillates~(librates) around the mean guiding radius even if the bar is rigidly rotating or adiabatically evolves on a long time scale. Such a second-order orbital effect, while being rather trivial, is often ignored in the measurements of the migration rate and the interpretation of the data. 

The reason for this is partially because it is impossible to access the oscillations of guiding radii of the MW stars without proper knowledge of 3D mass distribution in the inner Galaxy and the long-term evolution of the MW bar parameters~(strength and pattern speed), while the angular momentum of stars calculated in axisymmetric potential is constant. In simulations, it is also not a trivial exercise since it requires a high cadence of snapshot outputs and a proper reconstruction of the guiding radii evolution~(see \cite{2015A&A...578A..58H} for details). 

\begin{figure}
\includegraphics[width=9.5cm]{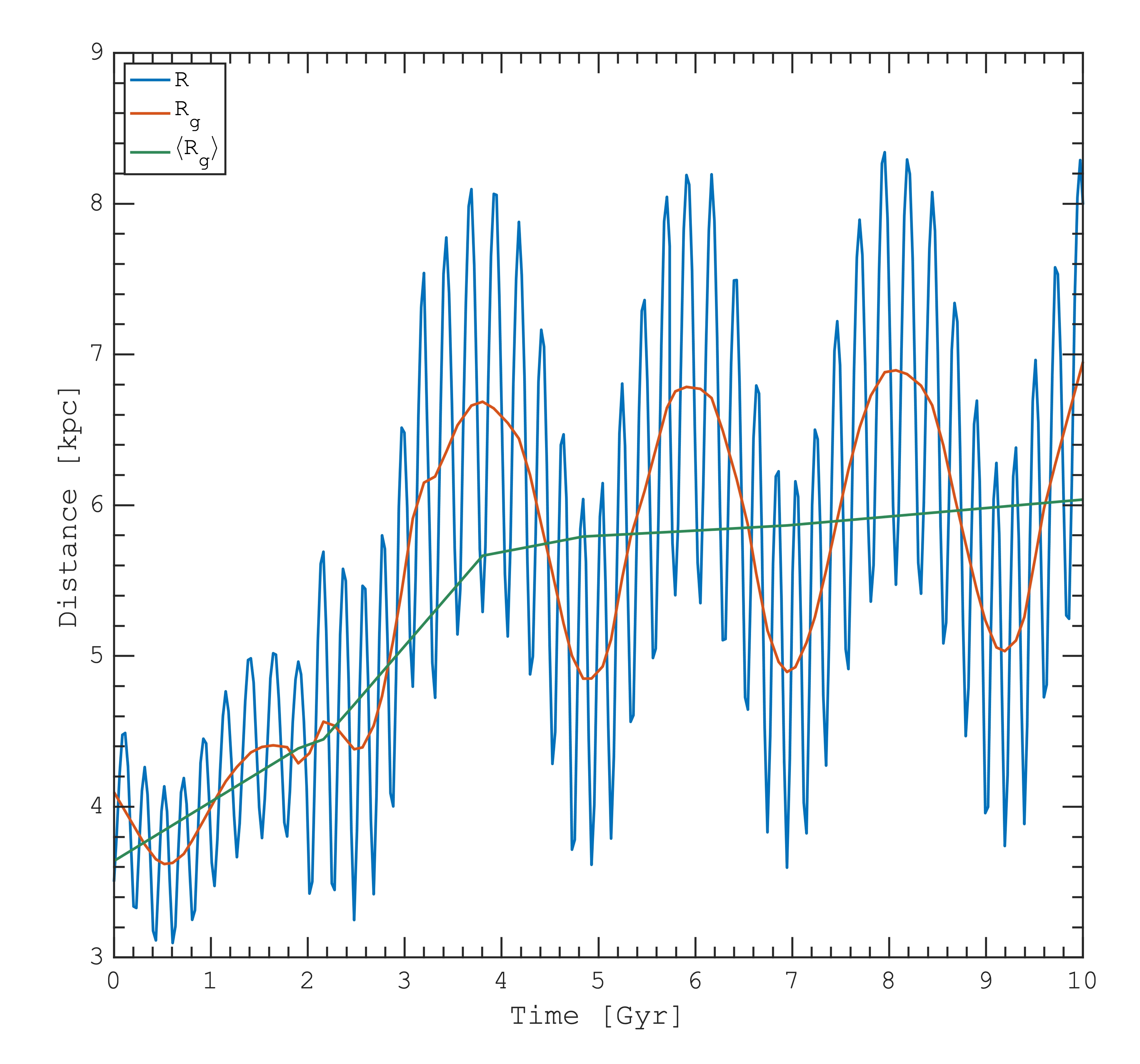}
\caption{Orbit (in blue) of a single particle in the simulation illustrating the libration, or oscillation of the angular momentum through its guiding radius (red curve), on a period of $\sim$ 2 Gyr. The particle is driven rapidly outwards during the intense episode of the bar growth at $2-3$~Gyr.  On a longer timescale, despite a significant oscillation~($\approx 2$~kpc) of the guiding radius in a non-axisymmetric bar potential, the mean guiding radius~(green line) shows only a small drift due a slow evolution of the bar~(see Fig. ~\ref{fig:simulation_characteristics}). These periodic oscillations of the angular momentum~(libration) thus can be misinterpreted as migration~(churning), while the net motion of stars in a rigid or slowly evolving barred potential, in fact, is close to zero.}
\label{fig:libration}
\end{figure}

To illustrate the motion of stars in evolving barred potential, in Fig. ~\ref{fig:libration} we show an orbit of a star from our simulation. The blue line corresponds to the orbit of the stars, while the red one shows the corresponding guiding radius, for each point calculated as the mean of $(R_{min}(t)+R_{max}(t))/2$~(see also Fig. 7 in \cite{2015A&A...578A..58H}). As expected, while the red line diminishes the epicyclic oscillations, it still librates around a certain value demonstrating the lack of angular momentum conservation. Once we repeat the same calculation for the guiding radius and calculate a `guiding' of the guiding radius evolution, we obtain the averaged guiding radius~(green line) which is nearly constant after the bar formation episode. The variation of the guiding radius around this value sets up the amplitude of the guiding radius libration, which is about 2~kpc for this particular star. 

Next, we estimate the contribution of the guiding radii libration in measured radial migration. With the black lines in Fig.~\ref{fig:migrations} we show the median libration amplitude for stars according to their birth positions. Since the calculation of the libration rate requires several periods of the guiding radii oscillation, we cannot do this properly for stars younger than $\approx 1$~Gyr. Interestingly, the black lines follow the mean migration rate shown with blue and red lines. This suggests that the migration rate we measure can be explained by the libration of the angular momentum of stars in the regions outside the bar. This means that although libration induces no net migration~(see also \cite{2020ApJ...889...81W}), it contributes to the apparent mixing of stars on a local scale (within 1-2~kpc of the mean orbit for the stars outside the corotation in the simulation) and hence to the observed metallicity dispersion. 

Another intriguing aspect is the discrepancy in the libration rate between old stars~($8-10$ Gyr old) that migrated from the inner region due to bar formation and those formed outside the bar. Despite being located within the same radial range, as illustrated in the top and middle panels of Fig.~\ref{fig:guiding_radii}, the former exhibit a somewhat lower libration rate. Such behaviour was illustrated in \cite{2020A&A...638A.144K} where stars that were trapped at the bar resonances and transferred outwards tend to have small orbital eccentricities~(see their Figs 6 and 7).

% The migration rate we measure corresponds to the difference between the birth position of stars and their instantaneous galactocentric distance or guiding radius. The change of the latter seems to correspond to the change in the angular momentum of stars, being often interpreted as churning. 

% In order to take into account the libration for each star we measure the amplitude of the guiding radii oscillation at the end of simulations, as shown in Fig.~\ref{fig:libration}.

% On all plots, the black curve shows the mean amplitude of the guiding radius change due to the particles librating around resonances \citep{2007MNRAS.379.1155C}. These particles have their angular momentum oscillating on top of the systematic change of the angular momentum when the stars migrate. One such orbit is shown in Fig.~\ref{fig:libration}. 
% This means that although libration induces no net migration, it participates in the mixing of stars on a local scale (within 1-2~kpc of the mean orbit for the stars outside the corotation in the simulation) and hence to the observed metallicity dispersion. 
% Fig. \ref{fig:migrations} shows that, except for stars initially inside corotation, the amplitude of this mixing is similar to the mean migration. 

%%%%%%%%%%%%%%%%%%%%%%%%%%%%%%%%%%%%%%%%
%%%%%%%%.    Discussion
%%%%%%%%%%%%%%%%%%%%%%%%%%%%%%%%%%%%%%%%

\section{Discussion}\label{sec:discussion}{}

Informed by the simulations, we are now in a position to propose an interpretation of the evolution of the metallicity profile with age, and in particular of the presence of a break at 10-12~kpc. We first discuss the consequences of these results for the history of stellar migration then and on the epoch of the bar formation in our Galaxy.

\subsection{Radial migration}\label{sec:migration}

\paragraph{The evolution of gradients}
The slowing down of the bar, as described in the previous section, induces a systematic migration of stars from the inner disc towards, and up to, the OLR. This is explained in \cite{2015A&A...578A..58H} and \cite{2020A&A...638A.144K}, where it is shown how the stars initially on non-circular orbits in the inner regions of the disc migrate to circular orbits at larger distances, but limited within the OLR. 
The simulations show that it is the early slowdown phase of the bar that triggers the episode of radial migration. Although some outer disc stars migrate to the inner disc, the most visible part of the migration is made by metal-rich stars that are moved towards the outer disc, up to the OLR, see Fig.~\ref{fig:low_alpha_xy_fehMax}. The accumulation of these stars at the OLR creates a break in the metallicity profile, or a knee in the metallicity-guiding radius distribution at the epoch corresponding to the age of the youngest stars now in the knee. The fading of the knee signature for stars younger than 6-7 Gyr is telling us that the slowdown phase (during which the bar-triggered migration occurs) terminated at this epoch.
This migration explains well the old metal-rich stars found at the solar vicinity, and up to the OLR: stars with metallicity up to 0.3-0.4~dex are found up to 11-12 kpc from the Galactic centre, see Fig. \ref{fig:rmean_FeH_age5-10}.

The knee observed in both observations and simulations (at ages $>$7 Gyr) shows a feature generated by the episode of radial migration triggered by the formation of the bar and that has been 'frozen' since then in the orbital structure of the disc. The similarity of metallicity - guiding radii distributions between the data and the simulation suggests that the simulation captures well the main evolution of the disc, and that a single episode of bar formation, occurring more than 7~Gyr ago, describes well the features observed here. 
Because this feature is fragile - a steep gradient (-0.4dex.kpc$^{-1}$) occurring over a limited radius interval, including also stars of low ($<$0.15) eccentricity, a second significant bar-triggered episode of radial migration, or any significant mixing occurring in the last 7 Gyr in the range of the distances discussed here would probably have significantly altered the knee observed here.

In the oldest age intervals of the simulation in Fig. \ref{fig:guiding_radii} (lower right plot), the radial metallicity gradient before the knee is flat (the metallicity ridge line is almost vertical between 2 and 11~kpc). This is due to the migration of inner disc stars to the OLR, but also to the mixing at corotation. The corotation of the bar is known to be the location where mixing is the most intense, both inwards and outwards \citep[see][Fig. 12]{2015A&A...578A..58H}. The flattening of the metallicity profile is also observed in older populations before the break (between 6 and 10~kpc), see our comments in Section \ref{sec:metallicity_profile}.
%compare for example the metallicity profile on Fig. \ref{fig:FeH_gradient_age} in the age interval 2-4, 4-6 Gyr, and then 6-8 Gyr. 

On the contrary, across the break, the gradient flattens in younger populations (at ages $<$ 5 Gyr). This flattening could not have been caused by the radial migration being the opposite of the expected trend -- stronger on older stars -- but it is well explained by the evolution of the bar and its resonances.
First, the OLR of the bar has created a steep break in the metallicity profile of old stars in the disc. After its slowdown, the bar may lose strength, weakening the effect that has kept the disc separated inside and outside the OLR. Gradually, at ages younger than about 7~Gyr, the gas and stars could have been allowed to mix, exchanging metals and reducing the metallicity offset between the two regions, thus flattening the metallicity gradient at younger ages. This effect could explain the gradient flattening observed for stars younger than 4~Gyr in Figs. \ref{fig:FeH_gradient_age} and \ref{fig:FeH_gradient_with_age}, and the gradient now bridging the two sides of the OLR in Fig. \ref{fig:rmean_alpha_and_FeH_age3Gyr}.

\paragraph{Skewness and dispersion}

Figure \ref{fig:dispersion_skewness} shows that there is a clear difference in the MADs observed on either side of the corotation (6.5-7.5~kpc). 
We assume that inside corotation, the gas is well homogenized due to small dynamical time scales. Thus, the dispersion observed on stars born inside corotation should be small or near the one observed in the ISM ($<$0.07 dex). There are no stars with age $<$ 3 Gyr inside corotation, but  Fig.~\ref{fig:dispersion_skewness} shows that the MAD in the age range 3-5 Gyr is small and similar to what it is in youngest age range outside corotation.  
The age range 5-7 and 7-10~Gyr show higher MAD.
This is expected, because as shown in Fig. \ref{fig:guiding_radii}, the bar-triggered migration also moves inwards metal-poorer stars that are initially at larger radii. In fact, Fig. \ref{fig:rmean_FeH_age} shows that a number of metal-poor stars exist inside corotation, and which are responsible for the significant dispersion observed. 
The skewness in the age intervals older than 3 Gyr in the inner regions is more negative than the one of the youngest intervals because of this contamination by the metal-poor stars. 

At and around corotation, the MAD begins to increase in the three intervals 3-5, 5-7 and 7-10~Gyr. This is expected because mixing is known to be maximal at corotation, decreasing with time but still remaining substantial for a long duration (see for example Fig. 12 from \cite{2015A&A...578A..58H}), bringing in more stars of different metallicities.

Beyond corotation, the dispersion differs strongly in the different age intervals. 
The age interval 7-10 Gyr contains the metal-poorest stars (because they are the oldest) of the outer disc and the metal-richest stars brought in by migration (from the inner disc), thus producing the maximum spread. 
In the age interval 5-7 Gyr, the dispersion is dominated by the same combination of effects, but the spread is smaller mainly because the distribution is more dominated by stars born in situ.
In the 3-5 Gyr interval, much fewer stars that come from the inner disc (those which reach or even cross the inclined line as observed in the 7-10 ~Gyr plot) are present, and the main population is getting more metal-rich, as can be seen in Fig. \ref{fig:rmean_FeH_age} (the metallicity at R$_{\rm guiding} \sim$ 12~kpc is increasing with time, and is metal-richer at ages 3-5~Gyr than it is at 7-10 or 5-7~Gyr).
Finally, the age interval 0-3~Gyr shows the smallest MAD, about 0.08-0.09~dex, above what is measured on the ISM or the youngest stars (0.04-0.07~dex, see section~\ref{sec:dispersion_and_skewness}. 
A gradient between -0.042 and -0.066 dex.kpc$^{-1}$ for stars younger than 3~Gyr combined with a libration effect mixing stars on the kpc scale - and which is maximum between corotation and the OLR, see fig.~\ref{fig:migrations} - would add a few hundredths of dex of dispersion to the ISM value, allowing it to reach the observed value.

Thus, we attribute most of the dispersion observed in the age range 7-10~Gyr to the migration triggered by the bar early slowdown. Likewise, the residual dispersion observed in the age range 5-7~Gyr is probably due to the fact that the migration episode may have prolonged a bit after 7 Gyr and to the contamination by older stars of lower age bins due to age uncertainty. 

Libration may also play a role by adding a dispersion of a few hundredths of dex to all age intervals. This could, in particular, explain the presence in the solar vicinity of metal-rich stars as young as 3-5 Gyr. 
Once trapped at resonances, stars have their 'instantaneous' guiding radius oscillating due to libration, deviating from the mean guiding radius by a significant distance. In the cases studied by \cite{2007MNRAS.379.1155C}, one of their models with a corotation radius close to 6~kpc shows stars trapped at this resonance wandering up to 10~kpc.
Thus, the mechanism could explain the presence of metal-rich stars between the corotation and the OLR (and thus in the solar vicinity) for stars too young to have migrated at the time of the bar formation. It should be noted that a star observed near the maximum of its angular momentum oscillation would have its measured instantaneous guiding radius significantly larger than its mean guiding radius, thus appearing as a star that has migrated from the inner regions when in fact its mean guiding radius would be significantly smaller.

Finally, it is striking to see that the MAD for the age intervals 3-5 and 5-7~Gyr decreases to larger guiding radii (at $\sim$12~kpc) to reach the value measured on the youngest stars. 
This clearly means that, beyond the OLR ($>$12~kpc), no extra-mixing is measured, even on stars significantly old (3-7~Gyr), strongly suggesting that no significant migration has taken place beyond this limit and in particular that spiral arms had no obvious effect.
Another indication suggesting that transient spiral arms may have had a negligible contribution to the scattering is that, as mentioned above, the break in the metallicity profile is still clearly visible 7~Gyr after its formation, limiting the amount of reshuffling by subsequent mixing.

We conclude this section by saying that the behaviour of the MAD and skewness observed on Fig.\ref{fig:dispersion_skewness} can be explained by the effect of the bar-triggered migration and libration of orbits, and that there is no compelling indication that some other migration mechanism is necessary to explain the present observations.

\subsection{Dating the epoch of the bar formation}\label{bar}

Simulations show that bars reach their maximum strength within about 2 Gyr of the beginning of their formation, with the resonances expanding outwards over the same period. The early slowdown triggers a radial migration episode that ends up when the bar stabilizes. 
The epoch of the formation of the bar is therefore bracketed by the following evidence.
Figure \ref{fig:rmean_FeH_age5-10} shows that the knee is not visible at age$<$6-7 Gyr, suggesting that the end of the radial migration episode occurred around this epoch. 
The ages of the super metal-rich stars observed in the solar vicinity \citep[e.g.][]{2023A&A...669A..96D}, which are the best candidates for migration \citep{2008MNRAS.388.1175H,2015ApJ...808..132H,2015MNRAS.447.3526K,2018A&A...609A..79H,2020MNRAS.493.2952H}, are also a constraint on the epoch of formation of the bar, since the end of the migration episode could not have occurred before these stars were born. These stars are 7-9 Gyr old, again constraining that the radial migration episode could not have ended before the time implied by these age estimates. A similar age is obtained for the super metal-rich, old, open cluster NGC 6791 \citep{2021A&A...649A.178B,2019ApJ...874..180M}, which imposes similar limits. 
The bar formation cannot be younger than the limit imposed by these ages. If the present simulation and the one presented in \cite{2020A&A...638A.144K} are representative of the formation timescales of bars, it suggests that the Milky Way must have started forming its bar 1-2 Gyr before the end of the radial migration episode, or 8-9 Gyr ago. Given the fact that age estimates of old stars may be underestimated in astroNN (see section \ref{sec:data}), it is safe to assume an age range of 8-10 Gyr for the bar formation, perhaps implying that it started during the thick disc phase \citep{2023A&A...674A.128G}. 

The formation timescale of bars have been discussed in a wider context in \cite{2018MNRAS.477.1451F} and \cite{2023ApJ...947...80B}.
These authors show that the timescale for the bar formation primarily depends on the fraction of the baryonic disc mass over the total mass in the inner disc being over a Hubble time when this fraction is below 0.35, and decreasing to less than about 1 Gyr when it is above about 0.5.
%In the case of the Milky Way,  the mass growth of the  Galaxy would possibly favor a rapid formation of the bar, because of the early formation of the thick disc, which, it has been argued, is massive \citep{2014ApJ...781L..31S,2014MmSAI..85..253H}.
In the Milky Way, the chemical characteristics of the thick disc favours a scenario where most of the gas has been acquired by the Milky Way early \citep{2016A&A...589A..66H}, and the thick disc formed from that reservoir, reaching a fraction that could reach half of the present day stellar mass of the disc of our Galaxy  \citep{2014ApJ...781L..31S,2014MmSAI..85..253H}
%These estimates for our Galaxy are also in agreement with the mass growth rate of Milky Way mass galaxies \citep{2013ApJ...771L..35V}, which are forming half their stellar content before redshift 1-1.5.
%, and now find a new echo with the more recent findings from the JWST \citep{2023ApJ...946L..15K}. 
Thus, these arguments point to a scenario where most of the baryonic disc mass has been in place early, more in favour of establishing a high baryonic to dark matter fraction and an early bar formation epoch.

It has previously been suggested that the formation of the bar may have been crucial in the transition from the thick to thin disc \citep{2013A&A...560A.109H, 2016A&A...589A..66H}, by quenching the star formation activity within the corotation region in less than 1 Gyr. Simulations \citep{2018A&A...609A..60K} have shown that this is possible. In particular, it would explain the sudden drop of the star formation rate, which reached a minimum 8-9 Gyr ago \citep{2015A&A...578A..87S,2016A&A...589A..66H}, again placing the formation of the bar in this time interval, 8-10 Gyr. It is difficult to establish a precise sequence of events, all of which seem to have occurred within 2-3 Gyr, but the timeline could be as follows.

The bar begins to form 9-10~Gyr ago, quenching star formation but allowing stars to migrate. Star formation resumes rapidly after reaching a minimum 8-9~Gyr. The most metal-rich stars form and migrate, accompanying the expansion of the bar resonances. The slowdown ends about 7 Gyr ago, halting the episode of radial migration.
We note that our estimate of the epoch of the bar formation is similar to those obtained by \cite{2019MNRAS.490.4740B} who found that the bar formed $\sim$ 8~Gyr ago.

Our result differs from \cite{2024A&A...681L...8N}, who found, from the analysis of a sample of stars in the solar neighbourhood, that the bar formed 3-4~Gyr ago. 
Assuming that the most-metal-rich stars come from the inner regions due to the action of the bar, they deduce from a steep decrease in the number of stars younger than 3 Gyr that the bar would have formed more than 3 Gyr ago. This assumes that star formation has been active in the last 3 Gyr in the inner regions. On the contrary, several studies find a lack of young stars and regions of star formation in the inner 4kpc from the Galactic centre \citep{2013ApJ...774..117J,2016MNRAS.462..414M,2019ApJ...883...58D,2019Sci...365..478S,2023ApJ...954..124I}, while regions inside bar corotation are known to host star formation 'desert' \citep{2018MNRAS.474.3101J}. 
They also claim that the bar formation would have started 4 Gyr ago because bars trigger star formation activity in galactic centres which typically lasts 1 Gyr.  They draw a parallel with an episode of star formation that occurred about 3 Gyr ago at the solar radius, implying that it would have been triggered by the bar formation. We note that \cite{2024MNRAS.530.2972S}, using the same argument that nuclear regions are fuelled by gas funneled at the epoch of the bar formation, date stars in the nuclear regions to find that the bar is more than 8 Gyr old, in agreement with our results.

\subsection{Compatibility with the solar neighbourhood evolution}\label{gradient}

In \cite{2019A&A...625A.105H}, we speculated that the bar may have been responsible for the dilution of the gas in the region of the disc outside the OLR. The angular momentum redistribution induced by the bar drives the (metal-rich) gas from corotation and the (metal-poor) gas from the outer disc to the OLR. If the OLR first appeared inside the solar orbit, we would expect the solar annulus (before the formation of the Sun) to have been diluted. This may be required since the metallicity of the ISM at the end of the formation of the thick disc (9-10~Gyr ago) had solar metallicity, so it is expected that the disc at the solar radius would have reached above solar metallicity 4 Gyr later, at the time of formation of the Sun, except if some kind of dilution had intervened. This picture still holds here. Given that the OLR is now located at 10.5-12.5~kpc, we need to understand whether the OLR could have expanded to this limit from 6~kpc, as advocated in  \cite{2019A&A...625A.105H}, or from a larger initial radius. An initial OLR radius could have been in the range 6-8~kpc without fundamentally changing our scenario.

%%%%%%%%%%%%%%%%%%%%%%%%%%%%%%%%%%%%%%%%
%%%% Conclusions
%%%%%%%%%%%%%%%%%%%%%%%%%%%%%%%%%%%%%%%%

\section{Conclusions}

We have analysed the contents of the astroNN catalogue in combination with APOGEE DR 17 and found strong evidence, supported by hydrodynamical simulations, for the formation of the MW bar 8-10 Gyr ago. This result is based on the following findings:\\

\begin{enumerate}
\item The radial metallicity profile shows a change in slope at $\sim$6 and $\sim$11 kpc, identified as the corotation and the OLR, respectively. The slope becomes steeper after the corotation and flatter after the OLR. The change in slope at the OLR is accompanied by a $\sim$0.4 dex break in the metallicity profile that occurs in less than 2~kpc and is visible in the old populations ($>$6~Gyr).
\item This step, or break, in the metallicity profile at the OLR is absent in populations younger than 4~Gyr, appears between 4 and 6~Gyr, and is maximal between 6 and 8 Gyr. It also appears as a knee in the metallicity-guiding radius distributions, and it is well reproduced by  simulations. We interpret this break as a consequence of the dynamical effects of the OLR, predicted in \cite{2015A&A...578A..58H} and \cite{2020A&A...638A.144K} to create a barrier that divides the disc into two regions, allowing their chemistry to evolve separately and limiting the migration of stars beyond this resonance. This effect is expected to occur when the bar becomes sufficiently strong, within about 1 Gyr of its formation. 

The distributions of guiding radii as a function of metallicity show very clearly that stars more metal-rich than about -0.3 are limited to at most 12~kpc from the Galactic centre, illustrating this effect. 

\item An episode of radial migration has occurred in the MW, at an age above 7-8~Gyr. This episode occurred during the phase of early slowdown of the bar which, according to simulations, lasts about 2 Gyr after the start of the formation of the bar. Stars have migrated from the inner Galaxy to a maximum distance from the Galactic centre of 11-12~kpc, the limit imposed by the OLR. This limit is clearly visible, especially in stars at low eccentricities or in very metal-rich stars (Fig.~\ref{fig:low_alpha_xy_fehMax}). Since both dispersion and skewness of the metallicity distribution are well explained by a single bar-driven migration episode that occurred in the MW, no other significant episode of radial migration can be inferred from the data. 

\item While most of the observed metallicity dispersion can be attributed to the effect of the slowing down of the bar, we suggest that the libration of orbits in a barred potential, which does not introduce a net radial migration, can have a significant effect on the metallicity dispersion between corotation and the OLR. The decrease in the dispersion down to the ISM values observed beyond the OLR suggests that no substantial stellar migration has occurred beyond this limit. Finally, because the knee is the signpost of a migration episode triggered by the bar, which together with the effect of libration can explain the observed metallicity dispersion, we see little room for significant mixing caused by transient spiral arms, which, in addition, would probably erase the metallicity break observed in the data. This implies that spiral arms are closely intertwined with the MW bar, making their appearance less transient and consequently reducing their potential to trigger radial migration.

\item The flattening of the gradient at younger ages between about 9 and 14~kpc can be interpreted as a consequence of the weakening of the barrier effect due to the decrease in the bar strength in the last gigayear, allowing gas and metals from both sides of the OLR to mix, flattening the gradient. 
\end{enumerate}

\begin{acknowledgements}
We thank the anonymous referee for helpful comments and suggestions that have improved the manuscript.
This study would not have been possible without the work of Henry Leung, Jo Bovy and Ted Mackereth in producing the astroNN catalogue. V. C., P. DM., M. H., D. K. acknowledge the support of the French Agence Nationale de la Recherche (ANR), under grant ANR-13-BS01-0005 (project ANR-20- CE31-0004-01 MWDisc). O.S. acknowledges support from an ERC Consolidator Grant (Grant Agreement ID 101003096) and STFC Consolidated Grant (ST/V000721/1).
We have made use of data from the European Space Agency (ESA) mission \textit{Gaia} (https://www.cosmos.esa.int/gaia), processed by the \textit{Gaia} Data Processing and Analysis Consortium (DPAC, https://www.cosmos.esa.int/web/gaia/dpac/consortium). Funding for the DPAC has been provided by national institutions, in particular the institutions participating in the \textit{Gaia} Multilateral Agreement.
This research made use of Astropy, a community-developed core Python package for Astronomy (Astropy Collaboration, 2018).
Funding for the Sloan Digital Sky 
Survey IV has been provided by the 
Alfred P. Sloan Foundation, the U.S. 
Department of Energy Office of 
Science, and the Participating 
Institutions. 

SDSS-IV acknowledges support and 
resources from the Center for High 
Performance Computing  at the 
University of Utah. The SDSS 
website is www.sdss.org.

SDSS-IV is managed by the 
Astrophysical Research Consortium 
for the Participating Institutions 
of the SDSS Collaboration including 
the Brazilian Participation Group, 
the Carnegie Institution for Science, 
Carnegie Mellon University, Center for 
Astrophysics | Harvard \& 
Smithsonian, the Chilean Participation 
Group, the French Participation Group, 
Instituto de Astrof\'isica de 
Canarias, The Johns Hopkins 
University, Kavli Institute for the 
Physics and Mathematics of the 
Universe (IPMU) / University of 
Tokyo, the Korean Participation Group, 
Lawrence Berkeley National Laboratory, 
Leibniz Institut f\"ur Astrophysik 
Potsdam (AIP),  Max-Planck-Institut 
f\"ur Astronomie (MPIA Heidelberg), 
Max-Planck-Institut f\"ur 
Astrophysik (MPA Garching), 
Max-Planck-Institut f\"ur 
Extraterrestrische Physik (MPE), 
National Astronomical Observatories of 
China, New Mexico State University, 
New York University, University of 
Notre Dame, Observat\'ario 
Nacional / MCTI, The Ohio State 
University, Pennsylvania State 
University, Shanghai 
Astronomical Observatory, United 
Kingdom Participation Group, 
Universidad Nacional Aut\'onoma 
de M\'exico, University of Arizona, 
University of Colorado Boulder, 
University of Oxford, University of 
Portsmouth, University of Utah, 
University of Virginia, University 
of Washington, University of 
Wisconsin, Vanderbilt University, 
and Yale University.
\end{acknowledgements}

\bibliographystyle{aa}
\bibliography{export-bibtex}

\begin{thebibliography}{152}
\expandafter\ifx\csname natexlab\endcsname\relax\def\natexlab#1{#1}\fi

\bibitem[{{Abdurro'uf} {et~al.}(2022){Abdurro'uf}, {Accetta}, {Aerts}, {Silva
  Aguirre}, {Ahumada}, {Ajgaonkar}, {Filiz Ak}, {Alam}, {Allende Prieto},
  {Almeida}, {Anders}, {Anderson}, {Andrews}, {Anguiano}, {Aquino-Ort{\'\i}z},
  {Arag{\'o}n-Salamanca}, {Argudo-Fern{\'a}ndez}, {Ata}, {Aubert},
  {Avila-Reese}, {Badenes}, {Barb{\'a}}, {Barger}, {Barrera-Ballesteros},
  {Beaton}, {Beers}, {Belfiore}, {Bender}, {Bernardi}, {Bershady}, {Beutler},
  {Bidin}, {Bird}, {Bizyaev}, {Blanc}, {Blanton}, {Boardman}, {Bolton},
  {Boquien}, {Borissova}, {Bovy}, {Brandt}, {Brown}, {Brownstein}, {Brusa},
  {Buchner}, {Bundy}, {Burchett}, {Bureau}, {Burgasser}, {Cabang}, {Campbell},
  {Cappellari}, {Carlberg}, {Wanderley}, {Carrera}, {Cash}, {Chen}, {Chen},
  {Cherinka}, {Chiappini}, {Choi}, {Chojnowski}, {Chung}, {Clerc}, {Cohen},
  {Comerford}, {Comparat}, {da Costa}, {Covey}, {Crane}, {Cruz-Gonzalez},
  {Culhane}, {Cunha}, {Dai}, {Damke}, {Darling}, {Davidson}, {Davies},
  {Dawson}, {De Lee}, {Diamond-Stanic}, {Cano-D{\'\i}az}, {S{\'a}nchez},
  {Donor}, {Duckworth}, {Dwelly}, {Eisenstein}, {Elsworth}, {Emsellem},
  {Eracleous}, {Escoffier}, {Fan}, {Farr}, {Feng}, {Fern{\'a}ndez-Trincado},
  {Feuillet}, {Filipp}, {Fillingham}, {Frinchaboy}, {Fromenteau}, {Galbany},
  {Garc{\'\i}a}, {Garc{\'\i}a-Hern{\'a}ndez}, {Ge}, {Geisler}, {Gelfand},
  {G{\'e}ron}, {Gibson}, {Goddy}, {Godoy-Rivera}, {Grabowski}, {Green},
  {Greener}, {Grier}, {Griffith}, {Guo}, {Guy}, {Hadjara}, {Harding},
  {Hasselquist}, {Hayes}, {Hearty}, {Hern{\'a}ndez}, {Hill}, {Hogg},
  {Holtzman}, {Horta}, {Hsieh}, {Hsu}, {Hsu}, {Huber}, {Huertas-Company},
  {Hutchinson}, {Hwang}, {Ibarra-Medel}, {Chitham}, {Ilha}, {Imig}, {Jaekle},
  {Jayasinghe}, {Ji}, {Johnson}, {Jones}, {J{\"o}nsson}, {Katkov}, {Khalatyan},
  {Kinemuchi}, {Kisku}, {Knapen}, {Kneib}, {Kollmeier}, {Kong}, {Kounkel},
  {Kreckel}, {Krishnarao}, {Lacerna}, {Lane}, {Langgin}, {Lavender}, {Law},
  {Lazarz}, {Leung}, {Leung}, {Lewis}, {Li}, {Li}, {Lian}, {Liang}, {Lin},
  {Lin}, {Lin}, {Lintott}, {Long}, {Longa-Pe{\~n}a}, {L{\'o}pez-Cob{\'a}},
  {Lu}, {Lundgren}, {Luo}, {Mackereth}, {de la Macorra}, {Mahadevan},
  {Majewski}, {Manchado}, {Mandeville}, {Maraston}, {Margalef-Bentabol},
  {Masseron}, {Masters}, {Mathur}, {McDermid}, {Mckay}, {Merloni},
  {Merrifield}, {Meszaros}, {Miglio}, {Di Mille}, {Minniti}, {Minsley},
  {Monachesi}, {Moon}, {Mosser}, {Mulchaey}, {Muna}, {Mu{\~n}oz}, {Myers},
  {Myers}, {Nadathur}, {Nair}, {Nandra}, {Neumann}, {Newman}, {Nidever},
  {Nikakhtar}, {Nitschelm}, {O'Connell}, {Garma-Oehmichen}, {Luan Souza de
  Oliveira}, {Olney}, {Oravetz}, {Ortigoza-Urdaneta}, {Osorio}, {Otter},
  {Pace}, {Padilla}, {Pan}, {Pan}, {Parikh}, {Parker}, {Peirani}, {Pe{\~n}a
  Ram{\'\i}rez}, {Penny}, {Percival}, {Perez-Fournon}, {Pinsonneault},
  {Poidevin}, {Poovelil}, {Price-Whelan}, {B{\'a}rbara de Andrade Queiroz},
  {Raddick}, {Ray}, {Rembold}, {Riddle}, {Riffel}, {Riffel}, {Rix}, {Robin},
  {Rodr{\'\i}guez-Puebla}, {Roman-Lopes}, {Rom{\'a}n-Z{\'u}{\~n}iga}, {Rose},
  {Ross}, {Rossi}, {Rubin}, {Salvato}, {S{\'a}nchez}, {S{\'a}nchez-Gallego},
  {Sanderson}, {Santana Rojas}, {Sarceno}, {Sarmiento}, {Sayres}, {Sazonova},
  {Schaefer}, {Schiavon}, {Schlegel}, {Schneider}, {Schultheis}, {Schwope},
  {Serenelli}, {Serna}, {Shao}, {Shapiro}, {Sharma}, {Shen}, {Shetrone}, {Shu},
  {Simon}, {Skrutskie}, {Smethurst}, {Smith}, {Sobeck}, {Spoo}, {Sprague},
  {Stark}, {Stassun}, {Steinmetz}, {Stello}, {Stone-Martinez},
  {Storchi-Bergmann}, {Stringfellow}, {Stutz}, {Su}, {Taghizadeh-Popp},
  {Talbot}, {Tayar}, {Telles}, {Teske}, {Thakar}, {Theissen}, {Tkachenko},
  {Thomas}, {Tojeiro}, {Hernandez Toledo}, {Troup}, {Trump}, {Trussler},
  {Turner}, {Tuttle}, {Unda-Sanzana}, {V{\'a}zquez-Mata}, {Valentini},
  {Valenzuela}, {Vargas-Gonz{\'a}lez}, {Vargas-Maga{\~n}a}, {Alfaro},
  {Villanova}, {Vincenzo}, {Wake}, {Warfield}, {Washington}, {Weaver},
  {Weijmans}, {Weinberg}, {Weiss}, {Westfall}, {Wild}, {Wilde}, {Wilson},
  {Wilson}, {Wilson}, {Wolf}, {Wood-Vasey}, {Yan}, {Zamora}, {Zasowski},
  {Zhang}, {Zhao}, {Zheng}, {Zheng}, \& {Zhu}}]{2022ApJS..259...35A}
{Abdurro'uf}, {Accetta}, K., {Aerts}, C., {et~al.} 2022, \apjs, 259, 35

\bibitem[{{Adibekyan} {et~al.}(2012){Adibekyan}, {Sousa}, {Santos}, {Delgado
  Mena}, {Gonz{\'a}lez Hern{\'a}ndez}, {Israelian}, {Mayor}, \&
  {Khachatryan}}]{2012A&A...545A..32A}
{Adibekyan}, V.~Z., {Sousa}, S.~G., {Santos}, N.~C., {et~al.} 2012, \aap, 545,
  A32

\bibitem[{{Anders} {et~al.}(2017){Anders}, {Chiappini}, {Minchev}, {Miglio},
  {Montalb{\'a}n}, {Mosser}, {Rodrigues}, {Santiago}, {Baudin}, {Beers}, {da
  Costa}, {Garc{\'\i}a}, {Garc{\'\i}a-Hern{\'a}ndez}, {Holtzman}, {Maia},
  {Majewski}, {Mathur}, {Noels-Grotsch}, {Pan}, {Schneider}, {Schultheis},
  {Steinmetz}, {Valentini}, \& {Zamora}}]{2017A&A...600A..70A}
{Anders}, F., {Chiappini}, C., {Minchev}, I., {et~al.} 2017, \aap, 600, A70

\bibitem[{{Anders} {et~al.}(2023){Anders}, {Gispert}, {Ratcliffe}, {Chiappini},
  {Minchev}, {Nepal}, {Queiroz}, {Amarante}, {Antoja}, {Casali}, {Casamiquela},
  {Khalatyan}, {Miglio}, {Perottoni}, \& {Schultheis}}]{2023A&A...678A.158A}
{Anders}, F., {Gispert}, P., {Ratcliffe}, B., {et~al.} 2023, \aap, 678, A158

\bibitem[{{Antoja} {et~al.}(2014){Antoja}, {Helmi}, {Dehnen}, {Bienaym{\'e}},
  {Bland-Hawthorn}, {Famaey}, {Freeman}, {Gibson}, {Gilmore}, {Grebel},
  {Kordopatis}, {Kunder}, {Minchev}, {Munari}, {Navarro}, {Parker}, {Reid},
  {Seabroke}, {Siebert}, {Steinmetz}, {Watson}, {Wyse}, \&
  {Zwitter}}]{2014A&A...563A..60A}
{Antoja}, T., {Helmi}, A., {Dehnen}, W., {et~al.} 2014, \aap, 563, A60

\bibitem[{{Antoja} {et~al.}(2009){Antoja}, {Valenzuela}, {Pichardo}, {Moreno},
  {Figueras}, \& {Fern{\'a}ndez}}]{2009ApJ...700L..78A}
{Antoja}, T., {Valenzuela}, O., {Pichardo}, B., {et~al.} 2009, \apjl, 700, L78

\bibitem[{{Arellano-C{\'o}rdova} {et~al.}(2020){Arellano-C{\'o}rdova},
  {Esteban}, {Garc{\'\i}a-Rojas}, \&
  {M{\'e}ndez-Delgado}}]{2020MNRAS.496.1051A}
{Arellano-C{\'o}rdova}, K.~Z., {Esteban}, C., {Garc{\'\i}a-Rojas}, J., \&
  {M{\'e}ndez-Delgado}, J.~E. 2020, \mnras, 496, 1051

\bibitem[{{Arellano-C{\'o}rdova} {et~al.}(2021){Arellano-C{\'o}rdova},
  {Esteban}, {Garc{\'\i}a-Rojas}, \&
  {M{\'e}ndez-Delgado}}]{2021MNRAS.502..225A}
{Arellano-C{\'o}rdova}, K.~Z., {Esteban}, C., {Garc{\'\i}a-Rojas}, J., \&
  {M{\'e}ndez-Delgado}, J.~E. 2021, \mnras, 502, 225

\bibitem[{{Asano} {et~al.}(2022){Asano}, {Fujii}, {Baba}, {B{\'e}dorf},
  {Sellentin}, \& {Portegies Zwart}}]{2022MNRAS.514..460A}
{Asano}, T., {Fujii}, M.~S., {Baba}, J., {et~al.} 2022, \mnras, 514, 460

\bibitem[{{Bensby} {et~al.}(2017){Bensby}, {Feltzing}, {Gould}, {Yee},
  {Johnson}, {Asplund}, {Mel{\'e}ndez}, {Lucatello}, {Howes}, {McWilliam},
  {Udalski}, {Szyma{\'n}ski}, {Soszy{\'n}ski}, {Poleski}, {Wyrzykowski},
  {Ulaczyk}, {Koz{\l}owski}, {Pietrukowicz}, {Skowron}, {Mr{\'o}z}, {Pawlak},
  {Abe}, {Asakura}, {Bhattacharya}, {Bond}, {Bennett}, {Hirao}, {Nagakane},
  {Koshimoto}, {Sumi}, {Suzuki}, \& {Tristram}}]{2017A&A...605A..89B}
{Bensby}, T., {Feltzing}, S., {Gould}, A., {et~al.} 2017, \aap, 605, A89

\bibitem[{{Bensby} {et~al.}(2013){Bensby}, {Yee}, {Feltzing}, {Johnson},
  {Gould}, {Cohen}, {Asplund}, {Mel{\'e}ndez}, {Lucatello}, {Han}, {Thompson},
  {Gal-Yam}, {Udalski}, {Bennett}, {Bond}, {Kohei}, {Sumi}, {Suzuki}, {Suzuki},
  {Takino}, {Tristram}, {Yamai}, \& {Yonehara}}]{2013A&A...549A.147B}
{Bensby}, T., {Yee}, J.~C., {Feltzing}, S., {et~al.} 2013, \aap, 549, A147

\bibitem[{{Binney} \& {Tremaine}(2008)}]{2008gady.book.....B}
{Binney}, J. \& {Tremaine}, S. 2008, {Galactic Dynamics: Second Edition}

\bibitem[{{Bissantz} \& {Gerhard}(2002)}]{2002MNRAS.330..591B}
{Bissantz}, N. \& {Gerhard}, O. 2002, \mnras, 330, 591

\bibitem[{{Bland-Hawthorn} {et~al.}(2023){Bland-Hawthorn}, {Tepper-Garcia},
  {Agertz}, \& {Freeman}}]{2023ApJ...947...80B}
{Bland-Hawthorn}, J., {Tepper-Garcia}, T., {Agertz}, O., \& {Freeman}, K. 2023,
  \apj, 947, 80

\bibitem[{{Blitz} \& {Spergel}(1991)}]{1991ApJ...379..631B}
{Blitz}, L. \& {Spergel}, D.~N. 1991, \apj, 379, 631

\bibitem[{{Bovy} \& {Hogg}(2010)}]{2010ApJ...717..617B}
{Bovy}, J. \& {Hogg}, D.~W. 2010, \apj, 717, 617

\bibitem[{{Bovy} {et~al.}(2019){Bovy}, {Leung}, {Hunt}, {Mackereth},
  {Garc{\'\i}a-Hern{\'a}ndez}, \& {Roman-Lopes}}]{2019MNRAS.490.4740B}
{Bovy}, J., {Leung}, H.~W., {Hunt}, J. A.~S., {et~al.} 2019, \mnras, 490, 4740

\bibitem[{{Brogaard} {et~al.}(2021){Brogaard}, {Grundahl}, {Sandquist},
  {Slumstrup}, {Jensen}, {Thomsen}, {J{\o}rgensen}, {Larsen}, {Bj{\o}rn},
  {S{\o}rensen}, {Bruntt}, {Arentoft}, {Frandsen}, {Jessen-Hansen}, {Orosz},
  {Mathieu}, {Geller}, {Ryde}, {Stello}, {Meibom}, \&
  {Platais}}]{2021A&A...649A.178B}
{Brogaard}, K., {Grundahl}, F., {Sandquist}, E.~L., {et~al.} 2021, \aap, 649,
  A178

\bibitem[{{Brook} {et~al.}(2005){Brook}, {Gibson}, {Martel}, \&
  {Kawata}}]{2005ApJ...630..298B}
{Brook}, C.~B., {Gibson}, B.~K., {Martel}, H., \& {Kawata}, D. 2005, \apj, 630,
  298

\bibitem[{{Brook} {et~al.}(2004){Brook}, {Kawata}, {Gibson}, \&
  {Freeman}}]{2004ApJ...612..894B}
{Brook}, C.~B., {Kawata}, D., {Gibson}, B.~K., \& {Freeman}, K.~C. 2004, \apj,
  612, 894

\bibitem[{{Buck}(2020)}]{2020MNRAS.491.5435B}
{Buck}, T. 2020, \mnras, 491, 5435

\bibitem[{{Buta} \& {Combes}(1996)}]{1996FCPh...17...95B}
{Buta}, R. \& {Combes}, F. 1996, \fcp, 17, 95

\bibitem[{{Cameron}(1968)}]{1968Obs....88..254C}
{Cameron}, M.~J. 1968, The Observatory, 88, 254

\bibitem[{{Carraro} {et~al.}(2004){Carraro}, {Bresolin}, {Villanova},
  {Matteucci}, {Patat}, \& {Romaniello}}]{2004AJ....128.1676C}
{Carraro}, G., {Bresolin}, F., {Villanova}, S., {et~al.} 2004, \aj, 128, 1676

\bibitem[{{Carraro} {et~al.}(2007){Carraro}, {Geisler}, {Villanova},
  {Frinchaboy}, \& {Majewski}}]{2007A&A...476..217C}
{Carraro}, G., {Geisler}, D., {Villanova}, S., {Frinchaboy}, P.~M., \&
  {Majewski}, S.~R. 2007, \aap, 476, 217

\bibitem[{{Cerqui} {et~al.}(2023){Cerqui}, {Haywood}, {Di Matteo}, {Katz}, \&
  {Royer}}]{2023A&A...676A.108C}
{Cerqui}, V., {Haywood}, M., {Di Matteo}, P., {Katz}, D., \& {Royer}, F. 2023,
  \aap, 676, A108

\bibitem[{{Ceverino} \& {Klypin}(2007)}]{2007MNRAS.379.1155C}
{Ceverino}, D. \& {Klypin}, A. 2007, \mnras, 379, 1155

\bibitem[{{Chakrabarty}(2007)}]{2007A&A...467..145C}
{Chakrabarty}, D. 2007, \aap, 467, 145

\bibitem[{{Chen} {et~al.}(2022){Chen}, {Zhao}, \&
  {Zhang}}]{2022ApJ...936L...7C}
{Chen}, Y., {Zhao}, G., \& {Zhang}, H. 2022, \apjl, 936, L7

\bibitem[{{Chiba} {et~al.}(2021){Chiba}, {Friske}, \&
  {Sch{\"o}nrich}}]{2021MNRAS.500.4710C}
{Chiba}, R., {Friske}, J. K.~S., \& {Sch{\"o}nrich}, R. 2021, \mnras, 500, 4710

\bibitem[{{Chiba} \& {Sch{\"o}nrich}(2021)}]{2021MNRAS.505.2412C}
{Chiba}, R. \& {Sch{\"o}nrich}, R. 2021, \mnras, 505, 2412

\bibitem[{{Ciambur} {et~al.}(2017){Ciambur}, {Graham}, \&
  {Bland-Hawthorn}}]{2017MNRAS.471.3988C}
{Ciambur}, B.~C., {Graham}, A.~W., \& {Bland-Hawthorn}, J. 2017, \mnras, 471,
  3988

\bibitem[{{Clarke} \& {Gerhard}(2022)}]{2022MNRAS.512.2171C}
{Clarke}, J.~P. \& {Gerhard}, O. 2022, \mnras, 512, 2171

\bibitem[{{da Silva} {et~al.}(2023){da Silva}, {D'Orazi}, {Palla}, {Bono},
  {Braga}, {Fabrizio}, {Lemasle}, {Spitoni}, {Matteucci}, {J{\"o}nsson},
  {Kovtyukh}, {Magrini}, {Bergemann}, {Dall'Ora}, {Ferraro}, {Fiorentino},
  {Fran{\c{c}}ois}, {Iannicola}, {Inno}, {Kudritzki}, {Matsunaga}, {Monelli},
  {Nonino}, {Sneden}, {Storm}, {Th{\'e}v{\'e}nin}, {Tsujimoto}, \&
  {Zocchi}}]{2023A&A...678A.195D}
{da Silva}, R., {D'Orazi}, V., {Palla}, M., {et~al.} 2023, \aap, 678, A195

\bibitem[{{Dantas} {et~al.}(2023){Dantas}, {Smiljanic}, {Boesso},
  {Rocha-Pinto}, {Magrini}, {Guiglion}, {Tautvai{\v{s}}ien{\.{e}}}, {Gilmore},
  {Randich}, {Bensby}, {Bragaglia}, {Bergemann}, {Carraro}, {Jofr{\'e}}, \&
  {Zaggia}}]{2023A&A...669A..96D}
{Dantas}, M.~L.~L., {Smiljanic}, R., {Boesso}, R., {et~al.} 2023, \aap, 669,
  A96

\bibitem[{{de Vaucouleurs}(1964)}]{1964IAUS...20..195D}
{de Vaucouleurs}, G. 1964, in The Galaxy and the Magellanic Clouds, ed. F.~J.
  {Kerr}, Vol.~20, 195

\bibitem[{{de Vaucouleurs} \& {Pence}(1978)}]{1978AJ.....83.1163D}
{de Vaucouleurs}, G. \& {Pence}, W.~D. 1978, \aj, 83, 1163

\bibitem[{{Dehnen}(2000)}]{2000AJ....119..800D}
{Dehnen}, W. 2000, \aj, 119, 800

\bibitem[{{D{\'e}k{\'a}ny} {et~al.}(2019){D{\'e}k{\'a}ny}, {Hajdu}, {Grebel},
  \& {Catelan}}]{2019ApJ...883...58D}
{D{\'e}k{\'a}ny}, I., {Hajdu}, G., {Grebel}, E.~K., \& {Catelan}, M. 2019,
  \apj, 883, 58

\bibitem[{{Di Matteo}(2016)}]{2016PASA...33...27D}
{Di Matteo}, P. 2016, \pasa, 33, e027

\bibitem[{{Di Matteo} {et~al.}(2015){Di Matteo}, {G{\'o}mez}, {Haywood},
  {Combes}, {Lehnert}, {Ness}, {Snaith}, {Katz}, \&
  {Semelin}}]{2015A&A...577A...1D}
{Di Matteo}, P., {G{\'o}mez}, A., {Haywood}, M., {et~al.} 2015, \aap, 577, A1

\bibitem[{{Di Matteo} {et~al.}(2014){Di Matteo}, {Haywood}, {G{\'o}mez}, {van
  Damme}, {Combes}, {Hall{\'e}}, {Semelin}, {Lehnert}, \&
  {Katz}}]{2014A&A...567A.122D}
{Di Matteo}, P., {Haywood}, M., {G{\'o}mez}, A., {et~al.} 2014, \aap, 567, A122

\bibitem[{{Donor} {et~al.}(2020){Donor}, {Frinchaboy}, {Cunha}, {O'Connell},
  {Allende Prieto}, {Almeida}, {Anders}, {Beaton}, {Bizyaev}, {Brownstein},
  {Carrera}, {Chiappini}, {Cohen}, {Garc{\'\i}a-Hern{\'a}ndez}, {Geisler},
  {Hasselquist}, {J{\"o}nsson}, {Lane}, {Majewski}, {Minniti}, {Bidin}, {Pan},
  {Roman-Lopes}, {Sobeck}, \& {Zasowski}}]{2020AJ....159..199D}
{Donor}, J., {Frinchaboy}, P.~M., {Cunha}, K., {et~al.} 2020, \aj, 159, 199

\bibitem[{{Englmaier} \& {Gerhard}(1999)}]{1999MNRAS.304..512E}
{Englmaier}, P. \& {Gerhard}, O. 1999, \mnras, 304, 512

\bibitem[{{Fragkoudi} {et~al.}(2017){Fragkoudi}, {Di Matteo}, {Haywood},
  {G{\'o}mez}, {Combes}, {Katz}, \& {Semelin}}]{2017A&A...606A..47F}
{Fragkoudi}, F., {Di Matteo}, P., {Haywood}, M., {et~al.} 2017, \aap, 606, A47

\bibitem[{{Fragkoudi} {et~al.}(2018){Fragkoudi}, {Di Matteo}, {Haywood},
  {Schultheis}, {Khoperskov}, {G{\'o}mez}, \& {Combes}}]{2018A&A...616A.180F}
{Fragkoudi}, F., {Di Matteo}, P., {Haywood}, M., {et~al.} 2018, \aap, 616, A180

\bibitem[{{Fragkoudi} {et~al.}(2019){Fragkoudi}, {Katz}, {Trick}, {White}, {Di
  Matteo}, {Sormani}, {Khoperskov}, {Haywood}, {Hall{\'e}}, \&
  {G{\'o}mez}}]{2019MNRAS.488.3324F}
{Fragkoudi}, F., {Katz}, D., {Trick}, W., {et~al.} 2019, \mnras, 488, 3324

\bibitem[{{Freeman} \& {Bland-Hawthorn}(2002)}]{2002ARA&A..40..487F}
{Freeman}, K. \& {Bland-Hawthorn}, J. 2002, \araa, 40, 487

\bibitem[{{Fujii} {et~al.}(2018){Fujii}, {B{\'e}dorf}, {Baba}, \& {Portegies
  Zwart}}]{2018MNRAS.477.1451F}
{Fujii}, M.~S., {B{\'e}dorf}, J., {Baba}, J., \& {Portegies Zwart}, S. 2018,
  \mnras, 477, 1451

\bibitem[{{Fukushige} {et~al.}(2005){Fukushige}, {Makino}, \&
  {Kawai}}]{2005PASJ...57.1009F}
{Fukushige}, T., {Makino}, J., \& {Kawai}, A. 2005, \pasj, 57, 1009

\bibitem[{{Fux}(2001)}]{2001A&A...373..511F}
{Fux}, R. 2001, \aap, 373, 511

\bibitem[{{Gaia Collaboration} {et~al.}(2023){Gaia Collaboration},
  {Recio-Blanco}, {Kordopatis}, {de Laverny}, {Palicio}, {Spagna}, {Spina},
  {Katz}, {Re Fiorentin}, {Poggio}, {McMillan}, {Vallenari}, {Lattanzi},
  {Seabroke}, {Casamiquela}, {Bragaglia}, {Antoja}, {Bailer-Jones},
  {Schultheis}, {Andrae}, {Fouesneau}, {Cropper}, {Cantat-Gaudin}, {Bijaoui},
  {Heiter}, {Brown}, {Prusti}, {de Bruijne}, {Arenou}, {Babusiaux}, {Biermann},
  {Creevey}, {Ducourant}, {Evans}, {Eyer}, {Guerra}, {Hutton}, {Jordi},
  {Klioner}, {Lammers}, {Lindegren}, {Luri}, {Mignard}, {Panem}, {Pourbaix},
  {Randich}, {Sartoretti}, {Soubiran}, {Tanga}, {Walton}, {Bastian}, {Drimmel},
  {Jansen}, {van Leeuwen}, {Bakker}, {Cacciari}, {Casta{\~n}eda}, {De Angeli},
  {Fabricius}, {Fr{\'e}mat}, {Galluccio}, {Guerrier}, {Masana}, {Messineo},
  {Mowlavi}, {Nicolas}, {Nienartowicz}, {Pailler}, {Panuzzo}, {Riclet}, {Roux},
  {Sordo}, {Th{\'e}venin}, {Gracia-Abril}, {Portell}, {Teyssier}, {Altmann},
  {Audard}, {Bellas-Velidis}, {Benson}, {Berthier}, {Blomme}, {Burgess},
  {Busonero}, {Busso}, {C{\'a}novas}, {Carry}, {Cellino}, {Cheek},
  {Clementini}, {Damerdji}, {Davidson}, {de Teodoro}, {Nu{\~n}ez Campos},
  {Delchambre}, {Dell'Oro}, {Esquej}, {Fern{\'a}ndez-Hern{\'a}ndez}, {Fraile},
  {Garabato}, {Garc{\'\i}a-Lario}, {Gosset}, {Haigron}, {Halbwachs}, {Hambly},
  {Harrison}, {Hern{\'a}ndez}, {Hestroffer}, {Hodgkin}, {Holl}, {Jan{\ss}en},
  {Jevardat de Fombelle}, {Jordan}, {Krone-Martins}, {Lanzafame},
  {L{\"o}ffler}, {Marchal}, {Marrese}, {Moitinho}, {Muinonen}, {Osborne},
  {Pancino}, {Pauwels}, {Reyl{\'e}}, {Riello}, {Rimoldini}, {Roegiers},
  {Rybizki}, {Sarro}, {Siopis}, {Smith}, {Sozzetti}, {Utrilla}, {van Leeuwen},
  {Abbas}, {{\'A}brah{\'a}m}, {Abreu Aramburu}, {Aerts}, {Aguado}, {Ajaj},
  {Aldea-Montero}, {Altavilla}, {{\'A}lvarez}, {Alves}, {Anders}, {Anderson},
  {Anglada Varela}, {Baines}, {Baker}, {Balaguer-N{\'u}{\~n}ez}, {Balbinot},
  {Balog}, {Barache}, {Barbato}, {Barros}, {Barstow}, {Bartolom{\'e}},
  {Bassilana}, {Bauchet}, {Becciani}, {Bellazzini}, {Berihuete}, {Bernet},
  {Bertone}, {Bianchi}, {Binnenfeld}, {Blanco-Cuaresma}, {Boch}, {Bombrun},
  {Bossini}, {Bouquillon}, {Bramante}, {Breedt}, {Bressan}, {Brouillet},
  {Brugaletta}, {Bucciarelli}, {Burlacu}, {Butkevich}, {Buzzi}, {Caffau},
  {Cancelliere}, {Carballo}, {Carlucci}, {Carnerero}, {Carrasco}, {Castellani},
  {Castro-Ginard}, {Chaoul}, {Charlot}, {Chemin}, {Chiaramida}, {Chiavassa},
  {Chornay}, {Comoretto}, {Contursi}, {Cooper}, {Cornez}, {Cowell}, {Crifo},
  {Crosta}, {Crowley}, {Dafonte}, {Dapergolas}, {David}, {De Luise}, {De
  March}, {De Ridder}, {de Souza}, {de Torres}, {del Peloso}, {del Pozo},
  {Delbo}, {Delgado}, {Delisle}, {Demouchy}, {Dharmawardena}, {Di Matteo},
  {Diakite}, {Diener}, {Distefano}, {Dolding}, {Edvardsson}, {Enke}, {Fabre},
  {Fabrizio}, {Faigler}, {Fedorets}, {Fernique}, {Figueras}, {Fournier},
  {Fouron}, {Fragkoudi}, {Gai}, {Garcia-Gutierrez}, {Garcia-Reinaldos},
  {Garc{\'\i}a-Torres}, {Garofalo}, {Gavel}, {Gavras}, {Gerlach}, {Geyer},
  {Giacobbe}, {Gilmore}, {Girona}, {Giuffrida}, {Gomel}, {Gomez},
  {Gonz{\'a}lez-N{\'u}{\~n}ez}, {Gonz{\'a}lez-Santamar{\'\i}a},
  {Gonz{\'a}lez-Vidal}, {Granvik}, {Guillout}, {Guiraud},
  {Guti{\'e}rrez-S{\'a}nchez}, {Guy}, {Hatzidimitriou}, {Hauser}, {Haywood},
  {Helmer}, {Helmi}, {Sarmiento}, {Hidalgo}, {H{\l}adczuk}, {Hobbs}, {Holland},
  {Huckle}, {Jardine}, {Jasniewicz}, {Jean-Antoine Piccolo},
  {Jim{\'e}nez-Arranz}, {Juaristi Campillo}, {Julbe}, {Karbevska}, {Kervella},
  {Khanna}, {Korn}, {K{\'o}sp{\'a}l}, {Kostrzewa-Rutkowska}, {Kruszy{\'n}ska},
  {Kun}, {Laizeau}, {Lambert}, {Lanza}, {Lasne}, {Le Campion}, {Lebreton},
  {Lebzelter}, {Leccia}, {Leclerc}, {Lecoeur-Taibi}, {Liao}, {Licata},
  {Lindstr{\o}m}, {Lister}, {Livanou}, {Lobel}, {Lorca}, {Loup}, {Madrero
  Pardo}, {Magdaleno Romeo}, {Managau}, {Mann}, {Manteiga}, {Marchant},
  {Marconi}, {Marcos}, {Marcos Santos}, {Mar{\'\i}n Pina}, {Marinoni},
  {Marocco}, {Marshall}, {Martin Polo}, {Mart{\'\i}n-Fleitas}, {Marton},
  {Mary}, {Masip}, {Massari}, {Mastrobuono-Battisti}, {Mazeh}, {Messina},
  {Michalik}, {Millar}, {Mints}, {Molina}, {Molinaro}, {Moln{\'a}r}, {Monari},
  {Mongui{\'o}}, {Montegriffo}, {Montero}, {Mor}, {Mora}, {Morbidelli},
  {Morel}, {Morris}, {Muraveva}, {Murphy}, {Musella}, {Nagy}, {Noval},
  {Oca{\~n}a}, {Ogden}, {Ordenovic}, {Osinde}, {Pagani}, {Pagano}, {Palaversa},
  {Pallas-Quintela}, {Panahi}, {Payne-Wardenaar}, {Pe{\~n}alosa Esteller},
  {Penttil{\"a}}, {Pichon}, {Piersimoni}, {Pineau}, {Plachy}, {Plum},
  {Pr{\v{s}}a}, {Pulone}, {Racero}, {Ragaini}, {Rainer}, {Raiteri}, {Ramos},
  {Ramos-Lerate}, {Regibo}, {Richards}, {Rios Diaz}, {Ripepi}, {Riva}, {Rix},
  {Rixon}, {Robichon}, {Robin}, {Robin}, {Roelens}, {Rogues}, {Rohrbasser},
  {Romero-G{\'o}mez}, {Rowell}, {Royer}, {Ruz Mieres}, {Rybicki}, {Sadowski},
  {S{\'a}ez N{\'u}{\~n}ez}, {Sagrist{\`a} Sell{\'e}s}, {Sahlmann}, {Salguero},
  {Samaras}, {Sanchez Gimenez}, {Sanna}, {Santove{\~n}a}, {Sarasso}, {Sciacca},
  {Segol}, {Segovia}, {S{\'e}gransan}, {Semeux}, {Shahaf}, {Siddiqui},
  {Siebert}, {Siltala}, {Silvelo}, {Slezak}, {Slezak}, {Smart}, {Snaith},
  {Solano}, {Solitro}, {Souami}, {Souchay}, {Spoto}, {Steele},
  {Steidelm{\"u}ller}, {Stephenson}, {S{\"u}veges}, {Surdej}, {Szabados},
  {Szegedi-Elek}, {Taris}, {Taylor}, {Teixeira}, {Tolomei}, {Tonello}, {Torra},
  {Torra}, {Torralba Elipe}, {Trabucchi}, {Tsounis}, {Turon}, {Ulla}, {Unger},
  {Vaillant}, {van Dillen}, {van Reeven}, {Vanel}, {Vecchiato}, {Viala},
  {Vicente}, {Voutsinas}, {Weiler}, {Wevers}, {Wyrzykowski}, {Yoldas}, {Yvard},
  {Zhao}, {Zorec}, {Zucker}, \& {Zwitter}}]{2023A&A...674A..38G}
{Gaia Collaboration}, {Recio-Blanco}, A., {Kordopatis}, G., {et~al.} 2023,
  \aap, 674, A38

\bibitem[{{Genovali} {et~al.}(2014){Genovali}, {Lemasle}, {Bono}, {Romaniello},
  {Fabrizio}, {Ferraro}, {Iannicola}, {Laney}, {Nonino}, {Bergemann},
  {Buonanno}, {Fran{\c{c}}ois}, {Inno}, {Kudritzki}, {Matsunaga}, {Pedicelli},
  {Primas}, \& {Th{\'e}venin}}]{2014A&A...566A..37G}
{Genovali}, K., {Lemasle}, B., {Bono}, G., {et~al.} 2014, \aap, 566, A37

\bibitem[{{Ghosh} {et~al.}(2023){Ghosh}, {Fragkoudi}, {Di Matteo}, \&
  {Saha}}]{2023A&A...674A.128G}
{Ghosh}, S., {Fragkoudi}, F., {Di Matteo}, P., \& {Saha}, K. 2023, \aap, 674,
  A128

\bibitem[{{Grand} {et~al.}(2019){Grand}, {van de Voort}, {Zjupa}, {Fragkoudi},
  {G{\'o}mez}, {Kauffmann}, {Marinacci}, {Pakmor}, {Springel}, \&
  {White}}]{2019MNRAS.490.4786G}
{Grand}, R. J.~J., {van de Voort}, F., {Zjupa}, J., {et~al.} 2019, \mnras, 490,
  4786

\bibitem[{{GRAVITY Collaboration} {et~al.}(2019){GRAVITY Collaboration},
  {Abuter}, {Amorim}, {Baub{\"o}ck}, {Berger}, {Bonnet}, {Brandner},
  {Cl{\'e}net}, {Coud{\'e} Du Foresto}, {de Zeeuw}, {Dexter}, {Duvert},
  {Eckart}, {Eisenhauer}, {F{\"o}rster Schreiber}, {Garcia}, {Gao}, {Gendron},
  {Genzel}, {Gerhard}, {Gillessen}, {Habibi}, {Haubois}, {Henning}, {Hippler},
  {Horrobin}, {Jim{\'e}nez-Rosales}, {Jocou}, {Kervella}, {Lacour},
  {Lapeyr{\`e}re}, {Le Bouquin}, {L{\'e}na}, {Ott}, {Paumard}, {Perraut},
  {Perrin}, {Pfuhl}, {Rabien}, {Rodriguez Coira}, {Rousset}, {Scheithauer},
  {Sternberg}, {Straub}, {Straubmeier}, {Sturm}, {Tacconi}, {Vincent}, {von
  Fellenberg}, {Waisberg}, {Widmann}, {Wieprecht}, {Wiezorrek}, {Woillez}, \&
  {Yazici}}]{2019A&A...625L..10G}
{GRAVITY Collaboration}, {Abuter}, R., {Amorim}, A., {et~al.} 2019, \aap, 625,
  L10

\bibitem[{{Halle} {et~al.}(2015){Halle}, {Di Matteo}, {Haywood}, \&
  {Combes}}]{2015A&A...578A..58H}
{Halle}, A., {Di Matteo}, P., {Haywood}, M., \& {Combes}, F. 2015, \aap, 578,
  A58

\bibitem[{{Halle} {et~al.}(2018){Halle}, {Di Matteo}, {Haywood}, \&
  {Combes}}]{2018A&A...616A..86H}
{Halle}, A., {Di Matteo}, P., {Haywood}, M., \& {Combes}, F. 2018, \aap, 616,
  A86

\bibitem[{{Hayden} {et~al.}(2020){Hayden}, {Bland-Hawthorn}, {Sharma},
  {Freeman}, {Kos}, {Buder}, {Anguiano}, {Asplund}, {Chen}, {De Silva},
  {Khanna}, {Lin}, {Horner}, {Martell}, {Ting}, {Wyse}, {Zucker}, \&
  {Zwitter}}]{2020MNRAS.493.2952H}
{Hayden}, M.~R., {Bland-Hawthorn}, J., {Sharma}, S., {et~al.} 2020, \mnras,
  493, 2952

\bibitem[{{Hayden} {et~al.}(2015){Hayden}, {Bovy}, {Holtzman}, {Nidever},
  {Bird}, {Weinberg}, {Andrews}, {Majewski}, {Allende Prieto}, {Anders},
  {Beers}, {Bizyaev}, {Chiappini}, {Cunha}, {Frinchaboy},
  {Garc{\'\i}a-Her{\'n}andez}, {Garc{\'\i}a P{\'e}rez}, {Girardi}, {Harding},
  {Hearty}, {Johnson}, {M{\'e}sz{\'a}ros}, {Minchev}, {O'Connell}, {Pan},
  {Robin}, {Schiavon}, {Schneider}, {Schultheis}, {Shetrone}, {Skrutskie},
  {Steinmetz}, {Smith}, {Wilson}, {Zamora}, \&
  {Zasowski}}]{2015ApJ...808..132H}
{Hayden}, M.~R., {Bovy}, J., {Holtzman}, J.~A., {et~al.} 2015, \apj, 808, 132

\bibitem[{{Hayden} {et~al.}(2018){Hayden}, {Recio-Blanco}, {de Laverny},
  {Mikolaitis}, {Guiglion}, {Hill}, {Gilmore}, {Randich}, {Bayo}, {Bensby},
  {Bergemann}, {Bragaglia}, {Casey}, {Costado}, {Feltzing}, {Franciosini},
  {Hourihane}, {Jofre}, {Koposov}, {Kordopatis}, {Lanzafame}, {Lardo}, {Lewis},
  {Lind}, {Magrini}, {Monaco}, {Morbidelli}, {Pancino}, {Sacco}, {Stonkute},
  {Worley}, \& {Zwitter}}]{2018A&A...609A..79H}
{Hayden}, M.~R., {Recio-Blanco}, A., {de Laverny}, P., {et~al.} 2018, \aap,
  609, A79

\bibitem[{{Haywood}(2008)}]{2008MNRAS.388.1175H}
{Haywood}, M. 2008, \mnras, 388, 1175

\bibitem[{{Haywood}(2014)}]{2014MmSAI..85..253H}
{Haywood}, M. 2014, \memsai, 85, 253

\bibitem[{{Haywood} {et~al.}(2013){Haywood}, {Di Matteo}, {Lehnert}, {Katz}, \&
  {G{\'o}mez}}]{2013A&A...560A.109H}
{Haywood}, M., {Di Matteo}, P., {Lehnert}, M.~D., {Katz}, D., \& {G{\'o}mez},
  A. 2013, \aap, 560, A109

\bibitem[{{Haywood} {et~al.}(2016{\natexlab{a}}){Haywood}, {Di Matteo},
  {Snaith}, \& {Calamida}}]{2016A&A...593A..82H}
{Haywood}, M., {Di Matteo}, P., {Snaith}, O., \& {Calamida}, A.
  2016{\natexlab{a}}, \aap, 593, A82

\bibitem[{{Haywood} {et~al.}(2015){Haywood}, {Di Matteo}, {Snaith}, \&
  {Lehnert}}]{2015A&A...579A...5H}
{Haywood}, M., {Di Matteo}, P., {Snaith}, O., \& {Lehnert}, M.~D. 2015, \aap,
  579, A5

\bibitem[{{Haywood} {et~al.}(2016{\natexlab{b}}){Haywood}, {Lehnert}, {Di
  Matteo}, {Snaith}, {Schultheis}, {Katz}, \&
  {G{\'o}mez}}]{2016A&A...589A..66H}
{Haywood}, M., {Lehnert}, M.~D., {Di Matteo}, P., {et~al.} 2016{\natexlab{b}},
  \aap, 589, A66

\bibitem[{{Haywood} {et~al.}(2019){Haywood}, {Snaith}, {Lehnert}, {Di Matteo},
  \& {Khoperskov}}]{2019A&A...625A.105H}
{Haywood}, M., {Snaith}, O., {Lehnert}, M.~D., {Di Matteo}, P., \&
  {Khoperskov}, S. 2019, \aap, 625, A105

\bibitem[{{Hilmi} {et~al.}(2020){Hilmi}, {Minchev}, {Buck}, {Martig},
  {Quillen}, {Monari}, {Famaey}, {de Jong}, {Laporte}, {Read}, {Sanders},
  {Steinmetz}, \& {Wegg}}]{2020MNRAS.497..933H}
{Hilmi}, T., {Minchev}, I., {Buck}, T., {et~al.} 2020, \mnras, 497, 933

\bibitem[{{Hunt} \& {Bovy}(2018)}]{2018MNRAS.477.3945H}
{Hunt}, J. A.~S. \& {Bovy}, J. 2018, \mnras, 477, 3945

\bibitem[{{Imig} {et~al.}(2023){Imig}, {Price}, {Holtzman}, {Stone-Martinez},
  {Majewski}, {Weinberg}, {Johnson}, {Allende Prieto}, {Beaton}, {Beers},
  {Bizyaev}, {Blanton}, {Brownstein}, {Cunha}, {Fern{\'a}ndez-Trincado},
  {Feuillet}, {Hasselquist}, {Hayes}, {J{\"o}nsson}, {Lane}, {Lian},
  {M{\'e}sz{\'a}ros}, {Nidever}, {Robin}, {Shetrone}, {Smith}, \&
  {Wilson}}]{2023ApJ...954..124I}
{Imig}, J., {Price}, C., {Holtzman}, J.~A., {et~al.} 2023, \apj, 954, 124

\bibitem[{{James} \& {Percival}(2018)}]{2018MNRAS.474.3101J}
{James}, P.~A. \& {Percival}, S.~M. 2018, \mnras, 474, 3101

\bibitem[{{Jones} {et~al.}(2013){Jones}, {Dickey}, {Dawson},
  {McClure-Griffiths}, {Anderson}, \& {Bania}}]{2013ApJ...774..117J}
{Jones}, C., {Dickey}, J.~M., {Dawson}, J.~R., {et~al.} 2013, \apj, 774, 117

\bibitem[{{Katz} {et~al.}(2021){Katz}, {G{\'o}mez}, {Haywood}, {Snaith}, \& {Di
  Matteo}}]{2021A&A...655A.111K}
{Katz}, D., {G{\'o}mez}, A., {Haywood}, M., {Snaith}, O., \& {Di Matteo}, P.
  2021, \aap, 655, A111

\bibitem[{{Kawata} {et~al.}(2021){Kawata}, {Baba}, {Hunt}, {Sch{\"o}nrich},
  {Ciuc{\u{a}}}, {Friske}, {Seabroke}, \& {Cropper}}]{2021MNRAS.508..728K}
{Kawata}, D., {Baba}, J., {Hunt}, J. A.~S., {et~al.} 2021, \mnras, 508, 728

\bibitem[{{Kerr}(1967)}]{1967IAUS...31..239K}
{Kerr}, F.~J. 1967, in Radio Astronomy and the Galactic System, ed. H.~{van
  Woerden}, Vol.~31, 239

\bibitem[{{Khoperskov} {et~al.}(2019){Khoperskov}, {Di Matteo}, {Gerhard},
  {Katz}, {Haywood}, {Combes}, {Berczik}, \& {Gomez}}]{2019A&A...622L...6K}
{Khoperskov}, S., {Di Matteo}, P., {Gerhard}, O., {et~al.} 2019, \aap, 622, L6

\bibitem[{{Khoperskov} {et~al.}(2018{\natexlab{a}}){Khoperskov}, {Di Matteo},
  {Haywood}, \& {Combes}}]{2018A&A...611L...2K}
{Khoperskov}, S., {Di Matteo}, P., {Haywood}, M., \& {Combes}, F.
  2018{\natexlab{a}}, \aap, 611, L2

\bibitem[{{Khoperskov} {et~al.}(2020{\natexlab{a}}){Khoperskov}, {Di Matteo},
  {Haywood}, {G{\'o}mez}, \& {Snaith}}]{2020A&A...638A.144K}
{Khoperskov}, S., {Di Matteo}, P., {Haywood}, M., {G{\'o}mez}, A., \& {Snaith},
  O.~N. 2020{\natexlab{a}}, \aap, 638, A144

\bibitem[{{Khoperskov} \& {Gerhard}(2022)}]{2022A&A...663A..38K}
{Khoperskov}, S. \& {Gerhard}, O. 2022, \aap, 663, A38

\bibitem[{{Khoperskov} {et~al.}(2020{\natexlab{b}}){Khoperskov}, {Gerhard}, {Di
  Matteo}, {Haywood}, {Katz}, {Khrapov}, {Khoperskov}, \&
  {Arnaboldi}}]{2020A&A...634L...8K}
{Khoperskov}, S., {Gerhard}, O., {Di Matteo}, P., {et~al.} 2020{\natexlab{b}},
  \aap, 634, L8

\bibitem[{{Khoperskov} {et~al.}(2018{\natexlab{b}}){Khoperskov}, {Haywood}, {Di
  Matteo}, {Lehnert}, \& {Combes}}]{2018A&A...609A..60K}
{Khoperskov}, S., {Haywood}, M., {Di Matteo}, P., {Lehnert}, M.~D., \&
  {Combes}, F. 2018{\natexlab{b}}, \aap, 609, A60

\bibitem[{{Khoperskov} {et~al.}(2021){Khoperskov}, {Haywood}, {Snaith}, {Di
  Matteo}, {Lehnert}, {Vasiliev}, {Naroenkov}, \&
  {Berczik}}]{2021MNRAS.501.5176K}
{Khoperskov}, S., {Haywood}, M., {Snaith}, O., {et~al.} 2021, \mnras, 501, 5176

\bibitem[{{Khoperskov} {et~al.}(2018{\natexlab{c}}){Khoperskov},
  {Mastrobuono-Battisti}, {Di Matteo}, \& {Haywood}}]{2018A&A...620A.154K}
{Khoperskov}, S., {Mastrobuono-Battisti}, A., {Di Matteo}, P., \& {Haywood}, M.
  2018{\natexlab{c}}, \aap, 620, A154

\bibitem[{{Khoperskov} {et~al.}(2014){Khoperskov}, {Vasiliev}, {Khoperskov}, \&
  {Lubimov}}]{2014JPhCS.510a2011K}
{Khoperskov}, S.~A., {Vasiliev}, E.~O., {Khoperskov}, A.~V., \& {Lubimov},
  V.~N. 2014, in Journal of Physics Conference Series, Vol. 510, Journal of
  Physics Conference Series, 012011

\bibitem[{{Kordopatis} {et~al.}(2015){Kordopatis}, {Binney}, {Gilmore}, {Wyse},
  {Belokurov}, {McMillan}, {Hatfield}, {Grebel}, {Steinmetz}, {Navarro},
  {Seabroke}, {Minchev}, {Chiappini}, {Bienaym{\'e}}, {Bland-Hawthorn},
  {Freeman}, {Gibson}, {Helmi}, {Munari}, {Parker}, {Reid}, {Siebert},
  {Siviero}, \& {Zwitter}}]{2015MNRAS.447.3526K}
{Kordopatis}, G., {Binney}, J., {Gilmore}, G., {et~al.} 2015, \mnras, 447, 3526

\bibitem[{{Kovtyukh} {et~al.}(2022){Kovtyukh}, {Lemasle}, {Bono}, {Usenko}, {da
  Silva}, {Kniazev}, {Grebel}, {Andronov}, {Shakun}, \&
  {Chinarova}}]{2022MNRAS.510.1894K}
{Kovtyukh}, V., {Lemasle}, B., {Bono}, G., {et~al.} 2022, \mnras, 510, 1894

\bibitem[{{Leung} \& {Bovy}(2019)}]{2019MNRAS.483.3255L}
{Leung}, H.~W. \& {Bovy}, J. 2019, \mnras, 483, 3255

\bibitem[{{Lucey} {et~al.}(2023){Lucey}, {Pearson}, {Hunt}, {Hawkins}, {Ness},
  {Petersen}, {Price-Whelan}, \& {Weinberg}}]{2023MNRAS.520.4779L}
{Lucey}, M., {Pearson}, S., {Hunt}, J. A.~S., {et~al.} 2023, \mnras, 520, 4779

\bibitem[{{Mackereth} {et~al.}(2019){Mackereth}, {Bovy}, {Leung}, {Schiavon},
  {Trick}, {Chaplin}, {Cunha}, {Feuillet}, {Majewski}, {Martig}, {Miglio},
  {Nidever}, {Pinsonneault}, {Aguirre}, {Sobeck}, {Tayar}, \&
  {Zasowski}}]{2019MNRAS.489..176M}
{Mackereth}, J.~T., {Bovy}, J., {Leung}, H.~W., {et~al.} 2019, \mnras, 489, 176

\bibitem[{{Mackereth} {et~al.}(2017){Mackereth}, {Bovy}, {Schiavon},
  {Zasowski}, {Cunha}, {Frinchaboy}, {Garc{\'\i}a Perez}, {Hayden}, {Holtzman},
  {Majewski}, {M{\'e}sz{\'a}ros}, {Nidever}, {Pinsonneault}, \&
  {Shetrone}}]{2017MNRAS.471.3057M}
{Mackereth}, J.~T., {Bovy}, J., {Schiavon}, R.~P., {et~al.} 2017, \mnras, 471,
  3057

\bibitem[{{Mackereth} {et~al.}(2018){Mackereth}, {Crain}, {Schiavon}, {Schaye},
  {Theuns}, \& {Schaller}}]{2018MNRAS.477.5072M}
{Mackereth}, J.~T., {Crain}, R.~A., {Schiavon}, R.~P., {et~al.} 2018, \mnras,
  477, 5072

\bibitem[{{Magrini} {et~al.}(2023){Magrini}, {Viscasillas V{\'a}zquez},
  {Spina}, {Randich}, {Romano}, {Franciosini}, {Recio-Blanco}, {Nordlander},
  {D'Orazi}, {Baratella}, {Smiljanic}, {Dantas}, {Pasquini}, {Spitoni},
  {Casali}, {Van der Swaelmen}, {Bensby}, {Stonkute}, {Feltzing}, {Sacco},
  {Bragaglia}, {Pancino}, {Heiter}, {Biazzo}, {Gilmore}, {Bergemann},
  {Tautvai{\v{s}}ien{\.{e}}}, {Worley}, {Hourihane}, {Gonneau}, \&
  {Morbidelli}}]{2023A&A...669A.119M}
{Magrini}, L., {Viscasillas V{\'a}zquez}, C., {Spina}, L., {et~al.} 2023, \aap,
  669, A119

\bibitem[{{Mart{\'\i}nez-Barbosa} {et~al.}(2016){Mart{\'\i}nez-Barbosa},
  {Brown}, {Boekholt}, {Portegies Zwart}, {Antiche}, \&
  {Antoja}}]{2016MNRAS.457.1062M}
{Mart{\'\i}nez-Barbosa}, C.~A., {Brown}, A.~G.~A., {Boekholt}, T., {et~al.}
  2016, \mnras, 457, 1062

\bibitem[{{Mart{\'\i}nez-Barbosa} {et~al.}(2015){Mart{\'\i}nez-Barbosa},
  {Brown}, \& {Portegies Zwart}}]{2015MNRAS.446..823M}
{Mart{\'\i}nez-Barbosa}, C.~A., {Brown}, A.~G.~A., \& {Portegies Zwart}, S.
  2015, \mnras, 446, 823

\bibitem[{{Matsunaga} {et~al.}(2016){Matsunaga}, {Feast}, {Bono}, {Kobayashi},
  {Inno}, {Nagayama}, {Nishiyama}, {Matsuoka}, \&
  {Nagata}}]{2016MNRAS.462..414M}
{Matsunaga}, N., {Feast}, M.~W., {Bono}, G., {et~al.} 2016, \mnras, 462, 414

\bibitem[{{McKeever} {et~al.}(2019){McKeever}, {Basu}, \&
  {Corsaro}}]{2019ApJ...874..180M}
{McKeever}, J.~M., {Basu}, S., \& {Corsaro}, E. 2019, \apj, 874, 180

\bibitem[{{Miglio} {et~al.}(2021){Miglio}, {Chiappini}, {Mackereth}, {Davies},
  {Brogaard}, {Casagrande}, {Chaplin}, {Girardi}, {Kawata}, {Khan}, {Izzard},
  {Montalb{\'a}n}, {Mosser}, {Vincenzo}, {Bossini}, {Noels}, {Rodrigues},
  {Valentini}, \& {Mandel}}]{2021A&A...645A..85M}
{Miglio}, A., {Chiappini}, C., {Mackereth}, J.~T., {et~al.} 2021, \aap, 645,
  A85

\bibitem[{{Minchev} \& {Famaey}(2010)}]{2010ApJ...722..112M}
{Minchev}, I. \& {Famaey}, B. 2010, \apj, 722, 112

\bibitem[{{Minchev} {et~al.}(2011){Minchev}, {Famaey}, {Combes}, {Di Matteo},
  {Mouhcine}, \& {Wozniak}}]{2011A&A...527A.147M}
{Minchev}, I., {Famaey}, B., {Combes}, F., {et~al.} 2011, \aap, 527, A147

\bibitem[{{Minchev} {et~al.}(2012){Minchev}, {Famaey}, {Quillen}, {Di Matteo},
  {Combes}, {Vlaji{\'c}}, {Erwin}, \& {Bland-Hawthorn}}]{2012A&A...548A.126M}
{Minchev}, I., {Famaey}, B., {Quillen}, A.~C., {et~al.} 2012, \aap, 548, A126

\bibitem[{{Minchev} {et~al.}(2009){Minchev}, {Quillen}, {Williams}, {Freeman},
  {Nordhaus}, {Siebert}, \& {Bienaym{\'e}}}]{2009MNRAS.396L..56M}
{Minchev}, I., {Quillen}, A.~C., {Williams}, M., {et~al.} 2009, \mnras, 396,
  L56

\bibitem[{{Miyamoto} \& {Nagai}(1975)}]{1975PASJ...27..533M}
{Miyamoto}, M. \& {Nagai}, R. 1975, \pasj, 27, 533

\bibitem[{{Monari} {et~al.}(2015){Monari}, {Famaey}, \&
  {Siebert}}]{2015MNRAS.452..747M}
{Monari}, G., {Famaey}, B., \& {Siebert}, A. 2015, \mnras, 452, 747

\bibitem[{{Monari} {et~al.}(2019){Monari}, {Famaey}, {Siebert}, {Wegg}, \&
  {Gerhard}}]{2019A&A...626A..41M}
{Monari}, G., {Famaey}, B., {Siebert}, A., {Wegg}, C., \& {Gerhard}, O. 2019,
  \aap, 626, A41

\bibitem[{{Moreno} {et~al.}(2015){Moreno}, {Pichardo}, \&
  {Schuster}}]{2015MNRAS.451..705M}
{Moreno}, E., {Pichardo}, B., \& {Schuster}, W.~J. 2015, \mnras, 451, 705

\bibitem[{{Myers} {et~al.}(2022){Myers}, {Donor}, {Spoo}, {Frinchaboy},
  {Cunha}, {Price-Whelan}, {Majewski}, {Beaton}, {Zasowski}, {O'Connell},
  {Ray}, {Bizyaev}, {Chiappini}, {Garc{\'\i}a-Hern{\'a}ndez}, {Geisler},
  {J{\"o}nsson}, {Lane}, {Longa-Pe{\~n}a}, {Minchev}, {Minniti}, {Nitschelm},
  \& {Roman-Lopes}}]{2022AJ....164...85M}
{Myers}, N., {Donor}, J., {Spoo}, T., {et~al.} 2022, \aj, 164, 85

\bibitem[{{Nakada} {et~al.}(1991){Nakada}, {Onaka}, {Yamamura}, {Deguchi},
  {Hashimoto}, {Izumiura}, \& {Sekiguchi}}]{1991Natur.353..140N}
{Nakada}, Y., {Onaka}, T., {Yamamura}, I., {et~al.} 1991, \nat, 353, 140

\bibitem[{{Nataf} {et~al.}(2010){Nataf}, {Udalski}, {Gould}, {Fouqu{\'e}}, \&
  {Stanek}}]{2010ApJ...721L..28N}
{Nataf}, D.~M., {Udalski}, A., {Gould}, A., {Fouqu{\'e}}, P., \& {Stanek},
  K.~Z. 2010, \apjl, 721, L28

\bibitem[{{Nepal} {et~al.}(2024){Nepal}, {Chiappini}, {Guiglion}, {Steinmetz},
  {P{\'e}rez-Villegas}, {Queiroz}, {Miglio}, {Dohme}, \&
  {Khalatyan}}]{2024A&A...681L...8N}
{Nepal}, S., {Chiappini}, C., {Guiglion}, G., {et~al.} 2024, \aap, 681, L8

\bibitem[{{Ness} {et~al.}(2014){Ness}, {Debattista}, {Bensby}, {Feltzing},
  {Ro{\v{s}}kar}, {Cole}, {Johnson}, \& {Freeman}}]{2014ApJ...787L..19N}
{Ness}, M., {Debattista}, V.~P., {Bensby}, T., {et~al.} 2014, \apjl, 787, L19

\bibitem[{{Ness} {et~al.}(2013{\natexlab{a}}){Ness}, {Freeman}, {Athanassoula},
  {Wylie-de-Boer}, {Bland-Hawthorn}, {Asplund}, {Lewis}, {Yong}, {Lane}, \&
  {Kiss}}]{2013MNRAS.430..836N}
{Ness}, M., {Freeman}, K., {Athanassoula}, E., {et~al.} 2013{\natexlab{a}},
  \mnras, 430, 836

\bibitem[{{Ness} {et~al.}(2013{\natexlab{b}}){Ness}, {Freeman}, {Athanassoula},
  {Wylie-de-Boer}, {Bland-Hawthorn}, {Asplund}, {Lewis}, {Yong}, {Lane},
  {Kiss}, \& {Ibata}}]{2013MNRAS.432.2092N}
{Ness}, M., {Freeman}, K., {Athanassoula}, E., {et~al.} 2013{\natexlab{b}},
  \mnras, 432, 2092

\bibitem[{{Ness} {et~al.}(2012){Ness}, {Freeman}, {Athanassoula},
  {Wylie-De-Boer}, {Bland-Hawthorn}, {Lewis}, {Yong}, {Asplund}, {Lane},
  {Kiss}, \& {Ibata}}]{2012ApJ...756...22N}
{Ness}, M., {Freeman}, K., {Athanassoula}, E., {et~al.} 2012, \apj, 756, 22

\bibitem[{{Netopil} {et~al.}(2022){Netopil}, {Oralhan}, {{\c{C}}akmak},
  {Michel}, \& {Karata{\c{s}}}}]{2022MNRAS.509..421N}
{Netopil}, M., {Oralhan}, {\.I}.~A., {{\c{C}}akmak}, H., {Michel}, R., \&
  {Karata{\c{s}}}, Y. 2022, \mnras, 509, 421

\bibitem[{{Plummer}(1911)}]{1911MNRAS..71..460P}
{Plummer}, H.~C. 1911, \mnras, 71, 460

\bibitem[{{Portail} {et~al.}(2017){Portail}, {Gerhard}, {Wegg}, \&
  {Ness}}]{2017MNRAS.465.1621P}
{Portail}, M., {Gerhard}, O., {Wegg}, C., \& {Ness}, M. 2017, \mnras, 465, 1621

\bibitem[{{Portail} {et~al.}(2015){Portail}, {Wegg}, {Gerhard}, \&
  {Martinez-Valpuesta}}]{2015MNRAS.448..713P}
{Portail}, M., {Wegg}, C., {Gerhard}, O., \& {Martinez-Valpuesta}, I. 2015,
  \mnras, 448, 713

\bibitem[{{Quillen}(2003)}]{2003AJ....125..785Q}
{Quillen}, A.~C. 2003, \aj, 125, 785

\bibitem[{{Raboud} {et~al.}(1998){Raboud}, {Grenon}, {Martinet}, {Fux}, \&
  {Udry}}]{1998A&A...335L..61R}
{Raboud}, D., {Grenon}, M., {Martinet}, L., {Fux}, R., \& {Udry}, S. 1998,
  \aap, 335, L61

\bibitem[{{Ratcliffe} {et~al.}(2023){Ratcliffe}, {Minchev}, {Anders},
  {Khoperskov}, {Guiglion}, {Buck}, {Cunha}, {Queiroz}, {Nitschelm},
  {Meszaros}, {Steinmetz}, {de Jong}, {Nepal}, {Lane}, \&
  {Sobeck}}]{2023MNRAS.525.2208R}
{Ratcliffe}, B., {Minchev}, I., {Anders}, F., {et~al.} 2023, \mnras, 525, 2208

\bibitem[{{Ripepi} {et~al.}(2022){Ripepi}, {Catanzaro}, {Clementini}, {De
  Somma}, {Drimmel}, {Leccia}, {Marconi}, {Molinaro}, {Musella}, \&
  {Poggio}}]{2022A&A...659A.167R}
{Ripepi}, V., {Catanzaro}, G., {Clementini}, G., {et~al.} 2022, \aap, 659, A167

\bibitem[{{Ritchey} {et~al.}(2023){Ritchey}, {Jenkins}, {Shull}, {Savage},
  {Federman}, \& {Lambert}}]{2023ApJ...952...57R}
{Ritchey}, A.~M., {Jenkins}, E.~B., {Shull}, J.~M., {et~al.} 2023, \apj, 952,
  57

\bibitem[{{Robin} {et~al.}(2012){Robin}, {Marshall}, {Schultheis}, \&
  {Reyl{\'e}}}]{2012A&A...538A.106R}
{Robin}, A.~C., {Marshall}, D.~J., {Schultheis}, M., \& {Reyl{\'e}}, C. 2012,
  \aap, 538, A106

\bibitem[{{Ro{\v{s}}kar} {et~al.}(2008){Ro{\v{s}}kar}, {Debattista}, {Quinn},
  {Stinson}, \& {Wadsley}}]{2008ApJ...684L..79R}
{Ro{\v{s}}kar}, R., {Debattista}, V.~P., {Quinn}, T.~R., {Stinson}, G.~S., \&
  {Wadsley}, J. 2008, \apjl, 684, L79

\bibitem[{{Ro{\v{s}}kar} {et~al.}(2012){Ro{\v{s}}kar}, {Debattista}, {Quinn},
  \& {Wadsley}}]{2012MNRAS.426.2089R}
{Ro{\v{s}}kar}, R., {Debattista}, V.~P., {Quinn}, T.~R., \& {Wadsley}, J. 2012,
  \mnras, 426, 2089

\bibitem[{{Saburova} {et~al.}(2018){Saburova}, {Chilingarian}, {Katkov},
  {Egorov}, {Kasparova}, {Khoperskov}, {Uklein}, \&
  {Vozyakova}}]{2018MNRAS.481.3534S}
{Saburova}, A.~S., {Chilingarian}, I.~V., {Katkov}, I.~Y., {et~al.} 2018,
  \mnras, 481, 3534

\bibitem[{{Sanders} {et~al.}(2024){Sanders}, {Kawata}, {Matsunaga}, {Sormani},
  {Smith}, {Minniti}, \& {Gerhard}}]{2024MNRAS.530.2972S}
{Sanders}, J.~L., {Kawata}, D., {Matsunaga}, N., {et~al.} 2024, \mnras, 530,
  2972

\bibitem[{{Sanders} {et~al.}(2019){Sanders}, {Smith}, \&
  {Evans}}]{2019MNRAS.488.4552S}
{Sanders}, J.~L., {Smith}, L., \& {Evans}, N.~W. 2019, \mnras, 488, 4552

\bibitem[{{Schwarz}(1981)}]{1981ApJ...247...77S}
{Schwarz}, M.~P. 1981, \apj, 247, 77

\bibitem[{{Sellwood} \& {Binney}(2002)}]{2002MNRAS.336..785S}
{Sellwood}, J.~A. \& {Binney}, J.~J. 2002, \mnras, 336, 785

\bibitem[{{Sestito} {et~al.}(2008){Sestito}, {Bragaglia}, {Randich},
  {Pallavicini}, {Andrievsky}, \& {Korotin}}]{2008A&A...488..943S}
{Sestito}, P., {Bragaglia}, A., {Randich}, S., {et~al.} 2008, \aap, 488, 943

\bibitem[{{Skowron} {et~al.}(2019){Skowron}, {Skowron}, {Mr{\'o}z}, {Udalski},
  {Pietrukowicz}, {Soszy{\'n}ski}, {Szyma{\'n}ski}, {Poleski}, {Koz{\l}owski},
  {Ulaczyk}, {Rybicki}, \& {Iwanek}}]{2019Sci...365..478S}
{Skowron}, D.~M., {Skowron}, J., {Mr{\'o}z}, P., {et~al.} 2019, Science, 365,
  478

\bibitem[{{Snaith} {et~al.}(2015){Snaith}, {Haywood}, {Di Matteo}, {Lehnert},
  {Combes}, {Katz}, \& {G{\'o}mez}}]{2015A&A...578A..87S}
{Snaith}, O., {Haywood}, M., {Di Matteo}, P., {et~al.} 2015, \aap, 578, A87

\bibitem[{{Snaith} {et~al.}(2014){Snaith}, {Haywood}, {Di Matteo}, {Lehnert},
  {Combes}, {Katz}, \& {G{\'o}mez}}]{2014ApJ...781L..31S}
{Snaith}, O.~N., {Haywood}, M., {Di Matteo}, P., {et~al.} 2014, \apjl, 781, L31

\bibitem[{{Soubiran} {et~al.}(2022){Soubiran}, {Brouillet}, \&
  {Casamiquela}}]{2022A&A...663A...4S}
{Soubiran}, C., {Brouillet}, N., \& {Casamiquela}, L. 2022, \aap, 663, A4

\bibitem[{{Spina} {et~al.}(2022){Spina}, {Magrini}, \&
  {Cunha}}]{2022Univ....8...87S}
{Spina}, L., {Magrini}, L., \& {Cunha}, K. 2022, Universe, 8, 87

\bibitem[{{Tissera} {et~al.}(2022){Tissera}, {Rosas-Guevara}, {Sillero},
  {Pedrosa}, {Theuns}, \& {Bignone}}]{2022MNRAS.511.1667T}
{Tissera}, P.~B., {Rosas-Guevara}, Y., {Sillero}, E., {et~al.} 2022, \mnras,
  511, 1667

\bibitem[{{Trick}(2022)}]{2022MNRAS.509..844T}
{Trick}, W.~H. 2022, \mnras, 509, 844

\bibitem[{{Twarog} {et~al.}(1997){Twarog}, {Ashman}, \&
  {Anthony-Twarog}}]{1997AJ....114.2556T}
{Twarog}, B.~A., {Ashman}, K.~M., \& {Anthony-Twarog}, B.~J. 1997, \aj, 114,
  2556

\bibitem[{{Valenti} {et~al.}(2013){Valenti}, {Zoccali}, {Renzini}, {Brown},
  {Gonzalez}, {Minniti}, {Debattista}, \& {Mayer}}]{2013A&A...559A..98V}
{Valenti}, E., {Zoccali}, M., {Renzini}, A., {et~al.} 2013, \aap, 559, A98

\bibitem[{{Vasiliev}(2019)}]{2019MNRAS.482.1525V}
{Vasiliev}, E. 2019, \mnras, 482, 1525

\bibitem[{{Vislosky} {et~al.}(2024){Vislosky}, {Minchev}, {Khoperskov},
  {Martig}, {Buck}, {Hilmi}, {Ratcliffe}, {Bland-Hawthorn}, {Quillen},
  {Steinmetz}, \& {de Jong}}]{2024MNRAS.528.3576V}
{Vislosky}, E., {Minchev}, I., {Khoperskov}, S., {et~al.} 2024, \mnras, 528,
  3576

\bibitem[{{Vorobyov}(2006)}]{2006MNRAS.370.1046V}
{Vorobyov}, E.~I. 2006, \mnras, 370, 1046

\bibitem[{{Wegg} \& {Gerhard}(2013)}]{2013MNRAS.435.1874W}
{Wegg}, C. \& {Gerhard}, O. 2013, \mnras, 435, 1874

\bibitem[{{Wegg} {et~al.}(2015){Wegg}, {Gerhard}, \&
  {Portail}}]{2015MNRAS.450.4050W}
{Wegg}, C., {Gerhard}, O., \& {Portail}, M. 2015, \mnras, 450, 4050

\bibitem[{{Weinberg}(1992)}]{1992ApJ...384...81W}
{Weinberg}, M.~D. 1992, \apj, 384, 81

\bibitem[{{Wheeler} {et~al.}(2022){Wheeler}, {Abril-Cabezas}, {Trick},
  {Fragkoudi}, \& {Ness}}]{2022ApJ...935...28W}
{Wheeler}, A., {Abril-Cabezas}, I., {Trick}, W.~H., {Fragkoudi}, F., \& {Ness},
  M. 2022, \apj, 935, 28

\bibitem[{{Willett} {et~al.}(2023){Willett}, {Miglio}, {Mackereth},
  {Chiappini}, {Lyttle}, {Elsworth}, {Mosser}, {Khan}, {Anders}, {Casali}, \&
  {Grisoni}}]{2023MNRAS.526.2141W}
{Willett}, E., {Miglio}, A., {Mackereth}, J.~T., {et~al.} 2023, \mnras, 526,
  2141

\bibitem[{{Wozniak}(2020)}]{2020ApJ...889...81W}
{Wozniak}, H. 2020, \apj, 889, 81

\bibitem[{{Xiang} \& {Rix}(2022)}]{2022Natur.603..599X}
{Xiang}, M. \& {Rix}, H.-W. 2022, \nat, 603, 599

\bibitem[{{Yong} {et~al.}(2005){Yong}, {Carney}, \& {Teixera de
  Almeida}}]{2005AJ....130..597Y}
{Yong}, D., {Carney}, B.~W., \& {Teixera de Almeida}, M.~L. 2005, \aj, 130, 597

\end{thebibliography}

\begin{appendix}

\section{The median or the mode}\label{A}

\begin{figure}
\includegraphics[width=9cm]{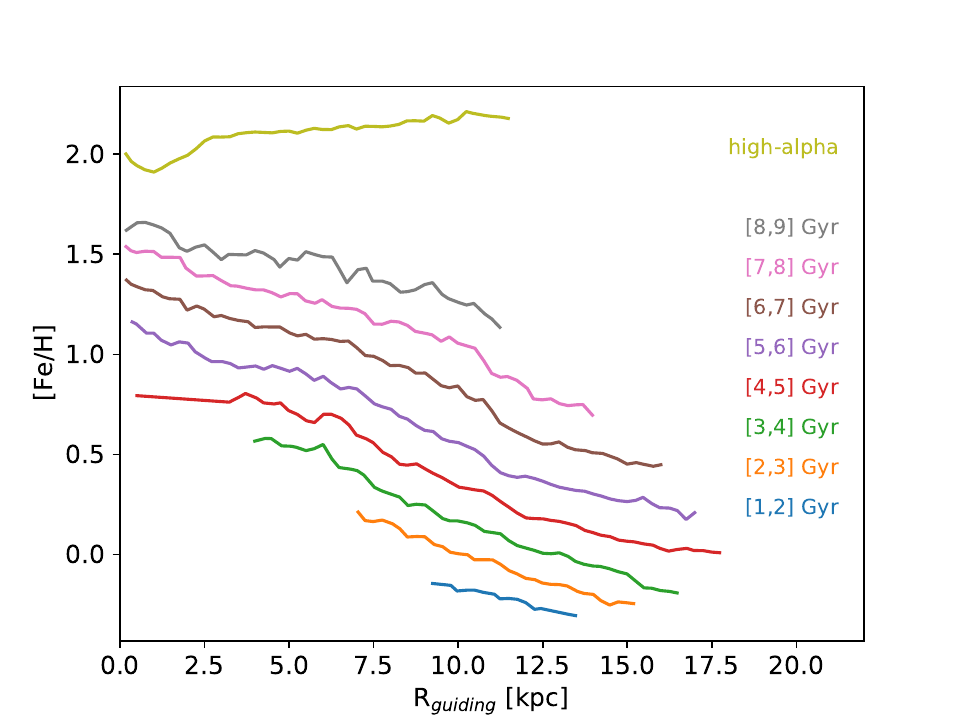}
\caption{Metallicity profiles as a function of the guiding radius and age range. This figure is the same as Fig.\ref{fig:FeH_gradient_with_age}, except that the metallicity indicator is the median, not the mode.
}
\label{fig:feh_gradient_age_mode}
\end{figure}

\begin{figure}
\includegraphics[width=8cm]{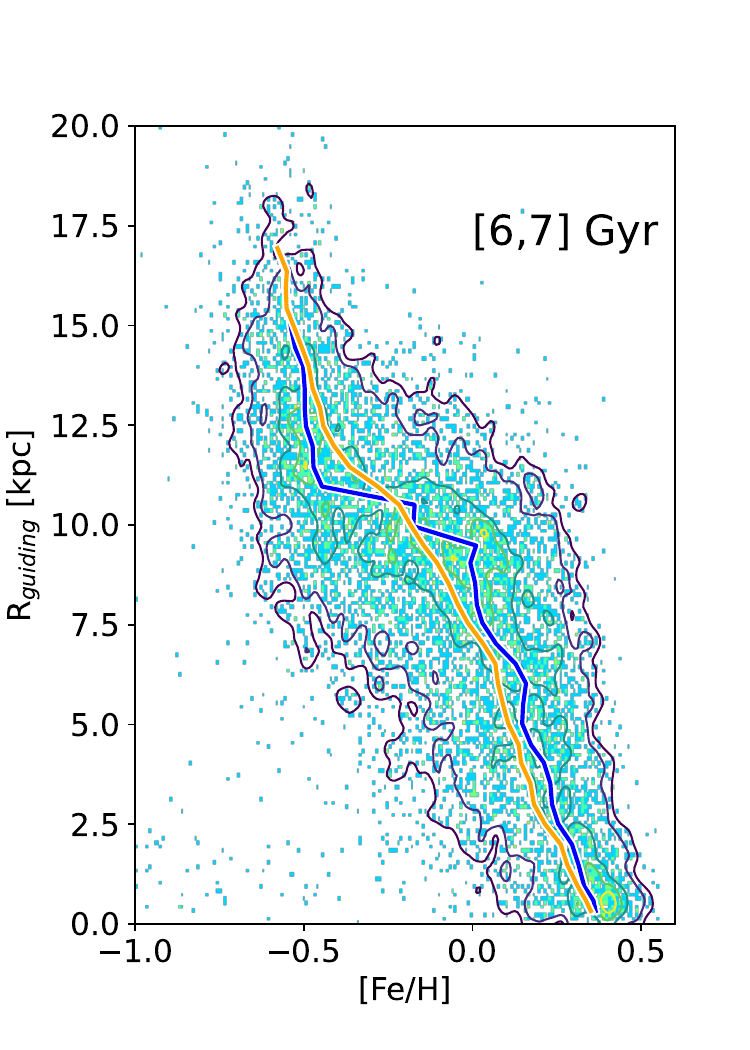}
\caption{Metallicity-guiding radius distribution for stars with ages between 6 and 7 Gyr. The blue curve shows the mode of the metallicity distribution as a function of the guiding radius, the orange curve shows the median.}
\label{fig:median_or_mode}
\end{figure}

Most studies use the median as the representative metallicity indicator at a given radius. We have chosen the mode of the metallicity distribution, because the former has a tendency to erase the features observed in the metallicity profile. This is illustrated by the fig. \ref{fig:median_or_mode}. The plot shows the metallicity-guiding radius distribution for stars in the age interval 6-7 Gyr, with the metallicity profile obtained by the median (in orange) and the mode (in blue) overplotted, and illustrating the difference between the two. Features in the distribution tend to be smoothed out with the median.

\section{The median absolute deviation of the metallicity of the youngest stars and ISM}\label{B}

\begin{figure}
\includegraphics[width=9cm]{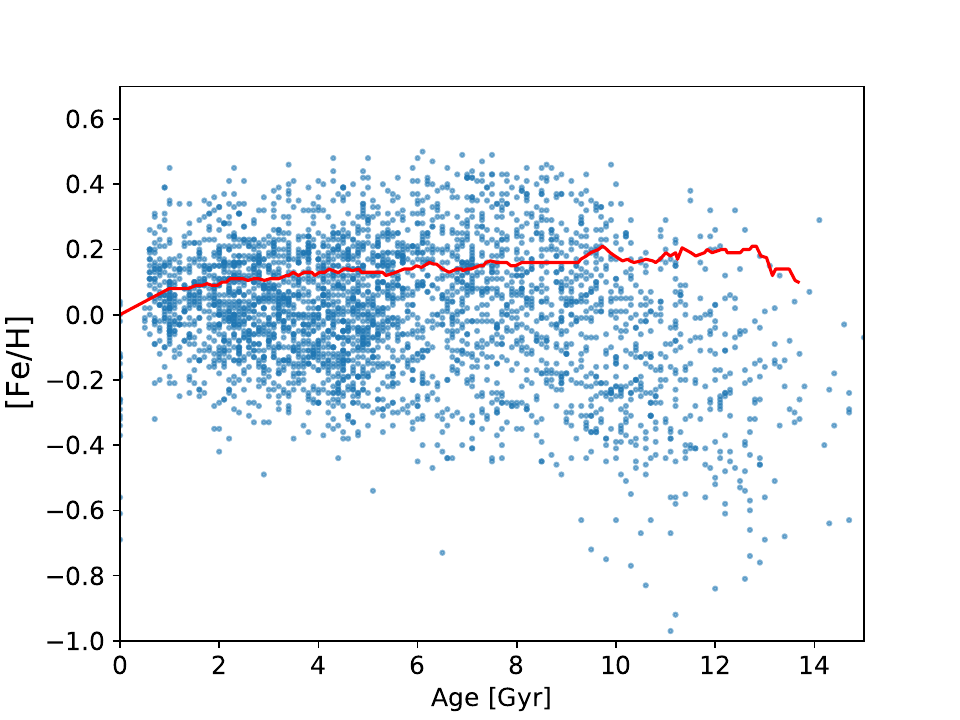}
\caption{Age-metallicity distribution for giants in the \textit{Kepler} sample from \cite{2021A&A...645A..85M}.}
\label{fig:age_met_kepler}
\end{figure}

Various tracers can be used to estimate the metallicity dispersion of stars for which mixing must be negligible. We can rely on known young stars, such as cepheids, young open clusters, field stars whose age can be determined, or metallicity of the gas in the ISM.
We start with a sample of red giants in the \textit{Kepler} field, which metallicities come from APOGEE and age estimates from \cite{2021A&A...645A..85M}.
Figure\ref{fig:age_met_kepler} shows the age-metallicity distribution for this sample, with metallicity coming from the APOGEE catalogue and the age from asteroseismic values.
The curve on the plot shows the MAD determined on stars in 1~Gyr age intervals. It evolves from 0.08~dex at age$<$2~Gyr to 0.2~dex at 13~Gyr. There are 82 stars with age younger than 1 Gyr, and have a MAD of 0.075~dex within a distance of 2~kpc. Enlarging the selection to 1.5~Gyr, we find 190 stars with a MAD of 0.08~dex. 

To estimate the dispersion from open clusters, we use the catalogue published by \cite{2023A&A...669A.119M}, based on the \textit{Gaia}-ESO survey.
It contains 24 clusters younger than 1.5 Gyr with guiding radii between 6 and 10~kpc yielding a MAD of 0.07 dex. 
An estimate can be obtained from Classical Cepheids, which have lifetimes shorter than about a few 10$^8$~yr.
The catalogue from \cite{2023A&A...678A.195D} contains 379 objects, 124 of which are within 2~kpc. The MAD of the metallicity of this sample is 0.044~dex.
Finally, \cite{2023ApJ...952...57R} measured the metallicity of the neutral gas in 84 lines of sight. Selecting those limited to less than 2~kpc, their number reduces to 54, giving a metallicity dispersion with a MAD of 0.069 dex. These last measurements are compatible with the oxygen abundance in 44 HII regions from \cite{2020MNRAS.496.1051A,2021MNRAS.502..225A}, which have a MAD of 0.067~dex.
Overall, all these measurements are consistent. In the case of APOGEE, the uncertainties amount to about 0.04 dex \citep{2022A&A...663A...4S}. For a measured dispersion of 0.08~dex, this means an intrinsic (MAD) dispersion of 0.07~dex.

\section{Simulation characteristics}\label{C}

\begin{figure}[ht]
\includegraphics[width=15cm]{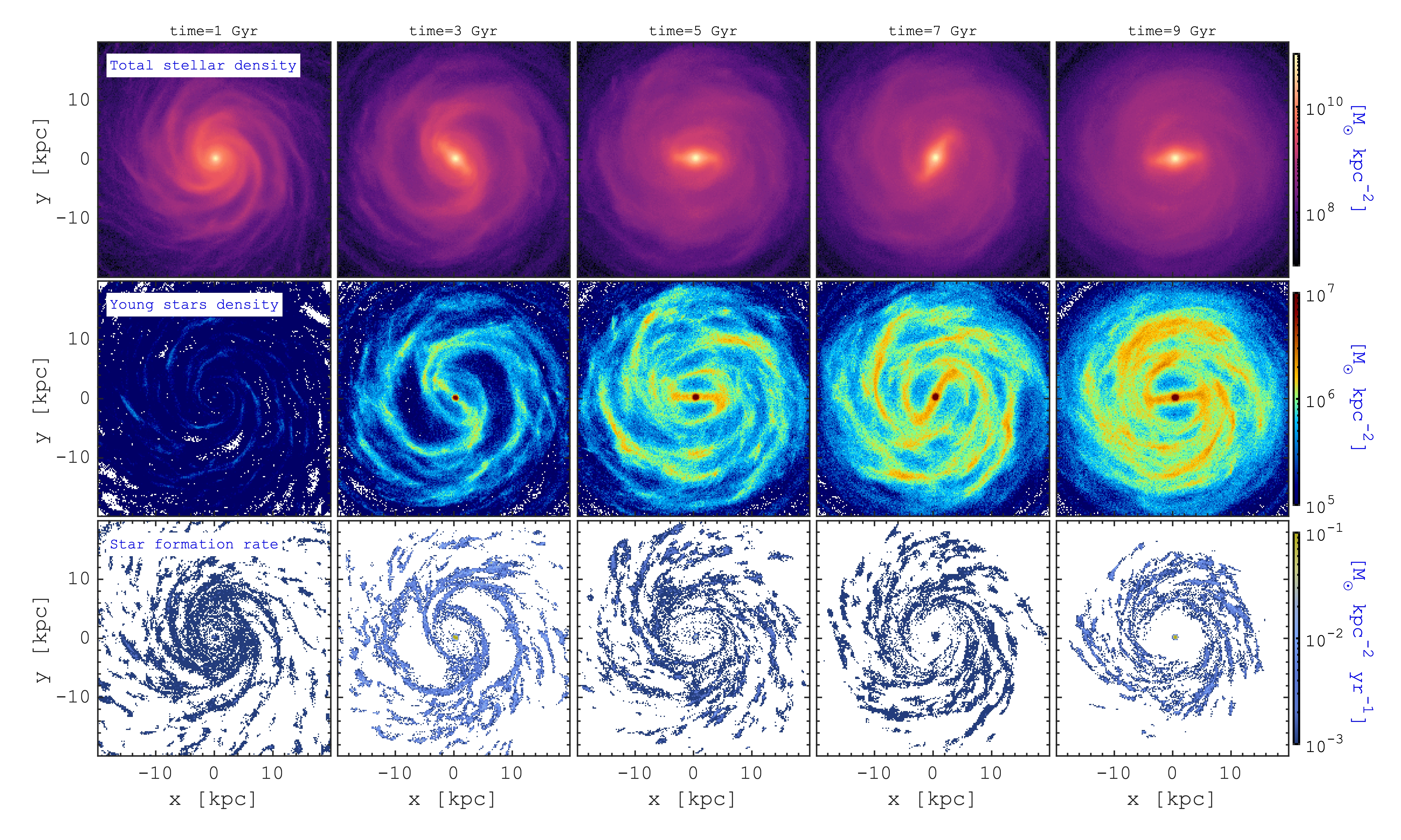}
\caption{Face-on density maps (M$_{\odot}$.kpc$^{-2}$) of the total stellar (top frames) young stars (middle frames) components and star formation rate (bottom frames) during the 10 Gyr of evolution.}
\label{fig:simulation_main}
\end{figure}

\begin{figure}[ht]
\includegraphics[width=15cm]{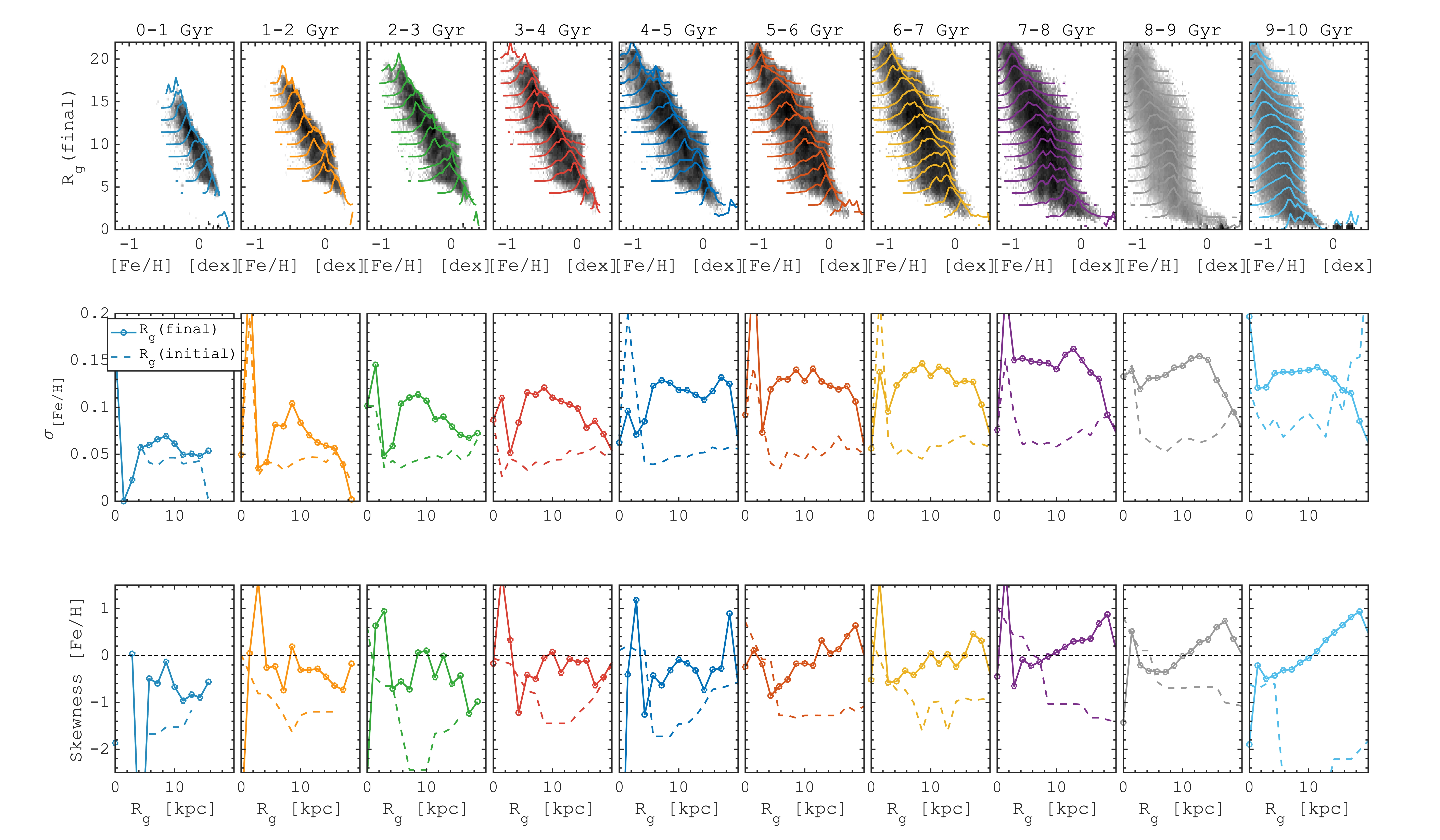}
\caption{Top row: Final guiding radius as a function of the metallicity of the stars, in different age intervals in the form of a scaled grey density distribution and histograms. Middle row: Metallicity dispersion as a function of the initial and final guiding radii. Bottom row: Skewness of the metallicity distribution as a function of the initial and final guiding radii.
}
\label{fig:simulation_mdf}
\end{figure}

\end{appendix}
\end{document}